%% file: qldpc_ghz_quantum_final.tex
\documentclass[a4paper,onecolumn,11pt,accepted=2024-01-11]{quantumarticle}
\pdfoutput=1

\usepackage[T1]{fontenc}

\usepackage{url}
\usepackage{ifthen}
\usepackage{braket}
\usepackage{cite}
\usepackage{doi}
\usepackage[cmex10]{amsmath} 
\usepackage{longtable}
\usepackage{makecell}
\usepackage{algpseudocode}
\usepackage[linesnumbered,ruled,vlined]{algorithm2e}
\usepackage{color}
\usepackage[normalem]{ulem}
                             
\usepackage{amssymb,amsthm}
\usepackage{mathtools,bbm}  
\usepackage{IEEEtrantools}                           
\usepackage{xifthen,xparse}
\usepackage{hyperref}  
\usepackage{float}
\usepackage{hhline}
\usepackage{caption}
\usepackage{subcaption}
\usepackage{stmaryrd}
\usepackage{booktabs, tabularx, ltablex, xltabular}
\newcolumntype{C}{>{\centering\arraybackslash}X}
\usepackage[table]{xcolor}
\usepackage{tikz,pgfplots}
\pgfplotsset{compat=newest}

\usepackage{rotating}

\usepackage[numbers]{natbib}

\usepackage{textcomp}

\allowdisplaybreaks

\newtheorem{theorem}{Theorem}

\newtheorem{remark}[theorem]{Remark}
\newtheorem{lemma}[theorem]{Lemma}

\newtheorem{example}{Example}

\newenvironment{example*}
  {\addtocounter{example}{-1}\example}
  {\endexample}

\newcommand{\ghzmap}[1]{\widehat{#1}}

\newcommand{\dket}[1]{\left\lvert #1 \right\rangle}
\newcommand{\dbra}[1]{\left\langle #1 \right\rvert}
\NewDocumentCommand\dketbra{+m+g}{%
  \IfNoValueTF{#2}
    {\left\lvert #1 \right\rangle \left\langle #1 \right\vert}
  {\left\lvert #1 \right\rangle \left\langle #2 \right\rvert}%
}
\NewDocumentCommand\dbraket{+m+g}{%
  \IfNoValueTF{#2}
    {\left\langle #1 \vert #1 \right\rangle}
  {\left\langle #1 \vert #2 \right\rangle}%
}

\newcommand{\etal}{\emph{et al.~}}
\newcommand{\MCC}{\mathcal{C}}

\newcommand{\syminn}[2]{\langle #1, #2 \rangle_{\text{s}}}
\newcommand{\llbr}{[\![}
\newcommand{\rrbr}{]\!]}

\def\mathcolor#1#{\@mathcolor{#1}}
\def\@mathcolor#1#2#3{%
  \protect\leavevmode
  \begingroup
    \color#1{#2}#3%
  \endgroup
}

\begin{document}

\title{Entanglement Purification with Quantum LDPC Codes and Iterative Decoding}

\author{Narayanan Rengaswamy}
\affiliation{Department of Electrical and Computer Engineering,
           University of Arizona, Tucson, Arizona 85721, USA}
\email{narayananr@arizona.edu}
\orcid{0000-0002-2369-3159}
\thanks{A shorter version of this work was presented at the 2023 International Symposium on Topics in Coding (\url{https://ieeexplore.ieee.org/abstract/document/10273456}).}
\author{Nithin Raveendran}
\affiliation{Department of Electrical and Computer Engineering,
           University of Arizona, Tucson, Arizona 85721, USA}
\email{nithin@arizona.edu}
\orcid{0000-0002-1024-8099}
\author{Ankur Raina}
\affiliation{Department of Electrical Engineering and Computer Sciences, Indian Institute of Science Education and Research, Bhopal, Madhya Pradesh 462066, India}
\email{ankur@iiserb.ac.in}
\orcid{0000-0002-9022-3595}
\author{Bane Vasi{\'c}}
\affiliation{Department of Electrical and Computer Engineering,
           University of Arizona, Tucson, Arizona 85721, USA}
\email{vasic@ece.arizona.edu}
\orcid{0000-0003-2365-4106}
\maketitle

\begin{abstract}
Recent constructions of quantum low-density parity-check (QLDPC) codes provide optimal scaling of the number of logical qubits and the minimum distance in terms of the code length, thereby opening the door to fault-tolerant quantum systems with minimal resource overhead.
However, the hardware path from nearest-neighbor-connection-based topological codes to long-range-interaction-demanding QLDPC codes is likely a challenging one.
Given the practical difficulty in building a monolithic architecture for quantum systems, such as computers, based on optimal QLDPC codes, it is worth considering a \emph{distributed} implementation of such codes over a network of interconnected medium-sized quantum processors.
In such a setting, all syndrome measurements and logical operations must be performed through the use of high-fidelity shared entangled states between the processing nodes.
Since probabilistic many-to-1 distillation schemes for purifying entanglement are inefficient, we investigate quantum error correction based entanglement purification in this work.
Specifically, we employ QLDPC codes to distill GHZ states, as the resulting high-fidelity logical GHZ states can interact directly with the code used to perform distributed quantum computing (DQC), e.g. for fault-tolerant Steane syndrome extraction.
This protocol is applicable beyond the application of DQC since entanglement distribution and purification is a quintessential task of any quantum network.
We use the min-sum algorithm (MSA) based iterative decoder with a sequential schedule for distilling $3$-qubit GHZ states using a rate $0.118$ family of lifted product QLDPC codes and obtain an input threshold of $\approx 0.7974$ under i.i.d. single-qubit depolarizing noise.
This represents the best threshold for a yield of $0.118$ for any GHZ purification protocol.
Our results apply to larger size GHZ states as well, where we extend our technical result about a measurement property of $3$-qubit GHZ states to construct a scalable GHZ purification protocol.
Our software is available at: \url{https://github.com/nrenga/ghz_distillation_qec/tree/main/qldpc-ghz_protocol_II} and \url{https://zenodo.org/record/8284903}.
\end{abstract}

\section{Introduction}
\label{sec:intro}

\IEEEPARstart{A}{dvances} in quantum technologies are happening at a breathtaking pace and these will lead to exciting applications in quantum computing, networking, sensing, security, and more.
Quantum networking is a common theme in all these applications, such as for employing quantum key distribution to enhance digital security, for connecting quantum sensors together to enable a quadratic gain in sensing precision, and for distributing quantum computation among multiple quantum processors to relax the burden of building enormous monolithic quantum computers.
This work is primarily motivated by the latter role of quantum networking.
Indeed, for fault-tolerant quantum computing (FTQC), the best codes for scalability are the recently proposed constructions of quantum low-density parity-check (QLDPC) codes~\cite{Hastings-stoc21,Panteleev-it21,Breuckmann-it21,Breuckmann-prxq21,Panteleev-stoc22,Leverrier-arxiv22}.
They provide optimal scaling of the code parameters, i.e., the number of logical qubits and the minimum distance, with respect to the length of the code, and thereby form promising candidates for FTQC with minimal resource overhead.
While topological codes such as the surface code are also QLDPC codes, they encode only a fixed number of logical qubits even with diverging code size and have suboptimal scaling of the minimum distance.
However, they just require nearest-neighbor connections to build in hardware, whereas these optimal QLDPC codes require many long-range connections.
Even though the LDPC property means that each stabilizer check involves only a fixed number of qubits and similarly each qubit is only involved in a fixed number of checks, both irrespective of the code size, there are a large number of connections between checks and qubits that are non-local geometrically~\cite{Baspin-quantum22}.
Thus, it becomes very challenging to build such codes in practice for several technologies such as superconducting qubits.

Given such practical constraints, it becomes very relevant and interesting to explore \emph{Distributed Quantum Computing (DQC)}: a distributed realization of these QLDPC codes where multiple interconnected medium-sized quantum processors each store a subset of qubits and coordinate processing through the means of a classical compute node.
Naturally, this means that all the logical operations and syndrome measurements on the coded qubits are now non-local, i.e., must involve multiple nodes.
Such an architecture was explored by Nickerson \etal~\cite{Nickerson-ncomms13} even a decade ago, but in the context of the surface code.
The solution to perform non-local operations is to share high-fidelity entangled Bell and GHZ states among the nodes, perform local gates between code qubits and these ancillary entangled qubits, and pool the classical measurement results across nodes to assess the state of the qubits.
For example, in the case of the surface code with each node possessing only one code qubit, each $4$-qubit syndrome measurement will involve one CNOT per node between the code qubit and one of the $4$ qubits of an ancillary GHZ state shared between the nodes; this is followed by a single-qubit Pauli measurement on the ancillary qubit and classical communication of the result with other nodes. 
The authors proposed to produce high-fidelity $4$-qubit GHZ states by generating Bell pairs between pairs of nodes and then ``fusing'' them to form the GHZ state.
The process involved multiple rounds of simple probabilistic purification of the entangled state, which is in general inefficient since the number of consumed Bell pairs can be very large (and uncertain).
While their hand-designed purification schemes have been extended by algorithmic procedures recently~\cite{Krastanov-quantum19,deBone-tqe20}, the approach still suffers from this inefficiency arising from the heralded nature of the protocol.
We will discuss comparisons to past work on GHZ purification after we present our results in the next section.

Our goal in this paper is to investigate a principled and systematic procedure to purify (or distill) GHZ states using quantum error correcting codes (QECCs).
If one can use the same QLDPC codes that DQC will employ for FTQC (``compute code'') to also store \emph{logical} GHZ ancillary states, then these can potentially be directly interacted with the compute code for performing fault-tolerant (e.g., Steane) syndrome extraction and measurement-based methods for logical operations.
Thus, it is very pertinent to develop a scalable GHZ purification protocol using these optimal QLDPC codes (``purification code'').
While DQC is a key motivation, such a protocol serves a much wider purpose, since entanglement generation, distribution and purification form the cornerstone of quantum networking.
For efficient and scalable quantum networks, one must necessarily deploy quantum repeaters whose primary function is to help entangle different subsets of parties in the network.
In the long-run, third generation quantum repeaters will use quantum error correction for entanglement purification~\cite{Muralidharan-scirep16}.
Such repeater nodes, and other nodes of the network that are not quantum computers, will still need to possess a  fault-tolerant quantum memory to generate and store (shares of) high-fidelity entangled states.
Therefore, if the compute nodes will deploy QLDPC codes, then QLDPC purification codes could potentially unify the functioning of different parts of the network and enable seamless integration.

\section{Main Results and Discussion}

Entanglement purification is a well-studied problem in quantum information, where one typically starts with $n$ copies of a noisy Bell pair, or a general mixed state, and distills $k$ Bell pairs of higher fidelity~\cite{Bennett-prl96}.
Several teams of researchers have worked on this problem, and the contributions range from fundamental limits~\cite{Bennett-prl96,Bennett-pra96,Miyake-prl05,Dur-rpp07,Leditzky-it17,Fang-it19} to simple and practical protocols~\cite{Bennett-pra96,Wilde-isit10,Rozpedek-pra18,Krastanov-quantum19}. 
Of course, if one can distill Bell pairs, then these can be ``fused'' in sequence to entangle multiple parties, but direct distillation of an entangled resource between all parties can be more efficient~\cite{Murao-pra98}.
Some purification schemes involve two-way communications between the involved parties while others only need one-way communication.
We focus on one-way schemes in this paper.
The connection between one-way entanglement purification protocols (1-EPPs) and QECCs was established by Bennett \etal in 1996~\cite{Bennett-pra96}.
They showed that any QECC can be converted into a 1-EPP (and vice-versa).
This framework enables systematic $n$-to-$k$ protocols where the rate and average output fidelity are directly a function of the QECC rate and decoding performance, respectively.
Since the recently constructed QLDPC codes have asymptotically constant rate and linear distance scaling with code size~\cite{Hastings-stoc21,Panteleev-it21,Breuckmann-it21,Breuckmann-prxq21,Panteleev-stoc22,Leverrier-arxiv22}, our work paves the way for high-rate high-fidelity entanglement distillation.

\subsection{Purifying Bell Pairs with QLDPC Codes}

\begin{figure}

\centering

\begin{subfigure}[t]{0.75\textwidth}
\centering
\scalebox{0.8}{%
\input{figures/LP118_12_16_20_MSAseqvars80.tex}
}
\end{subfigure}

\begin{subfigure}[t]{0.75\textwidth}
\centering
\scalebox{0.8}{%
\input{figures/LP118_12_16_20_MSAseqvars80_threshold.tex}
}
\end{subfigure}

\caption{\label{fig:LP118_MSAseqvars80}(top) The performance of a family of lifted product QLDPC codes with asymptotic rate $0.118$ using the sequential schedule of the min-sum algorithm (MSA) based decoder. Each data point is obtained by counting $100$ logical errors. (bottom) The threshold is about $10.6$-$10.7\%$. These results apply to Bell pair purification, up to a rescaling of the depolarizing probabilities.}
\end{figure}

In 2007, Wilde \etal \cite{Wilde-isit10} showed that any classical convolutional code can be used to distill Bell pairs via their entanglement assisted 1-EPP scheme.
In the development of this scheme, they mention a potentially different method to use a QECC for performing 1-EPP~\cite[Section II-D]{Wilde-isit10} (without entanglement assistance), compared to the protocol by Bennett \etal 
Initially, Alice generates $n$ perfect Bell pairs locally, marks one qubit of each pair as `A' and the other as `B', and measures the stabilizers of her chosen $\llbr n,k \rrbr$ code on qubits `A'.
Due to the ``transpose'' property of Bell states, this simultaneously projects qubits `B' onto an equivalent code (see Appendix~\ref{sec:bell_state_identity}).
Then, she performs a local Pauli operation on qubits `A' to fix her obtained syndrome, shares her code stabilizers and syndrome with Bob through a noiseless classical channel, and sends qubits `B' to Bob over a noisy Pauli channel.
Using the transpose property, Bob measures the appropriate code stabilizers on qubits `B', and corrects channel errors by combining his syndrome with Alice's syndrome.
Finally, Alice and Bob decode their respective qubits, i.e., invert the encoding unitary (see Appendix~\ref{sec:stabilizer_codes}), to convert the $k$ logical Bell pairs into $k$ physical Bell pairs.
Since the code corrects some errors, on average the output Bell pairs are of higher fidelity than the initial $n$ noisy ones.

As our first contribution, we elucidate this protocol for general stabilizer codes~\cite{Gottesman-phd97,Calderbank-it98} through the lens of the stabilizer formalism~\cite{Gottesman-icgtmp98}, using the $5$-qubit perfect code~\cite{Laflamme-prl96,Gottesman-phd97} as an example.
This approach clarifies many details of the protocol, especially from an error correction standpoint, and helps adapt it to different scenarios.
For the performance of the protocol, note that any error on Alice's qubits can be mapped into an equivalent error on Bob's qubits using the transpose property, in effect increasing the error rate on Bob's qubits.
Therefore, since only Bob corrects errors in this protocol, the failure rate of the protocol is the same as the logical error rate (LER) of the code on the depolarizing channel, with an effective channel error rate that accounts for errors on Alice's qubits as well as Bob's qubits (as long as they do not amount to a Bell state stabilizer together).
If errors only happen on Bob's qubits, then the failure rate of the protocol is identical to the logical error rate of the code.
For all simulations in this work, we consider a rate $0.118$ family of lifted product (LP118) QLDPC codes decoded using the sequential schedule of the min-sum algorithm (MSA) based iterative decoder with normalization factor $0.8$ and maximum number of iterations set to $100$~\cite{Raveendran-quantum21,Raveendran-qce22}.
The LER of this code-decoder pair is shown in Fig.~\ref{fig:LP118_MSAseqvars80}, where we see that the threshold is about $10.6$-$10.7\%$.
Since the fidelity is one minus the depolarizing probability, this translates to an input fidelity threshold of about $89.3$-$89.4\%$.
Also, even with $n=544$, the LER is $\approx 10^{-6}$ at depolarizing rate $10^{-2}$.
Again, note that these curves can be interpreted as the performance of Bell pair purification when only Bob's qubits are affected by errors.

\subsection{New Protocols to Purify GHZ States with QLDPC Codes}

\begin{sidewaysfigure}
\centering

\includegraphics[scale=0.85,keepaspectratio]{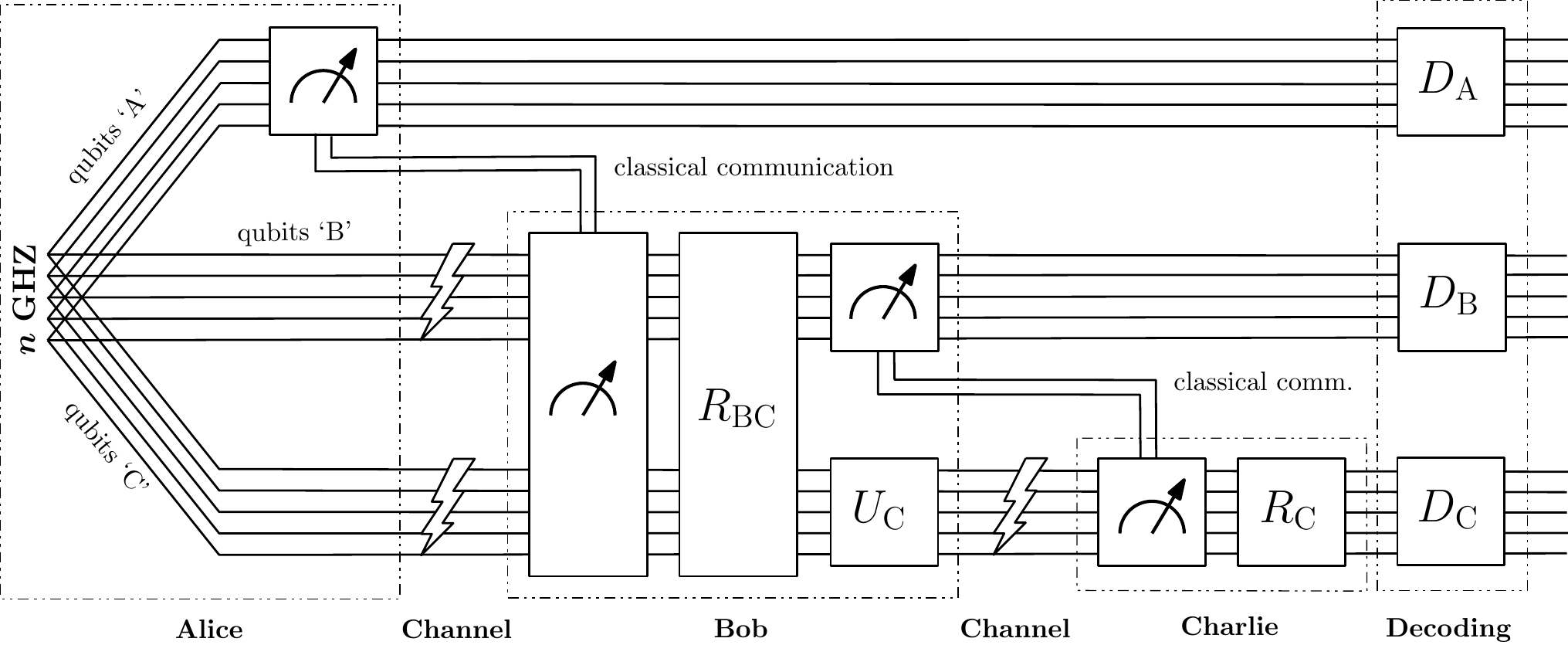}

\caption{\label{fig:GHZprotocol1} Protocol I for GHZ purification using stabilizer codes. Alice generates $n$ copies of the ideal $3$-qubit GHZ state and marks one qubit of each triple as `A', another as `B', and the third as `C'. She measures the stabilizers of the code on the $n$ qubits marked `A' and classically communicates the results through a noiseless channel to Bob. She also sends all the remaining $2n$ qubits to him. First, Bob uses the ``GHZ-map'' to measure stabilizers of the $2n$-qubit code induced by Alice's code and uses the results to correct errors on qubits `B' and `C'. Second, if the code is not CSS, then he performs a suitable diagonal Clifford $U_C$ on qubits `C'. Thirdly, he measures the stabilizers of Alice's $n$-qubit code on qubits `B' in order to impose the same code on qubits `B' and `C'. Lastly, he communicates the results to Charlie over a noiseless channel and also sends qubits `C' to him. Charlie measures the stabilizers of Alice's code to detect and correct errors on qubits `C'. Finally, all three parties invert the encoding to convert the logical GHZ states to physical GHZ states. The scheme is suited for a linear network of three parties, but notice the asymmetrically larger burden on Bob, which makes the protocol less scalable to larger GHZ states.}

\end{sidewaysfigure}

\textbf{Protocol I:}\ \ 
Given these insights, we proceed to investigate the purification of GHZ states.
As in the Wilde \etal protocol, we consider only local operations and one-way classical communications (LOCC), and assume that these are noiseless.
The key technical insight necessary to construct the protocol is the GHZ-equivalent of the transpose property of Bell pairs.
Given $n$ copies of the GHZ state, whose three subsystems are marked `A', `B' and `C', we find that applying a matrix on qubits `A' is equivalent to applying a ``stretched'' version of the matrix on qubits `B' and `C' together (see Lemma~\ref{lem:ghz_state_identity}).
We call this mapping to the stretched version of the matrix the \emph{GHZ-map}, and prove that it is an \emph{algebra homomorphism}~\cite{Dummit-2004}, i.e., linear, multiplicative, and hence projector-preserving.
Recollect from the Bell pair purification setting that we are interested in measuring stabilizers on qubits `A' and understanding their effect on the remaining qubits.
Using the properties of the GHZ-map, we show that it suffices to consider only the simple case of a single stabilizer.
With this great simplification, we prove that imposing a given $\llbr n,k,d \rrbr$ stabilizer code on qubits `A' simultaneously imposes a certain $\llbr 2n, k, d' \rrbr$ stabilizer code jointly on qubits `B' and `C'.
By performing diagonal Clifford operations on qubits `C', which commutes with any operations on the other qubits, one can vary the distance $d'$ of the induced `BC' code.
Then, we use this core technical result to devise a natural protocol that purifies GHZ states using any stabilizer code (``Protocol I'', see Fig.~\ref{fig:GHZprotocol1} and Algorithm~\ref{algo:algo_ghz}).

We perform simulations on the $\llbr 5,1,3 \rrbr$ perfect code and compare the protocol failure rate to the LER of the code on the depolarizing channel, both using a maximum-likelihood decoder. In terms of error exponents, we show that it is always better for Bob to perform a local diagonal Clifford operation on Charlie's qubits, rather than Alice doing the same.
We support the empirical observation with an analytical argument on the induced BC code and Charlie's code.
Finally, we finish by showing that the average output $k$-qubit density matrix of the protocol is diagonal in the GHZ-basis, and its fidelity is directly dictated by the protocol's failure rate.
While the scheme is suited for a linear network of three parties, it is obvious that there is an asymmetrically larger burden on Bob, which makes the protocol less scalable to larger GHZ states.
Nevertheless, we think that this protocol still has pedagogical value in understanding the implications of the new insight on GHZ states.

\vspace{5pt}

\begin{figure}
\centering

\includegraphics[scale=1.1,keepaspectratio]{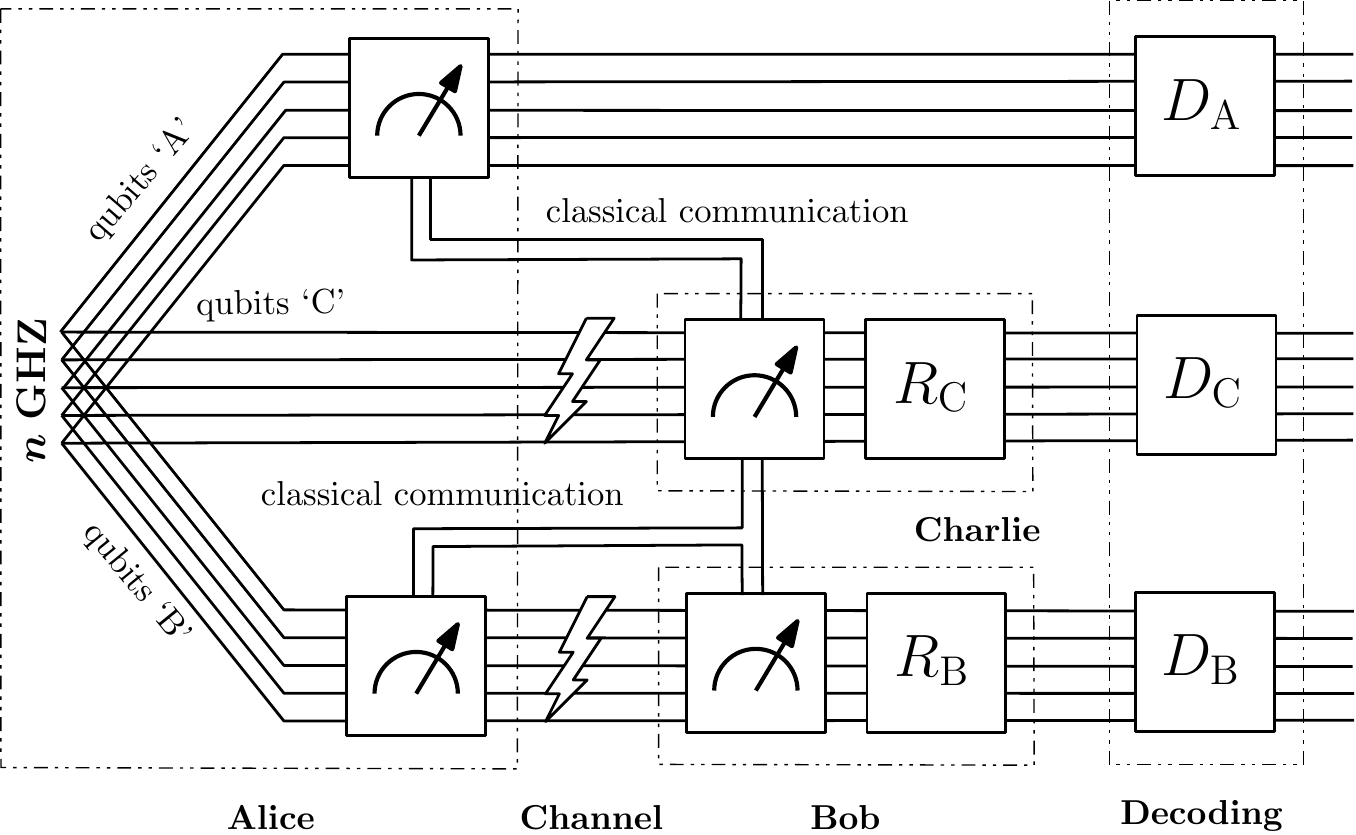}

\caption{\label{fig:GHZprotocol2} Protocol II for GHZ purification using CSS codes. The protocol can be extended to general stabilizer codes through additional diagonal Clifford operations as in Protocol I. Alice generates $n$ copies of the ideal $3$-qubit GHZ state and marks one qubit of each triple as `A', another as `B', and the third as `C'. She measures the stabilizers of the QLDPC code on qubits `A' and classically communicates the results through a noiseless channel to Charlie. She also uses the results to appropriately measure stabilizers of (a potentially equivalent) QLDPC code on qubits 'B' and communicates these results to both Bob and Charlie, again through noiseless classical channels. She sends qubits `B' to Bob and qubits `C' to Charlie. Finally, both Bob and Charlie make stabilizer measurements, correct errors, and then all three parties invert the encoding to convert the logical GHZ states to physical GHZ states. Note that Bob and Charlie can perform their operations asynchronously.}

\end{figure}

\textbf{Protocol II:}\ \ 
Motivated by this drawback, we devise an improved protocol that avoids the additional $2n$-qubit measurements of Protocol I.
The new scheme is depicted in Fig.~\ref{fig:GHZprotocol2} and described in Algorithm~\ref{algo:algo_ghz_2} for CSS codes.
The protocol can be extended to general stabilizer codes through additional diagonal Clifford operations as in Protocol I but, for simplicity, we focus on CSS (QLDPC) codes here.
It will also be interesting to investigate if there are any potential gains from employing non-CSS stabilizer codes in entanglement purification, because CSS codes are known to be optimal for certain aspects of fault-tolerant quantum computing~\cite{Rengaswamy-jsait20}.
When Alice measures stabilizers on qubits `A', the new GHZ property still implies that there is a $2n$-qubit code automatically induced on qubits `B' and `C' together.
In order to split that code into individual codes on qubits `B' and `C', Alice performs a second round of the same ($n$-qubit) stabilizer measurements but this time on qubits `B'.
This enables Bob and Charlie to measure the same stabilizers on their respective qubits and correct errors induced by the channel on qubits `B' and `C', respectively.
The flow of the protocol is naturally applicable when Alice is connected to both Bob and Charlie but those two parties are not connected directly.
But we emphasize that the protocol is scalable and we summarize its extension to larger GHZ states with larger number of parties connected by any network topology; \emph{the key requirement is that the qubits of a recipient over a network edge have already been measured and projected to the code subspace before those qubits are sent over the edge.}

In Fig.~\ref{fig:GHZsimple_LP118_MSAseqvars80} we report simulation results for Protocol II on $3$-qubit GHZ states using the same LP118 code family and MSA decoder as in Fig.~\ref{fig:LP118_MSAseqvars80}.
All data points except the first one on each curve (for depolarizing rate $0.09$) were computed by collecting close to $10^4$ logical errors.
We observe that the threshold ($\approx 10.7\%$) is very close to the single decoder case in Fig.~\ref{fig:LP118_MSAseqvars80}, which is reassuring since the GHZ protocol needs both Bob and Charlie to run decoders.
In terms of fidelity, unlike the Bell pair case, \emph{two} qubits of each GHZ state (i.e., those marked `B' and `C') undergo depolarizing noise, so the input fidelity threshold is $(1-p)^2 \approx 0.7974$ where $p \approx 10.7\%$.
Note that this is for a yield of $0.118$, which is the asymptotic rate of the LP118 QLDPC code family.
Technically, one must multiply the code rate with one minus the protocol failure rate to get the exact yield, but we assume that in practice we operate well away from the threshold where failure rates are orders of magnitude smaller (see Fig.~\ref{fig:LP118_MSAseqvars80} for reference).
However, the logical error rates are significantly higher than those in Fig.~\ref{fig:LP118_MSAseqvars80}.
This is likely due to the fact that both decoders must succeed for the protocol to not fail.
Note that there can be situations where an error on Alice's qubit cancels the errors on Bob's and Charlie's due to the new GHZ property.
But it is unclear whether these have a significant effect on the protocol performance.
We plan to study this carefully in future work because it is undesirable for failure rates to increase as we scale the protocol to larger number of parties.

The implementation of our protocol is available on GitHub and archived on Zenodo~\cite{Rengaswamy-zenodo23}.


\begin{figure}
\centering

\begin{subfigure}[b]{0.75\textwidth}
\centering
\scalebox{0.8}{%
\input{figures/GHZsimple_LP118_12_16_20_MSAseqvars80.tex}
}
\end{subfigure}

\begin{subfigure}[b]{0.75\textwidth}
\centering
\scalebox{0.8}{%
\input{figures/GHZsimple_LP118_12_16_20_MSAseqvars80_threshold.tex}
}
\end{subfigure}

\caption{\label{fig:GHZsimple_LP118_MSAseqvars80}(top) Protocol II performance of a family of lifted product QLDPC codes with asymptotic rate $0.118$ using the sequential schedule of the min-sum algorithm (MSA) based decoder. Each data point is obtained by counting almost $10^4$ logical errors except depolarizing probability $0.09$, which was obtained from $10^3$ logical errors. (bottom) The threshold is about $10.7\%$.}
\end{figure}

\subsection{Discussion and Connections to Existing GHZ Purification Protocols}

We are interested in comparing our protocols to past work on GHZ purification to judge the effectiveness of our work.
However, based on our knowledge of the literature and the differences in the settings of purification protocols, this appears to be challenging and is likely a work on its own.
Nevertheless, let us address this in some detail here.
In the process, we will make comparisons and show that \emph{our protocol has the best fidelity threshold for $3$-GHZ purification at a yield of $0.118$}.

\begin{enumerate}

\vspace{3pt}

\item Most protocols in the literature with numerical results perform \emph{heralded} purification where both the protocol success probability and the output fidelity are not ideal.
In our error correction based protocol, as long as the decoder succeeds in correcting the error, we always obtain $k$ perfect entangled states as the output (assuming perfect local operations and classical communication).
It then seems natural to model this setting as another probabilistic protocol, conditioned on the probability of successful decoding, but with unit output fidelity, ignoring for now the additional fact that $k \gg 1$ in our case whereas $k=1$ in most of the literature.
However, this is not quite true since (iterative) decoder success does not come with a heralding signal.
In general, there are three possible scenarios: the decoder succeeds in correcting the error, the decoder miscorrects the error (i.e., causes a logical error), or the decoder reaches the maximum number of iterations and returns a failure.
In the first two cases, the decoder does find an estimated error pattern that matches the syndrome obtained from stabilizer measurements, whereas in the last case, the decoder is unable to even find an error pattern matching the syndrome.
It is clear that this last case heralds a failure, but there is no way to distinguish the first two scenarios.
Let us mention here that in the particular case of the Lifted Product family of codes that we consider, most of the protocol failure events are due to the decoder declaring a failure (i.e., the third case above) and not due to miscorrections.
However, this is only a preliminary observation that we are investigating in more detail.
If we are able to design good codes for this iterative decoder where decoding success can be heralded, then we can model the protocol similar to other existing non-error-correction-based protocols.
Currently, this is an important bottleneck that hinders making a fair and useful comparison with existing protocols.

\vspace{3pt}

\item Note that if the middle case (i.e., miscorrections) happens with non-negligible probability, then there are two ways to model the protocol:
either the output fidelity is always unity and the success probability is dictated by the decoding success rate, or the protocol always succeeds whenever decoder doesn't declare failure (i.e., the third case) but the output fidelity is non-trivial and dictated by a mixed state accounting for all possible logical errors arising out of miscorrections.
The former seems more straightforward and especially appropriate if miscorrections hardly occur, but this is another modeling decision that we must make when using error correction for purification.

\vspace{3pt}

It is interesting to note that Chau and Ho~\cite{Chau-qip10} have thought about the case of an iterative decoding failure for quantum LDPC codes.
The paper is about purifying Bell pairs by concatenating recurrence with an outer QLDPC code rather than hashing, since it is more practical.
The authors use the final bitwise posterior probabilities of the iterative decoder to find an appropriate unencoding circuit, at the end of which they can throw away some Bell pairs with confidence that the decoding failure most likely only affected them.
They only provide one QLDPC code as an example, but the method seems quite computationally challenging because this must happen in runtime.
Since they do not consider a code family, there is no relevant threshold for their protocol and their work is restricted to Bell pairs.

\vspace{3pt}

\item Hashing was introduced by Bennett et al. in their seminal paper~\cite{Bennett-pra96} and it has become the go-to tool for obtaining finite yield (i.e., ratio of number of purified output states to noisy input states) from a mixture of imperfect noisy entangled states.
The threshold input fidelity for purifying Werner (Bell) states through hashing is about $0.8107$.
By first performing recurrence and then feeding the output into hashing brings the threshold down to $0.5$.
However, recurrence needs two-way communication and has zero yield by itself, whereas hashing needs one-way communication but infinite copies to produce finite yield.
Since hashing effectively depends on random codes, it is impractical because decoding random linear codes is NP-complete~\cite{Berlekamp-it78,Fang-cta88}.

\vspace{3pt}

\item Nevertheless, hashing has been extended to multipartite states such as GHZ states, first by Maneva and Smolin~\cite{Maneva-contmath02}.
They extract entropy from the bits representing the signs of the different stabilizers of multiple copies of the multipartite entangled state.
For Werner-type $3$-qubit GHZ states, their threshold is effectively about $0.8075$.
If we equate their yield to the rate of the Lifted Product quantum LDPC code family that we use in our simulations, which is about $0.118$ asymptotically, then the threshold fidelity of the Maneva-Smolin protocol is $0.8401$.
In our setting, where each `B' and `C' qubit of each GHZ state goes through an i.i.d. depolarizing noise channel, the resulting state is diagonal in the GHZ basis but not exactly of Werner type.
Nevertheless, the fidelity for the noisy state is simply given by the probability that both qubits are not affected by noise, i.e., $(1-p)^2$ if $p$ is the depolarizing rate.
Using this, our threshold of $10.7\%$ for $3$-qubit GHZ purification maps to a fidelity threshold of about $0.7974$, which is very encouraging.
Note that both hashing and our protocol assume ideal LOCC.
In fact, the Maneva-Smolin protocol appears to need several rounds of hashing-style broadcast, whereas our protocol only needs one-way communication, devoid of randomness.

\vspace{3pt}

\item In~\cite{Ho-pra08}, Ho and Chau generalize the Maneva-Smolin protocol for multipartite entanglement purification and produce three new protocols, based on concatenating inner repetition codes with outer random hashing codes.
For the case of three-qubit GHZ states, their best protocol has a fidelity threshold of $0.7074$ (assuming an inner repetition code of length $15$).
If we look at Figure 4 of this paper, which plots fidelity against yield for different size GHZ states for their best protocol, the curve for repetition length $7$ (the maximum that they consider in the plot) produces a yield of $0.118$ (the asymptotic rate of our QLDPC code family) only far above input fidelity of $0.95$.
These are the best thresholds that we could find for purifying GHZ states.
A recent paper on GHZ purification~\cite{Li-optlett23} also uses the Maneva-Smolin protocol as their reference, so our judgment appears to be justified.

\vspace{3pt}

It is encouraging to see that the same authors, Ho and Chau, were the ones who showed the use of a degenerate quantum (LDPC) code to purify Bell pairs as mentioned in point 2) above.
Besides, such hashing based methods are not resilient to noise unless implemented in a measurement-based way~\cite{Zwerger-pra14}, which itself needs preparations of highly entangled cluster states.
Therefore, our new protocol with good QLDPC codes serves as the state-of-the-art for purifying GHZ states.

\vspace{3pt}

\item Most existing protocols based on recurrence or hashing or other related methods involve deep circuits that appear to require interactions between arbitrary pairs of qubits.
This is extremely challenging in a fault-tolerant setting.
However, when our protocol is used in conjunction with good quantum LDPC codes, the circuits are deterministic as they only involve stabilizer measurements, and stabilizers are low-weight due to the LDPC property.
Therefore, these are much more conducive to fault-tolerant entanglement purification in quantum networks.

\vspace{3pt}

\item In recent protocols on purifying GHZ states, such as in~\cite{deBone-tqe20}, the setting is to use Bell pairs that are purified and fused to form one GHZ state.
The performance curves plot input fidelity of each Bell pair versus output fidelity of the single purified GHZ state.
We think that our setting is quite different, once again because our output fidelity is ideal conditioned on decoder success, but also because we do not use Bell pairs as inputs.
Even this particular work only compares their results with that of a single past work, which is that of Nickerson et al.~\cite{Nickerson-ncomms13} where they adopt a similar approach.
Other works, such as~\cite{Krastanov-quantum19}, consider Bell pair purification using optimized protocols under the practical setting where the purification circuits are imperfect and noisy.
We emphasize that our error correction based approach potentially offers fault tolerance but our current setting introduces noise only in the quantum communication channel and assumes perfect local operations.
We leave the investigation of a fully fault-tolerant setting for our protocol to future work.

\end{enumerate}

\subsection{Decoding QLDPC Codes under Realistic Noise Models}

While our main results are relevant to the ``code capacity'' error model, where there are only qubit errors and all operations are assumed noiseless, in a separate work a subset of the authors considered decoding this family of QLDPC codes under a ``phenomenological'' noise model, i.e., with an additional (classical) error model on the syndromes~\cite{Raveendran-qce22}.
In that setting, motivated by practical situations, the syndromes extracted from a measurement circuit are assumed to have an additional random Gaussian noise, thereby yielding ``soft'' syndromes.
It was shown then that the MSA decoder can be modified appropriately such that the decoding performance is almost as good as the above ideal syndrome scenario.
Therefore, by reinterpreting that work in the context of entanglement purification, we highlight that the protocol can be applied to more realistic settings as well.

Since we are constructing a new GHZ purification protocol based on this new insight about GHZ states, we have considered this simple model of noiseless LOCC and noisy qubit communications.
We emphasize here that, to the best of our knowledge, this is the first protocol to use quantum error correction for purifying GHZ states, and we also report simulation results of state-of-the-art QLDPC codes with an efficient iterative decoder.
Moreover, by comparison to past works, we have shown that our scheme has the best fidelity threshold of $0.7974$ for i.i.d. single-qubit depolarizing noise, at a yield of $0.118$.
While the problem of noisy local operations is important and has received attention~\cite{Pan-nature01,Dur-rpp07,Nickerson-ncomms13,Krastanov-quantum19}, we leave this to future work.

\subsection{Purification-Inspired Algorithm to Generate Logical Pauli Operators}

In the process, inspired by stabilizer measurements on Bell/GHZ states, we have developed a new algorithm to generate logical Pauli operators for any stabilizer code (see Algorithm~\ref{algo:logical_paulis_ghz_msmt} and its explanation in Appendix~\ref{sec:logical_paulis_ghz_msmt}).
The core idea is to first simulate the generation on $n$ Bell/GHZ states by creating a table of their $2n$ stabilizers. 
It turns out that we only need the $ZZI$- and $XXX$-type stabilizers for the GHZ case, which is why we ignore the $n$ $IZZ$-type stabilizers.
Then we simulate the measurement of each code stabilizer on qubits `A' using the stabilizer formalism.
At the end of this process, it can be shown that the non-code-stabilizer rows in the table must be a combination of logical Pauli operators on multiple subsystems.
Finally, we carefully identify the logical Pauli operators on qubits `A' and return those as the desired operators on the given code.

\section{Notation and Background}

The Pauli group on $n$ qubits is denoted by $\mathcal{P}_n$.
We denote Pauli matrices $I, X, Y, Z$ and their tensor products using the notation $E(a,b)$, where $a,b \in \{0,1\}^n$ denote respectively the $X$- and $Z$-components of the $n$-qubit Pauli operator.
The \emph{weight} of a Pauli operator is the number of qubits on which it acts nontrivially (i.e., does not apply $I$).
For example, $E([0,1,0,1],[0,0,1,1]) = I \otimes X \otimes Z \otimes Y \equiv IXZY$ has weight $3$ and we dropped the tensor product symbol $\otimes$ for brevity.
Two Pauli operators $E(a,b), E(c,d)$ either commute or anticommute, and this is dictated by the symplectic inner product in the binary vector space.
If $\syminn{[a,b]}{[c,d]} \coloneqq ad^T + bc^T = 0$ (resp. $1$) (mod $2$), then they commute (resp. anticommute).

A stabilizer group $S$ is generated by commuting Pauli operators $\varepsilon_i E(a_i, b_i), i = 1,2,\ldots,r$, where $\varepsilon_i \in \{ \pm 1 \}$ and $-I_{2^n} \notin S$.
The $\llbr n,k,d \rrbr$ stabilizer code defined by $S$ is given by $\mathcal{Q}(S) = \{ \ket{\psi} \in \mathbb{C}^{2^n} \colon g \ket{\psi} = \ket{\psi} \ \forall\ g \in S \}$, where $k = n-r$.
The logical Pauli operators of the code commute with all stabilizers but do not belong to $S$, and their minimum weight is $d$.
The code is completely defined by its stabilizers and logical operators, or equivalently by an encoding circuit $\mathcal{U}_{\text{Enc}}(S)$.
The projector onto the code subspace is given by $\Pi_S = \prod_{i=1}^r \frac{1}{2} \left[I_{2^n} + \varepsilon_i E(a_i,b_i) \right]$.

A CSS (Calderbank-Shor-Steane) code is a special type of stabilizer code for which there exists a set of stabilizer generators such that each generator is purely $X$-type, i.e., of the form $E(a_i,0)$, or purely $Z$-type, i.e., of the form $E(0,b_j)$.
Such a code can be described by a pair of classical binary linear codes $\MCC_X$ and $\MCC_Z$, where the rows of the parity-check matrix $H_X$ (resp. $H_Z$) for $\MCC_X$ (resp. $\MCC_Z$) are $a_i \in \{0,1\}^n$ (resp. $b_j \in \{0,1\}^n$).
Since $E(a_i,0)$ and $E(0,b_j)$ must commute, the symplectic inner product constraint leads to the condition $a_i b_j^T = 0$ for all $i,j$ or, equivalently, $H_X H_Z^T = 0$.

A quantum (CSS) low-density parity-check (QLDPC) code is described by a pair $(\MCC_X,\MCC_Z)$ of classical LDPC codes, which implies that $H_X$ and $H_Z$ are sparse, i.e., each stabilizer involves few qubits and each qubit is involved in few stabilizers.
It is very challenging to construct good QLDPC codes due to the constraint $H_X H_Z^T = 0$ on two sparse matrices, but recent exciting work has developed optimal QLDPC codes where $k$ and $d$ scale linearly with $n$~\cite{Hastings-stoc21,Panteleev-it21,Breuckmann-it21,Breuckmann-prxq21,Panteleev-stoc22,Leverrier-arxiv22}.
For our simulations, we chose a specific family of lifted product QLDPC codes from~\cite[Table II]{Raveendran-quantum21} that have asymptotic rate $k/n=0.118$.
To decode these codes, we use the computationally efficient min-sum algorithm (MSA) based iterative decoder under the sequential schedule~\cite{Chen-tcomm05,Hocevar-sps04}, with a normalization factor of $0.8$ and maximum number of iterations set to $100$ (also see the description in~\cite{Raveendran-quantum21}).

A stabilizer state corresponds to a code with dimension $k=0$, and can equivalently be represented by a maximal stabilizer group, i.e., with $r=n$.
Any Pauli measurement on the state can be simulated by a well-defined set of rules to update this stabilizer group.
These rules are given by the stabilizer formalism for measurements~\cite{Gottesman-icgtmp98,Aaronson-pra04}.

For any matrix $M$, the Bell state $\ket{\Phi}_{\text{AB}} = \frac{\ket{00}_{\text{AB}} + \ket{11}_{\text{AB}}}{\sqrt{2}}$ satisfies the property 
$(M_{\text{A}} \otimes I_{\text{B}}) \ket{\Phi}_{\text{AB}} = (I_{\text{A}} \otimes M_{\text{B}}^T) \ket{\Phi}_{\text{AB}}$.
This property extends to $n$ copies of the Bell state as well.
When $M$ is a projector, which is the case when we perform stabilizer measurements on qubits `A', i.e., $M = \Pi_S$, using the fact that $M^2 = M$ we conclude from the above property that projecting qubits `A' automatically projects qubits `B' as well according to $M^T$.
Therefore, imposing a code on qubits `A' simultaneously imposes the ``transpose'' code on qubits `B'.

A more detailed discussion of these background concepts can be found in Appendix~\ref{sec:background}.

\section{Revisiting the Bell Pair Distillation Protocol}
\label{sec:bell_distillation}

\begin{figure}
 \centering
\includegraphics[scale=0.8,keepaspectratio]{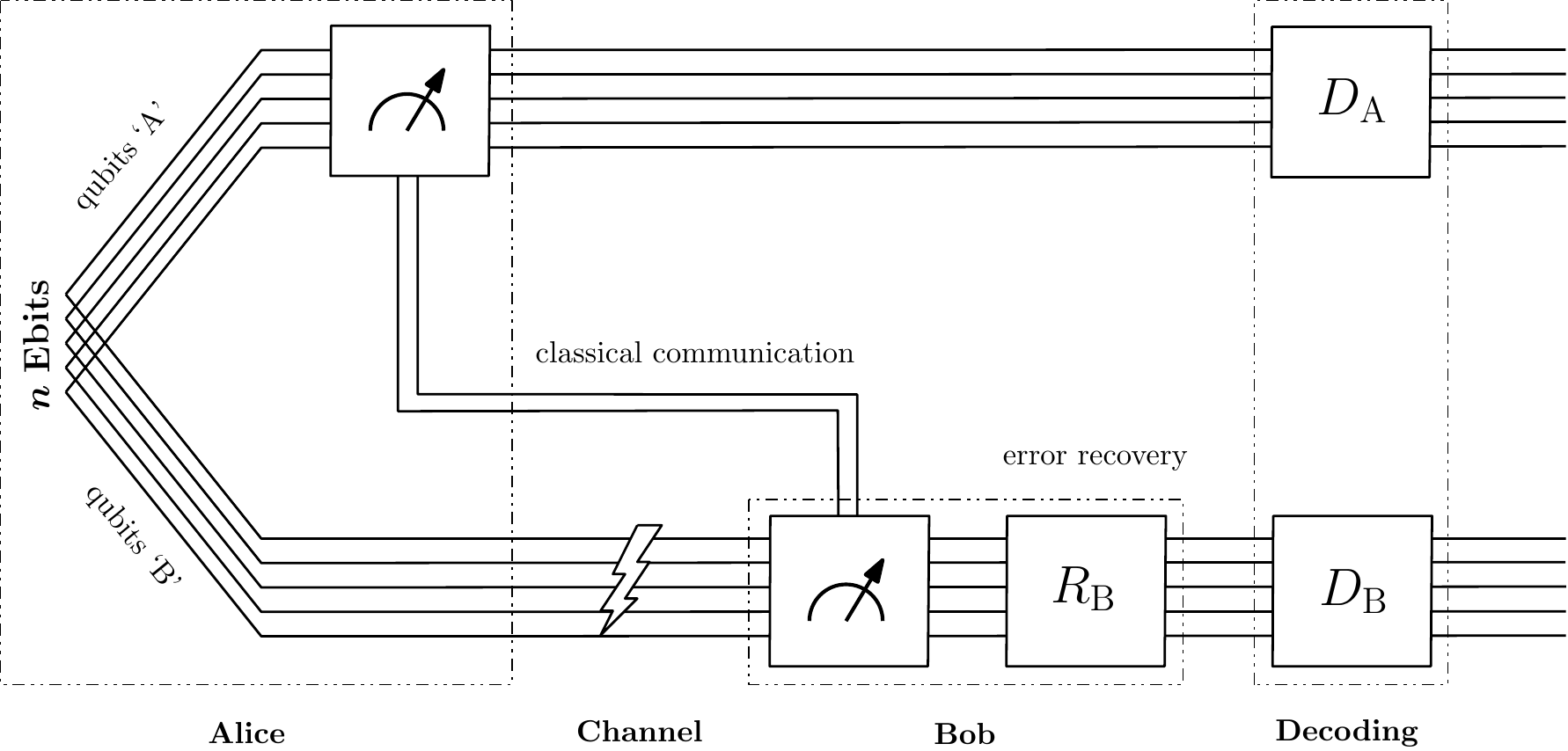}
  \caption{The QEC-based entanglement distillation protocol of Wilde \etal \cite{Wilde-isit10}. Figure adapted from~\cite{Wilde-isit10}. 
   }
   \label{fig:bell-pair-entanglement}
\end{figure}

\begin{algorithm}[h] 
\DontPrintSemicolon
\SetAlgoLined
\KwResult{Alice and Bob share $k$ perfect Bell pairs or at least one of the $k$ pairs has an unknown Pauli error}
\SetKwInOut{Input}{Input}
\SetKwInOut{Output}{Output}
\Input{$n$ Bell pairs $\ket{\Phi^{+}}^{\otimes n}$ at Alice, 
$\llbr n,k,d \rrbr$ stabilizer code $\mathcal{Q}(S)$ defined by a stabilizer group $S$}
\Output{$k$ Bell pairs of higher quality shared between Alice and Bob if channel introduces a correctable error}
 Initialization: Rearrange the $2n$ qubits in $\ket{\Phi^{+}}^{\otimes n}$ to obtain $\ket{\Phi^{+}_n}$~\eqref{eq:bell_state_rearranged} for processing by Alice and Bob respectively\;
  \;
 Alice \;
 (a) measures all the stabilizer generators $\{ \varepsilon_i E(a_i,b_i) \, ; \, i = 1,2,\ldots,r=n-k \}$ on her $n$ qubits, obtains syndrome, \; 
 (b) sends the remaining $n$ qubits to Bob over a noisy quantum channel, \;
 (c) sends the stabilizers and syndrome (which together define $\mathcal{Q}(S)$) 
 to Bob over a noiseless classical channel. \;
 \;
{  Bob \;
 (a) measures all the stabilizer generators $\{ \varepsilon_i E(a_i,b_i) \, ; \, i = 1,2,\ldots,r=n-k \}$ on his $n$ qubits, \;
(b) combines the syndrome information from Alice as well as his measurements and interprets using Section~\ref{sec:bell_state_identity}, \;
(c) performs necessary Pauli corrections on his qubits to bring them to the code space of $\mathcal{Q}(S)$. \;
  \; 
  If the channel error was correctable, pairs of logical qubits of Alice's and Bob's codes form $k$ Bell states \;
 // If channel error was NOT correctable, some pair of logical qubits form a Bell state with an unknown Pauli error \;
 Alice and Bob respectively apply the inverse of the encoding unitary for their code on their $n$ qubits \;
 // The encoding unitary is determined by the logical Pauli operators for $\mathcal{Q}(S)$ obtained from Algorithm~\ref{algo:logical_paulis_ghz_msmt}
}
  \caption{Algorithm to convert $n$ Bell pairs into $k$ Bell pairs of higher quality, using an $\llbr n,k,d \rrbr$ stabilizer code}
 \label{algo:algo_bell}
\end{algorithm}

In Ref.~\cite{Wilde-isit10}, Wilde \etal described a protocol to distill Bell pairs using an arbitrary quantum stabilizer code.
We reiterate this protocol here and provide more clarity on the reasons behind its working.
Then, in the next section, we will generalize this protocol to distill GHZ states, i.e., the $\ell$-qubit entangled state $\dket{\text{GHZ}^{\ell}} = \frac{1}{\sqrt{2}} \left( \dket{00 \cdots 0} + \dket{11 \cdots 1} \right)$.

Initially, Alice generates $n$ copies of the Bell state $\dket{\Phi^+}$ ($n$ ``ebits''), rearranges the qubits as described above, and sends Bob's set of $n$ qubits to him over a noisy channel. 
It is not necessary that Alice must prepare Bell pairs locally and then transmit half the qubits to Bob. 
Indeed, the protocol is applicable as long as Alice and Bob share some initial (noisy) Bell pairs.
Then, Alice measures the stabilizers of a quantum stabilizer code defined by $S = \langle \varepsilon_i E(a_i,b_i); \ i = 1,\ldots,r \rangle$ on her qubits, with $\varepsilon_i = \pm 1$.
Let her measurement results be $(-1)^{m_i}, m_i \in \{0,1\}$.
This projects her qubits onto the codespace fixed by the stabilizers $S' = \langle (-1)^{m_i} \varepsilon_i E(a_i,b_i); \ i = 1,\ldots,r \rangle$. 
Alice applies some suitable Pauli ``correction'' to bring her qubits back to the code subspace $\mathcal{Q}(S)$ (rather than $\mathcal{Q}(S')$), if that is the code she desires to use.
She classically communicates the chosen stabilizers, $S$, the measurements $\{ m_i \}_{i=1}^r$, and the Pauli correction to Bob.

Although we use the term ``correction'', there is really no error on Alice's qubits.
Instead, the terminology is used to indicate that Alice brings the qubits to her desired code space.
Furthermore, even if there is some error on Alice's qubits, one can map it to an equivalent error on Bob's qubits using the Bell state matrix identity.

Note that the authors of Ref.~\cite{Wilde-isit10} do not explicitly mention that the Pauli correction needs to be communicated, but it could be necessary in situations where Alice's and Bob's decoders are not identical or have some randomness embedded in them.
For the code, though any appropriate definition of logical Pauli generators works with the protocol, we employ Algorithm~\ref{algo:logical_paulis_ghz_msmt} to obtain generators that are ``compatible'' with our way of analyzing the protocol (using the stabilizer formalism).
This phenomenon will become more clear after the $\llbr 5,1,3 \rrbr$ code example in this section.
While the algorithm simulates measurements on GHZ states to define logical Paulis, an equivalent algorithm can be constructed that only simulates Bell measurements.

\begin{remark}
In this protocol, whenever the syndrome of Alice is non-trivial, i.e., at least one $m_i$ equals $1$, she can either perform a Pauli correction or just define her code to be $\mathcal{Q}(S')$ and not perform any correction.
If the protocol is defined so that she always does the latter, as depicted in Fig.~\ref{fig:bell-pair-entanglement} where there is no `Recovery' block on Alice's qubits, then Bob can adjust his processing accordingly based on the syndrome information from Alice.
\end{remark}


Without loss of generality, we can assume that Alice sends Bob's qubits to him only after performing her measurements and any Pauli correction. 
So, the channel applies a Pauli error only after Bob's qubits got projected according to $S'' = \langle (-1)^{m_i + a_i b_i^T} \varepsilon_i E(a_i,b_i); \ i = 1,\ldots,r \rangle$. 
Now, Bob measures the stabilizers $\varepsilon_i E(a_i,b_i)$ and applies corrections on his qubits using his syndromes as well as Alice's syndromes (and the Bell matrix identity, which in particular involves the transpose). 
This projects his qubits to the same codespace as Alice.
Finally, Alice and Bob locally apply the inverse of the encoding unitary for their code, $\mathcal{U}_{\text{Enc}}(S)^\dagger$. 
If Bob's correction was successful, this converts the $k$ logical Bell pairs into $k$ physical Bell pairs that are, on average, of higher quality than the $n$ noisy Bell pairs initially shared between them.
This protocol is shown in Figure~\ref{fig:bell-pair-entanglement} and summarized in Algorithm~\ref{algo:algo_bell}. 

While the steps of the protocol are clear, it is worth considering why the logical qubits of Alice and Bob must be $k$ copies of the Bell pair, assuming all errors were corrected successfully.
To get some intuition, let us quickly consider the example of the $3$-qubit bit-flip code defined by $S = \langle ZZI, IZZ \rangle$.
According to~\eqref{eq:stabilizer_projector}, the projector onto $\mathcal{Q}(S)$ is $\Pi_S = \frac{(I_8 + ZZI)}{2} \frac{(I_8 + IZZ)}{2}$.
The encoding unitary, as described in Appendix~\ref{sec:stabilizer_codes}, is $\mathcal{U}_{\text{Enc}} = \text{CNOT}_{1 \rightarrow 2}\, \text{CNOT}_{1 \rightarrow 3}$.
Since $Z^T = Z$, Alice's measurements will project Bob's qubits onto the same code subspace as her's.
For convenience, assume that Alice obtains the trivial syndrome $(+1,+1)$ and that the channel does not introduce any error.
Then, according to~\eqref{eq:projector_bell_pair}, the resulting (unnormalized) state after Alice's measurements is $(\Pi_S \otimes \Pi_S) \dket{\Phi_3^+}$.

Consider the action of $(I_8 + ZZI)$ on a computational basis state $\dket{x}, x = [x_1,x_2,x_3]$: 
\begin{align}
(I_8 + ZZI) \dket{x} = \dket{x} + E(000,110) \dket{x} = \dket{x} + (-1)^{[1,1,0] x^T} \dket{x} = 
\begin{cases}
2 \dket{x} & \ \text{if}\ x_1 \oplus x_2 = 0, \\
0          & \ \text{otherwise}.
\end{cases}
\end{align}
Hence, after the action of $(\Pi_S \otimes \Pi_S)$ and inversion of the encoding unitary by Alice and Bob, we obtain
\begin{align}
(\Pi_S \otimes \Pi_S) \dket{\Phi_3^+} & = \frac{1}{16 \cdot \sqrt{2^3}} \sum_{x \in \{000,111\}} 4\dket{x}_{\text{A}} \otimes 4\dket{x}_{\text{B}} \\
  & \propto \dket{000}_{\text{A}} \dket{000}_{\text{B}} + \dket{111}_{\text{A}} \dket{111}_{\text{B}} \\
  & \xrightarrow{(\mathcal{U}_{\text{Enc}}^{\dagger})_{\text{A}} \otimes (\mathcal{U}_{\text{Enc}}^{\dagger})_{\text{B}}} \dket{000}_{\text{A}} \dket{000}_{\text{B}} + \dket{100}_{\text{A}} \dket{100}_{\text{B}} \\
  & = \dket{00}_{\text{AB}} \otimes \dket{00}_{\text{AB}} \otimes \dket{00}_{\text{AB}} + \dket{11}_{\text{AB}} \otimes \dket{00}_{\text{AB}} \otimes \dket{00}_{\text{AB}} \\
  & = \left( \dket{00}_{\text{AB}} + \dket{11}_{\text{AB}} \right) \otimes \dket{00}_{\text{AB}} \otimes \dket{00}_{\text{AB}}.
\end{align}
Thus, the output is a single Bell pair and ancillary qubits on Alice and Bob.
In Appendix~\ref{sec:logical_bell_CSS}, we show this phenomenon for arbitrary CSS codes by generalizing the state vector approach used above.

\subsection{Bell Pair Distillation using the 5-Qubit Code}

In the remainder of this section, with the $\llbr 5,1,3 \rrbr$ code~\cite{Bennett-pra96,Laflamme-prl96} as an example, we use the stabilizer formalism to show that the above phenomenon is true for any stabilizer code.
Recall that this code is defined by
\begin{align}
S = \langle XZZXI, \ IXZZX, \ XIXZZ, \ ZXIXZ \rangle.
\end{align}
As described in Appendix~\ref{sec:stabilizer_codes}, the corresponding binary stabilizer matrix is given by
\begin{align}
G_S = 
\left[
\begin{array}{ccccc|ccccc|c}
1 & 0 & 0 & 1 & 0   &   0 & 1 & 1 & 0 & 0  &  +1 \\
0 & 1 & 0 & 0 & 1   &   0 & 0 & 1 & 1 & 0  &  +1 \\
1 & 0 & 1 & 0 & 0   &   0 & 0 & 0 & 1 & 1  &  +1 \\
0 & 1 & 0 & 1 & 0   &   1 & 0 & 0 & 0 & 1  &  +1
\end{array}
\right].
\end{align}
Initially, Alice starts with $5$ copies of the standard Bell state $\dket{\Phi^+}^{\otimes 5}$, and marks one qubit of each copy as Bob's.
She does not yet send Bob's qubits to him.
The stabilizer group for this joint state of 5 ``ebits'' (or ``EPR pairs'') is
\begin{align}
\mathcal{E}_5 & = \langle X_{\text{A}_1} X_{\text{B}_1}, \, Z_{\text{A}_1} Z_{\text{B}_1}, \ X_{\text{A}_2} X_{\text{B}_2}, \, Z_{\text{A}_2} Z_{\text{B}_2}, \ X_{\text{A}_3} X_{\text{B}_3}, \, Z_{\text{A}_3} Z_{\text{B}_3}, \ X_{\text{A}_4} X_{\text{B}_4}, \, Z_{\text{A}_4} Z_{\text{B}_4}, \nonumber \\ 
  & \hspace{10.5cm} X_{\text{A}_5} X_{\text{B}_5}, \, Z_{\text{A}_5} Z_{\text{B}_5} \rangle \\
  & = \langle X_{\text{A}_i} X_{\text{B}_i} =  E([e_i^{\text{A}}, e_i^{\text{B}}], [0^{\text{A}}, 0^{\text{B}}]), \ Z_{\text{A}_i} Z_{\text{B}_i} = E([0^{\text{A}}, 0^{\text{B}}],[e_i^{\text{A}}, e_i^{\text{B}}]) \, ; \, i = 1,\ldots,5  \rangle,
\end{align}
where $e_i \in \mathbb{F}_2^5$ is the standard basis vector with a $1$ in position $i$ and zeros elsewhere, $0 \in \mathbb{F}_2^5$ is the all-zeros vector, and the $X$- and $Z$- components in the $E(a,b)$ notation have been split into Alice's qubits and Bob's qubits.
Observe that this is a maximal stabilizer group on $10$ qubits and hence, there are no non-trivial logical operators associated with this group, i.e., the normalizer of $\mathcal{E}_5$ in $\mathcal{P}_{10}$ is itself.

It will be convenient to adopt a tabular format for these generators, where the first column of each row gives the sign of the generator, the next two columns give the $X$-components of Alice and Bob in that generator, the subsequent two columns give the $Z$-components of Alice and Bob in that generator, and the last column gives the Pauli representation of that generator for clarity.
Hence, the above generators are written as follows.

\begin{center}
\begin{tabularx}{.9\linewidth}{c || *{2}{C} | *{2}{C} || c}
\toprule
Sign & \multicolumn{2}{c|}{$X$-Components} & \multicolumn{2}{c||}{$Z$-Components} & Pauli Representation \\
     &   A & B   &   A & B   &   \\
\midrule
\midrule
$+1$ &    $e_i$ & $e_i$   &   $0$ & $0$    & $X_{\text{A}_i} X_{\text{B}_i} =  E([e_i^{\text{A}}, e_i^{\text{B}}], [0^{\text{A}}, 0^{\text{B}}])$ \\
\midrule
$+1$ &    $0$ & $0$   & $e_i$ & $e_i$   &   $Z_{\text{A}_i} Z_{\text{B}_i} =  E([0^{\text{A}}, 0^{\text{B}}], [e_i^{\text{A}}, e_i^{\text{B}}])$ \\
\bottomrule
\bottomrule
\end{tabularx}
\end{center}

{\footnotesize
\begin{xltabular}{.8\linewidth}{c c || *{2}{c} | *{2}{c} || c}
\caption{\label{tab:bell_protocol}
Steps of the Bell-pair distillation protocol based on the $\llbr 5,1,3 \rrbr$ code. Any `$0$' that is not part of a string represents $00000$, and $e_i \in \mathbb{F}_2^5$ is the standard basis vector with a $1$ in the $i$-th position and zeros elsewhere. Code stabilizers are typeset in boldface. An additional left arrow indicates which row is being replaced with a code stabilizer, i.e., the first row that anticommutes with the stabilizer. Other updated rows are highlighted in gray. Classical communications: A $\rightarrow$ B.} \\
\toprule
Step & Sign & \multicolumn{2}{c|}{$X$-Components} & \multicolumn{2}{c||}{$Z$-Components} & Pauli Representation \\
     &      &   A & B   &   A & B   &   \\
\midrule
\midrule
     &     &        &       &         &          &                   \\
$(0)$ & $+1$ &    $e_1$ & $e_1$   &   $0$ & $0$    & $X_{\text{A}_1} X_{\text{B}_1}$ \\
     & $+1$ &    $e_2$ & $e_2$   &   $0$ & $0$    & $X_{\text{A}_2} X_{\text{B}_2}$ \\
     & $+1$ &    $e_3$ & $e_3$   &   $0$ & $0$    & $X_{\text{A}_3} X_{\text{B}_3}$ \\
     & $+1$ &    $e_4$ & $e_4$   &   $0$ & $0$    & $X_{\text{A}_4} X_{\text{B}_4}$ \\
     & $+1$ &    $e_5$ & $e_5$   &   $0$ & $0$    & $X_{\text{A}_5} X_{\text{B}_5}$ \\
\cmidrule(lr){2-7}
     & $+1$ &    $0$ & $0$   &   $e_1$ & $e_1$    & $Z_{\text{A}_1} Z_{\text{B}_1}$ \\
     & $+1$ &    $0$ & $0$   &   $e_2$ & $e_2$    & $Z_{\text{A}_2} Z_{\text{B}_2}$ \\
     & $+1$ &    $0$ & $0$   &   $e_3$ & $e_3$    & $Z_{\text{A}_3} Z_{\text{B}_3}$ \\
     & $+1$ &    $0$ & $0$   &   $e_4$ & $e_4$    & $Z_{\text{A}_4} Z_{\text{B}_4}$ \\
     & $+1$ &    $0$ & $0$   &   $e_5$ & $e_5$    & $Z_{\text{A}_5} Z_{\text{B}_5}$ \\
     &     &        &       &         &          &                   \\
\midrule
     &     &        &       &         &          &                   \\
$(1)$ & $+1$ &    $e_1$ & $e_1$   &   $0$ & $0$    & $X_{\text{A}_1} X_{\text{B}_1}$ \\
\rowcolor{lightgray}
     & $+1$ &    $e_2$ & $e_2$   &   $e_4$ & $e_4$    & $X_{\text{A}_2} X_{\text{B}_2} Z_{\text{A}_4} Z_{\text{B}_4}$ \\
\rowcolor{lightgray}
     & $+1$ &    $e_3$ & $e_3$   &   $e_4$ & $e_4$    & $X_{\text{A}_3} X_{\text{B}_3} Z_{\text{A}_4} Z_{\text{B}_4}$ \\
     & $+1$ &    $e_4$ & $e_4$   &   $0$ & $0$    & $X_{\text{A}_4} X_{\text{B}_4}$ \\
     & $+1$ &    $e_5$ & $e_5$   &   $0$ & $0$    & $X_{\text{A}_5} X_{\text{B}_5}$ \\
\cmidrule(lr){2-7}
\rowcolor{lightgray}
     & $+1$ &    $0$ & $0$   &   $e_1+e_4$ & $e_1+e_4$    & $Z_{\text{A}_1} Z_{\text{B}_1} Z_{\text{A}_4} Z_{\text{B}_4}$ \\
     & $+1$ &    $0$ & $0$   &   $e_2$ & $e_2$    & $Z_{\text{A}_2} Z_{\text{B}_2}$ \\
     & $+1$ &    $0$ & $0$   &   $e_3$ & $e_3$    & $Z_{\text{A}_3} Z_{\text{B}_3}$ \\
     & $\boldsymbol{ \varepsilon_1 }$ &    $\boldsymbol{ 10010 }$ & $\boldsymbol{ 00000 }$   &   $\boldsymbol{ 01100 }$ & $\boldsymbol{ 00000 }$    & \qquad \ \  $\boldsymbol{ \varepsilon_1 \, X_{\text{A}_1} Z_{\text{A}_2} Z_{\text{A}_3} X_{\text{A}_4} } \ \ \boldsymbol{\longleftarrow} $ \\
     & $+1$ &    $0$ & $0$   &   $e_5$ & $e_5$    & $Z_{\text{A}_5} Z_{\text{B}_5}$ \\
     &     &        &       &         &          &                   \\
\midrule
     &     &        &       &         &          &                   \\
$(2)$ & $+1$ &    $e_1$ & $e_1$   &   $0$ & $0$    & $X_{\text{A}_1} X_{\text{B}_1}$ \\
     & $+1$ &    $e_2$ & $e_2$   &   $e_4$ & $e_4$    & $X_{\text{A}_2} X_{\text{B}_2} Z_{\text{A}_4} Z_{\text{B}_4}$ \\
\rowcolor{lightgray}
     & $+1$ &    $e_3$ & $e_3$   &   $e_4+e_5$ & $e_4+e_5$    & $X_{\text{A}_3} X_{\text{B}_3} Z_{\text{A}_4} Z_{\text{B}_4} Z_{\text{A}_5} Z_{\text{B}_5}$ \\
     & $+1$ &    $e_4$ & $e_4$   &   $e_5$ & $e_5$    & $X_{\text{A}_4} X_{\text{B}_4} Z_{\text{A}_5} Z_{\text{B}_5}$ \\
     & $+1$ &    $e_5$ & $e_5$   &   $0$ & $0$    & $X_{\text{A}_5} X_{\text{B}_5}$ \\
\cmidrule(lr){2-7}
     & $+1$ &    $0$ & $0$   &   $e_1+e_4$ & $e_1+e_4$    & $Z_{\text{A}_1} Z_{\text{B}_1} Z_{\text{A}_4} Z_{\text{B}_4}$ \\
\rowcolor{lightgray}
     & $+1$ &    $0$ & $0$   &   $e_2+e_5$ & $e_2+e_5$    & $Z_{\text{A}_2} Z_{\text{B}_2} Z_{\text{A}_5} Z_{\text{B}_5}$ \\
     & $+1$ &    $0$ & $0$   &   $e_3$ & $e_3$    & $Z_{\text{A}_3} Z_{\text{B}_3}$ \\
     & $\boldsymbol{ \varepsilon_1 }$ &    $\boldsymbol{ 10010 }$ & $\boldsymbol{ 00000 }$   &   $\boldsymbol{ 01100 }$ & $\boldsymbol{ 00000 }$    & $\boldsymbol{ \varepsilon_1 \, X_{\text{A}_1} Z_{\text{A}_2} Z_{\text{A}_3} X_{\text{A}_4} }$ \\
     & $\boldsymbol{ \varepsilon_2 }$ &    $\boldsymbol{ 01001 }$ & $\boldsymbol{ 00000 }$   &   $\boldsymbol{ 00110 }$ & $\boldsymbol{ 00000 }$    & \qquad \ \  $\boldsymbol{ \varepsilon_2 \, X_{\text{A}_2} Z_{\text{A}_3} Z_{\text{A}_4} X_{\text{A}_5} } \ \ \boldsymbol{\longleftarrow}$ \\
     &     &        &       &         &          &                   \\
\midrule
     &     &        &       &         &          &                   \\
$(3)$ & $+1$ &    $e_1$ & $e_1$   &   $0$ & $0$    & $X_{\text{A}_1} X_{\text{B}_1}$ \\
     & $+1$ &    $e_2$ & $e_2$   &   $e_4$ & $e_4$    & $X_{\text{A}_2} X_{\text{B}_2} Z_{\text{A}_4} Z_{\text{B}_4}$ \\
     & $+1$ &    $e_3$ & $e_3$   &   $e_4+e_5$ & $e_4+e_5$    & $X_{\text{A}_3} X_{\text{B}_3} Z_{\text{A}_4} Z_{\text{B}_4} Z_{\text{A}_5} Z_{\text{B}_5}$ \\
\rowcolor{lightgray}
     & $-1$ &    $e_4+e_5$ & $e_4+e_5$   &   $e_5$ & $e_5$    & $- \, X_{\text{A}_4} X_{\text{B}_4} Z_{\text{A}_5} Z_{\text{B}_5} X_{\text{A}_5} X_{\text{B}_5}$ \\
     & $\boldsymbol{ \varepsilon_3 }$ &    $\boldsymbol{ 10100 }$ & $\boldsymbol{ 00000 }$   &   $\boldsymbol{ 00011 }$ & $\boldsymbol{ 00000 }$    & \qquad \ \  $\boldsymbol{ \varepsilon_3 \, X_{\text{A}_1} X_{\text{A}_3} Z_{\text{A}_4} Z_{\text{A}_5} } \ \ \boldsymbol{\longleftarrow}$ \\
\cmidrule(lr){2-7}
\rowcolor{lightgray}
     & $+1$ &    $e_5$ & $e_5$   &   $e_1+e_4$ & $e_1+e_4$    & $Z_{\text{A}_1} Z_{\text{B}_1} Z_{\text{A}_4} Z_{\text{B}_4} X_{\text{A}_5} X_{\text{B}_5}$ \\
     & $+1$ &    $0$ & $0$   &   $e_2+e_5$ & $e_2+e_5$    & $Z_{\text{A}_2} Z_{\text{B}_2} Z_{\text{A}_5} Z_{\text{B}_5}$ \\
\rowcolor{lightgray}
     & $+1$ &    $e_5$ & $e_5$   &   $e_3$ & $e_3$    & $Z_{\text{A}_3} Z_{\text{B}_3} X_{\text{A}_5} X_{\text{B}_5}$ \\
     & $\boldsymbol{ \varepsilon_1 }$ &    $\boldsymbol{ 10010 }$ & $\boldsymbol{ 00000 }$   &   $\boldsymbol{ 01100 }$ & $\boldsymbol{ 00000 }$    & $\boldsymbol{ \varepsilon_1 \, X_{\text{A}_1} Z_{\text{A}_2} Z_{\text{A}_3} X_{\text{A}_4} }$ \\
     & $\boldsymbol{ \varepsilon_2 }$ &    $\boldsymbol{ 01001 }$ & $\boldsymbol{ 00000 }$   &   $\boldsymbol{ 00110 }$ & $\boldsymbol{ 00000 }$    & $\boldsymbol{ \varepsilon_2 \, X_{\text{A}_2} Z_{\text{A}_3} Z_{\text{A}_4} X_{\text{A}_5} }$ \\
     &     &        &       &         &          &                   \\
\midrule
     &     &        &       &         &          &                   \\
%
$(4)$ & $\boldsymbol{ \varepsilon_4 }$ &    $\boldsymbol{ 01010 }$ & $\boldsymbol{ 00000 }$   &   $\boldsymbol{ 10001 }$ & $\boldsymbol{ 00000 }$    & \qquad \ \  $\boldsymbol{ \varepsilon_4 \, Z_{\text{A}_1} X_{\text{A}_2} X_{\text{A}_4} Z_{\text{A}_5} } \ \ \boldsymbol{\longleftarrow}$ \\
\rowcolor{lightgray}
     & $+1$ &    $e_1+e_2$ & $e_1+e_2$   &   $e_4$ & $e_4$    & $X_{\text{A}_2} X_{\text{B}_2} Z_{\text{A}_4} Z_{\text{B}_4} X_{\text{A}_1} X_{\text{B}_1}$ \\
\rowcolor{lightgray}
     & $+1$ &    $e_1+e_3$ & $e_1+e_3$   &   $e_4+e_5$ & $e_4+e_5$    & $X_{\text{A}_3} X_{\text{B}_3} Z_{\text{A}_4} Z_{\text{B}_4} Z_{\text{A}_5} Z_{\text{B}_5} X_{\text{A}_1} X_{\text{B}_1}$ \\
\rowcolor{lightgray}
     & $-1$ &    $e_1+e_4+e_5$ & $e_1+e_4+e_5$   &   $e_5$ & $e_5$    & $- \, X_{\text{A}_4} X_{\text{B}_4} Z_{\text{A}_5} Z_{\text{B}_5} X_{\text{A}_5} X_{\text{B}_5} X_{\text{A}_1} X_{\text{B}_1}$ \\
     & $\boldsymbol{ \varepsilon_3 }$ &    $\boldsymbol{ 10100 }$ & $\boldsymbol{ 00000 }$   &   $\boldsymbol{ 00011 }$ & $\boldsymbol{ 00000 }$    & $\boldsymbol{ \varepsilon_3 \, X_{\text{A}_1} X_{\text{A}_3} Z_{\text{A}_4} Z_{\text{A}_5} }$ \\
\cmidrule(lr){2-7}
     & $+1$ &    $e_5$ & $e_5$   &   $e_1+e_4$ & $e_1+e_4$    & $Z_{\text{A}_1} Z_{\text{B}_1} Z_{\text{A}_4} Z_{\text{B}_4} X_{\text{A}_5} X_{\text{B}_5}$ \\
\rowcolor{lightgray}
     & $+1$ &    $e_1$ & $e_1$   &   $e_2+e_5$ & $e_2+e_5$    & $Z_{\text{A}_2} Z_{\text{B}_2} Z_{\text{A}_5} Z_{\text{B}_5} X_{\text{A}_1} X_{\text{B}_1}$ \\
\rowcolor{lightgray}
     & $+1$ &    $e_1+e_5$ & $e_1+e_5$   &   $e_3$ & $e_3$    & $Z_{\text{A}_3} Z_{\text{B}_3} X_{\text{A}_5} X_{\text{B}_5} X_{\text{A}_1} X_{\text{B}_1}$ \\
     & $\boldsymbol{ \varepsilon_1 }$ &    $\boldsymbol{ 10010 }$ & $\boldsymbol{ 00000 }$   &   $\boldsymbol{ 01100 }$ & $\boldsymbol{ 00000 }$    & $\boldsymbol{ \varepsilon_1 \, X_{\text{A}_1} Z_{\text{A}_2} Z_{\text{A}_3} X_{\text{A}_4} }$ \\
     & $\boldsymbol{ \varepsilon_2 }$ &    $\boldsymbol{ 01001 }$ & $\boldsymbol{ 00000 }$   &   $\boldsymbol{ 00110 }$ & $\boldsymbol{ 00000 }$    & $\boldsymbol{ \varepsilon_2 \, X_{\text{A}_2} Z_{\text{A}_3} Z_{\text{A}_4} X_{\text{A}_5} }$ \\
     &     &        &       &         &          &                   \\
\bottomrule
\bottomrule
\end{xltabular}
}


Given this ``initialization'', let us track these $10$ stabilizers through each step of the protocol, as shown in Table~\ref{tab:bell_protocol}. 

\begin{enumerate}

\item[(1)] Alice measures the first stabilizer generator $X_{\text{A}_1} Z_{\text{A}_2} Z_{\text{A}_3} X_{\text{A}_4}$, and 
assume that the measurement result is $\varepsilon_1 \in \{ \pm 1 \}$. 
We apply the stabilizer formalism for measurements from Section~\ref{sec:stabilizer_formalism} to update $\mathcal{E}_5$.
Since there are several elements of $\mathcal{E}_5$ that anticommute with this generator, we choose to remove%
\footnote{Later, in the GHZ protocol, we restrict this choice to be the first element in the table that anticommutes with the measured stabilizer.}
$Z_{\text{A}_4} Z_{\text{B}_4} = E([0^{\text{A}},0^{\text{B}}], [e_4^{\text{A}},e_4^{\text{B}}])$ and replace all other anticommuting elements by their product with $Z_{\text{A}_4} Z_{\text{B}_4}$.
Let this updated group in Step (1) of Table~\ref{tab:bell_protocol} be denoted as $\mathcal{E}_5^{(1)}$.
For visual clarity, code stabilizer rows are boldfaced and binary vectors are written out in full. \\

Now, we observe that if Bob measures the same generator $X_{\text{B}_1} Z_{\text{B}_2} Z_{\text{B}_3} X_{\text{B}_4}$ on his qubits, then it is trivial because it commutes with all elements in $\mathcal{E}_5^{(1)}$ and 
hence is already contained in $\mathcal{E}_5^{(1)}$.
This is a manifestation of the Bell state matrix identity discussed in Section~\ref{sec:bell_state_identity}.
Indeed, Bob's generator can be obtained by multiplying $X_{\text{A}_1} X_{\text{B}_1}, X_{\text{A}_4} X_{\text{B}_4}, Z_{\text{A}_2} Z_{\text{B}_2}$, $Z_{\text{A}_3} Z_{\text{B}_3},$ and $X_{\text{A}_1} Z_{\text{A}_2} Z_{\text{A}_3} X_{\text{A}_4}$ in Step (1) of Table~\ref{tab:bell_protocol}. \\

\item[(2)] Alice measures the second stabilizer generator $X_{\text{A}_2} Z_{\text{A}_3} Z_{\text{A}_4} X_{\text{A}_5}$, and 
assume that the measurement result is $\varepsilon_2 \in \{ \pm 1 \}$. 
Then, the new joint stabilizer group, $\mathcal{E}_5^{(2)}$, is given in Step (2) of Table~\ref{tab:bell_protocol}.
This stabilizer generator anticommutes with the third row of the top block and the second and fifth rows of the bottom block.
We have replaced $Z_{\text{A}_5} Z_{\text{B}_5}$ (fifth row of the bottom block) with this generator and multiplied the other anticommuting elements with $Z_{\text{A}_5} Z_{\text{B}_5}$.
It can be verified that the second stabilizer generator of Bob is already in $\mathcal{E}_5^{(2)}$. \\

\item[(3)] Alice measures the third stabilizer generator $X_{\text{A}_1} X_{\text{A}_3} Z_{\text{A}_4} Z_{\text{A}_5}$, and 
assume that the measurement result is $\varepsilon_3 \in \{ \pm 1 \}$. 
Then, the new joint stabilizer group, $\mathcal{E}_5^{(3)}$, is given in Step (3) of Table~\ref{tab:bell_protocol}.
Once again, it can be verified that the third stabilizer generator of Bob is already in $\mathcal{E}_5^{(3)}$.
The minus sign in the fourth row of the top block gets introduced when we apply the multiplication rule for $E(a,b)$ from Lemma~\ref{lem:Eab}(b). \\

\item[(4)] Alice measures the final stabilizer generator $Z_{\text{A}_1} X_{\text{A}_2} X_{\text{A}_4} Z_{\text{A}_5}$, and 
assume that the measurement result is $\varepsilon_4 \in \{ \pm 1 \}$. 
Then, the new joint stabilizer group, $\mathcal{E}_5^{(4)}$, is given in Step (4) of Table~\ref{tab:bell_protocol}.
As before, it can be verified that the final stabilizer generator of Bob is already in $\mathcal{E}_5^{(4)}$. 
This completes all measurements of Alice, and she now sends Bob's qubits over the channel.
To understand the working of the protocol in the ideal scenario, assume that no errors occur. \\

\end{enumerate}

Since we know that all stabilizer generators of Bob are in $\mathcal{E}_5^{(4)}$, we conveniently perform the following replacements:
\begin{align}
E([e_1^{\text{A}}, e_1^{\text{B}}],[(e_2+e_5)^{\text{A}}, (e_2+e_5)^{\text{B}}]) & \mapsto X_{\text{B}_1} Z_{\text{B}_2} Z_{\text{B}_3} X_{\text{B}_4}, \nonumber \\
E([(e_1+e_2)^{\text{A}}, (e_1+e_2)^{\text{B}}], [e_4^{\text{A}}, e_4^{\text{B}}]) & \mapsto X_{\text{B}_2} Z_{\text{B}_3} Z_{\text{B}_4} X_{\text{B}_5}, \nonumber \\
E([(e_1+e_3)^{\text{A}}, (e_1+e_3)^{\text{B}}], [(e_4+e_5)^{\text{A}}, (e_4+e_5)^{\text{B}}]) & \mapsto X_{\text{B}_1} X_{\text{B}_3} Z_{\text{B}_4} Z_{\text{B}_5}, \nonumber \\
E([e_5^{\text{A}}, e_5^{\text{B}}],[(e_1+e_4)^{\text{A}}, (e_1+e_4)^{\text{B}}]) & \mapsto Z_{\text{B}_1} X_{\text{B}_2} X_{\text{B}_4} Z_{\text{B}_5}.
\end{align}
Recollect that for the $\llbr 5,1,3 \rrbr$ code, the logical Pauli operators are $\overline{X} = X_1 X_2 X_3 X_4 X_5 = E([11111,00000])$ and $\overline{Z} = Z_1 Z_2 Z_3 Z_4 Z_5 = E([00000,11111])$.
If we used Algorithm~\ref{algo:logical_paulis_ghz_msmt}, we would obtain the same $\overline{Z}$ and $\overline{X} = - Y_1 Z_3 Z_4$.
Then, by grouping Alice's code stabilizers and Bob's code stabilizers, the group $\mathcal{E}_5^{(4)}$ can be rewritten as
\begin{align}
\mathcal{E}_5^{(4)} & = \langle \varepsilon_1 \, X_{\text{A}_1} Z_{\text{A}_2} Z_{\text{A}_3} X_{\text{A}_4} , \ \varepsilon_2 \,  X_{\text{A}_2} Z_{\text{A}_3} Z_{\text{A}_4} X_{\text{A}_5} , \ \varepsilon_3 \, X_{\text{A}_1} X_{\text{A}_3} Z_{\text{A}_4} Z_{\text{A}_5} , \ \varepsilon_4 \, Z_{\text{A}_1} X_{\text{A}_2} X_{\text{A}_4} Z_{\text{A}_5} , \nonumber \\
 & \hspace{-0.5cm} E([(e_1+e_5)^{\text{A}}, (e_1+e_5)^{\text{B}}],[e_3^{\text{A}}, e_3^{\text{B}}]) , \ -E([(e_1+e_4+e_5)^{\text{A}}, (e_1+e_4+e_5)^{\text{B}}], [e_5^{\text{A}}, e_5^{\text{B}}]) , \nonumber \\
  & \hspace{-0.5cm} \varepsilon_1 \, X_{\text{B}_1} Z_{\text{B}_2} Z_{\text{B}_3} X_{\text{B}_4} , \ \varepsilon_2 \, X_{\text{B}_2} Z_{\text{B}_3} Z_{\text{B}_4} X_{\text{B}_5} , \ \varepsilon_3 \, X_{\text{B}_1} X_{\text{B}_3} Z_{\text{B}_4} Z_{\text{B}_5}, \ \varepsilon_4 \, Z_{\text{B}_1} X_{\text{B}_2} X_{\text{B}_4} Z_{\text{B}_5} \rangle.
\end{align}
Using some manipulations, we see that the two 
operators on the second line in $\mathcal{E}_5^{(4)}$ are
\begin{align}
E([(e_1+e_5)^{\text{A}}, (e_1+e_5)^{\text{B}}],[e_3^{\text{A}}, e_3^{\text{B}}]) = (X_{\text{A}_1} Z_{\text{A}_3} X_{\text{A}_5}) (X_{\text{B}_1} Z_{\text{B}_3} X_{\text{B}_5}) & \equiv \overline{Z}_{\text{A}} \overline{Z}_{\text{B}}, \nonumber \\
-E([(e_1+e_4+e_5)^{\text{A}}, (e_1+e_4+e_5)^{\text{B}}], [e_5^{\text{A}}, e_5^{\text{B}}]) = (\imath X_{\text{A}_1} X_{\text{A}_4} Y_{\text{A}_5}) (\imath X_{\text{B}_1} X_{\text{B}_4} Y_{\text{B}_5}) & \equiv \overline{X}_{\text{A}} \overline{X}_{\text{B}}.
\end{align}
Thus, $\mathcal{E}_5^{(4)}$ can be interpreted as having $8$ stabilizer generators (Alice and Bob combined) and a pair of logical $X_{\text{A}} X_{\text{B}}$ and logical $Z_{\text{A}} Z_{\text{B}}$ operators, which implies that the pair of logical qubits shared between Alice and Bob forms a Bell pair. 
This can be converted into a physical Bell pair by performing the inverse of the encoding unitary on both Alice's and Bob's qubits locally. 
Note that this encoding unitary must be compatible with the above definition of the logical Paulis for the $\llbr 5,1,3 \rrbr$ code, i.e., when the physical $X$ and $Z$ on the input (logical) qubit to the encoder is conjugated by the chosen encoding unitary, the result must be the above logical Paulis $\overline{X}$ and $\overline{Z}$, respectively, potentially multiplied by some stabilizer element.

\begin{remark} \label{rem:Bell_local_msmt}
In this example, we have assumed that Bob's qubits do not suffer any error, so that we can clearly show the existence of the correct logical Bell stabilizers.
If, however, the channel introduced an error, then Alice and Bob can \textbf{jointly} deduce the error by measuring the signs of all generators of $\mathcal{E}_5^{(1)}$ and applying the necessary Pauli correction.
Since there are no non-trivial logical Pauli operators, any syndrome-matched correction can differ from the true error only by a stabilizer, so any error is correctable by the joint action of Alice and Bob.
But, since we prohibit \textbf{non-local} measurements between Alice and Bob, our error correction capability is limited to that of the code (on Bob's side).
If the channel introduces a correctable Pauli error for the chosen code and Bob's decoder, then the protocol will output $k$ perfect Bell pairs.
However, if the Pauli error is miscorrected by Bob's decoder, then there will be a logical error on the code, and hence at least one of the $k$ output Bell pairs will suffer from an unknown Pauli error.
\end{remark}

We can arrive at the above conclusion without knowing the specific logical operators for the code.
After Alice measures all her stabilizer generators, we know that Bob's stabilizer generators will also be present in the group, simply based on the Bell state matrix identity from Section~\ref{sec:bell_state_identity}.
For this example, the transpose in that identity did not make a difference, but for other codes this can only introduce an additional minus sign since $Y^T = -Y$.
For an $\llbr n,k,d \rrbr$ code, we now have a $2n$-qubit stabilizer group $\mathcal{E}_n^{(n-k)}$ where $2(n-k)$ generating elements are Alice's and Bob's stabilizer generators.
We are left with $2n - 2(n-k) = 2k$ elements in the generators, each of which \emph{must} jointly involve Alice's \emph{and} Bob's qubits.
These commute with each other and with the $2(n-k)$ stabilizer generators of Alice and Bob, and are independent, so we can rename them as the logical $X_{\text{A}_j} X_{\text{B}_j}$ and logical $Z_{\text{A}_j} Z_{\text{B}_j}$ for $j=1,2,\ldots,k$.
Thus, \emph{by definition}, the $k$ pairs of logical qubits form $k$ logical Bell pairs.
Alice and Bob can produce physical Bell pairs by simultaneously inverting the (same) encoding unitary for the code locally.
This is the key idea behind the working of the Bell pair distillation protocol employed by Wilde \etal in~\cite{Wilde-isit10}.

\section{Distillation of Greenberger-Horne-Zeilinger (GHZ) States}
\label{sec:ghz_distillation}

In this section, we extend the above Bell pair distillation protocol to distill GHZ states, $\dket{\text{GHZ}^{\ell}} = \frac{(\dket{00 \cdots 0} + \dket{11 \cdots 1})}{\sqrt{2}}$.
For clarity, we will specifically discuss the standard case of $\ell=3$, but the results and analysis extend to larger $\ell$ as well.
Let $n$ GHZ states be shared between Alice, Bob, and Charlie. 
We rearrange all the qubits to keep Alice's, Bob's and Charlie's qubits together respectively. 
Hence, this joint state can be expressed as
%
\begin{align}    
\label{eq:ghz_state_rearranged}
\dket{\text{GHZ}_n}_{\text{ABC}} & = \left(\frac{\dket{000}_{\text{ABC}} + \dket{111}_{\text{ABC}}}{\sqrt{2}}\right)^{\otimes n} = \frac{1}{\sqrt{2^n}} \sum_{x \in \mathbb{F}_2^n} \dket{x}_{\text{A}} \dket{x}_{\text{B}} \dket{x}_{\text{C}}. 
\end{align}
Since the GHZ state has stabilizers $S_{\text{GHZ}} = \langle Z_{\text{A}} Z_{\text{B}} I_{\text{C}}, \ I_{\text{A}} Z_{\text{B}} Z_{\text{C}}, \ X_{\text{A}} X_{\text{B}} X_{\text{C}} \rangle$, the stabilizers for $\dket{\text{GHZ}_n}_{\text{ABC}}$ are
\begin{align}
\label{eq:ghz_n_stabilizers}
S_{\text{GHZ}}^{\otimes n} = \langle \ Z_{\text{A}_i} Z_{\text{B}_i} I_{\text{C}_i}, \ I_{\text{A}_i} Z_{\text{B}_i} Z_{\text{C}_i}, \ X_{\text{A}_i} X_{\text{B}_i} X_{\text{C}_i} \ ; \ i = 1,2,\ldots,n \ \rangle.
\end{align}
Thus, we have identified the GHZ version of the basic properties of Bell states that was needed in the Bell pair distillation protocol.
However, the critical part of the Wilde \etal protocol was the transpose trick that formed the Bell matrix identity in Appendix~\ref{sec:bell_state_identity}.
When applied to stabilizer codes, this implied that each stabilizer generator $\varepsilon E(a,b)$ of Alice is transformed into the generator $\varepsilon E(a,b)^T = \varepsilon (-1)^{ab^T} E(a,b)$ (using Lemma~\ref{lem:Eab}(a)) for Bob.
Naturally, we need to determine the equivalent phenomenon for GHZ states before we can proceed to constructing a distillation protocol.

\subsection{GHZ State Matrix Identity}
\label{sec:ghz_state_identity}

In the following lemma, we generalize the Bell state matrix identity in Appendix~\ref{sec:bell_state_identity} to the GHZ state.

\begin{lemma}
\label{lem:ghz_state_identity}
Let $M = \sum_{x,y \in \mathbb{F}_2^n} M_{xy} \dketbra{x}{y} \in \mathbb{C}^{2^n \times 2^n}$ be any matrix acting on Alice's qubits.
Then, 
\begin{align*}
(M_{\text{A}} \otimes I_{\text{BC}}) \dket{\text{GHZ}_n}_{\text{ABC}} &= \left( I_{\text{A}} \otimes \left(\ghzmap{M^T}\right)_{\text{BC}} \right) \dket{\text{GHZ}_n}_{\text{ABC}} \ ; \\ 
\text{`GHZ-map'} \colon M \mapsto \ghzmap{M} & \coloneqq \sum_{x,y \in \mathbb{F}_2^n} M_{xy} \dketbra{x,x}{y,y} \in \mathbb{C}^{2^{2n} \times 2^{2n}}.
\end{align*}
\end{lemma}
\begin{IEEEproof}
Similar to the Bell case, we calculate
\begin{align}
(M_{\text{A}} \otimes I_{\text{BC}}) \dket{\text{GHZ}_n}_{\text{ABC}} & = \frac{1}{\sqrt{2^n}} \sum_{x,y \in \mathbb{F}_2^n} M_{xy} \dket{x}_{\text{A}} \dket{y}_{\text{B}} \dket{y}_{\text{C}} \\
  & = \frac{1}{\sqrt{2^n}} \sum_{x,y \in \mathbb{F}_2^n} \dket{x}_{\text{A}} (M^T)_{yx} \dket{y}_{\text{B}} \dket{y}_{\text{C}} \\
  & = \left( I_{\text{A}} \otimes \left(\ghzmap{M^T}\right)_{\text{BC}} \right) \dket{\text{GHZ}_n}_{\text{ABC}}.
\end{align}
This completes the proof and establishes the identity.
\end{IEEEproof}

The above property generalizes naturally to larger $\ell$-qubit GHZ states, $\dket{\text{GHZ}^{\ell}} = \frac{(\dket{00 \cdots 0} + \dket{11 \cdots 1})}{\sqrt{2}}$.

\begin{lemma}
\label{lem:ghz_state_identity_l}
Let $M = \sum_{x,y \in \mathbb{F}_2^n} M_{xy} \dketbra{x}{y} \in \mathbb{C}^{2^n \times 2^n}$ be any matrix acting on qubits `A'.
Then, 
\begin{align*}
(M_{\text{A}_1} \otimes I) \dket{\text{GHZ}_n^{\ell}}_{\text{A}_1 \cdots \text{A}_{\ell}} &= \left( I_{\text{A}_1} \otimes \left(\ghzmap{M^T}\right) \right) \dket{\text{GHZ}_n^{\ell}}_{\text{A}_1 \cdots \text{A}_{\ell}} \ ; \\ 
\text{`GHZ-map'} \colon M \mapsto \ghzmap{M} & \coloneqq \sum_{x,y \in \mathbb{F}_2^n} M_{xy} \dketbra{x}{y}^{\otimes (\ell-1)}. 
\end{align*}
\end{lemma}

As our next result, we prove some properties of the GHZ-map defined in the above lemma.

\begin{lemma}
The GHZ-map $M \in \mathbb{C}^{2^n \times 2^n} \mapsto \ghzmap{M} \in \mathbb{C}^{2^{2n} \times 2^{2n}}$ in Lemma~\ref{lem:ghz_state_identity} is an algebra homomorphism~\cite{Dummit-2004}: 
\begin{enumerate}

\item[(a)] Linear: If $M = \alpha A + \beta B$, where $\alpha, \beta \in \mathbb{C}$, then $\ghzmap{M} = \alpha \ghzmap{A} + \beta \ghzmap{B}$.

\item[(b)] Multiplicative: If $M = A B$, then $\ghzmap{M} = \ghzmap{A} \ghzmap{B}$.

\item[(c)] Projector-preserving: If $M$ is a projector, then $\ghzmap{M}$ is also a projector.

\end{enumerate}
\end{lemma}
\begin{IEEEproof}
We prove these properties via the definition of the mapping.
\begin{enumerate}
    
\item[(a)] Since $M_{xy} = \dbra{x} M \dket{y} = \alpha \dbra{x} A \dket{y} + \beta \dbra{x} B \dket{y} = \alpha A_{xy} + \beta B_{xy}$, the property follows.

\item[(b)] We observe that
\begin{align}
\ghzmap{A} \ghzmap{B} & = \sum_{x,y \in \mathbb{F}_2^n} A_{xy} \dketbra{x,x}{y,y} \cdot \sum_{x',y' \in \mathbb{F}_2^n} B_{x'y'} \dketbra{x',x'}{y',y'} \\
  & = \sum_{x,y' \in \mathbb{F}_2^n} \left[ \sum_{y \in \mathbb{F}_2^n} A_{xy} B_{yy'} \right] \dketbra{x,x}{y',y'} \\
  & = \sum_{x,y' \in \mathbb{F}_2^n} (AB)_{xy'} \dketbra{x,x}{y',y'} \\
  & = \ghzmap{AB} = \ghzmap{M}.
\end{align}

\item[(c)] This simply follows from the multiplicative property via the special case $A = B = M$.
    
\end{enumerate}
This completes the proof and establishes the said properties of the GHZ-map.
\end{IEEEproof}

We are interested in performing stabilizer measurements at Alice and deducing the effect on Bob's and Charlie's qubits.
The above properties greatly simplify the analysis, given that the code projector for a stabilizer code~\eqref{eq:stabilizer_projector} is a product of sums.
Due to the multiplicativity of the GHZ-map $M \mapsto \ghzmap{M}$, we only have to analyze the case where Alice's code has a single stabilizer generator $\varepsilon E(a,b)$, i.e., her code projector is simply $M = \frac{I_N + \varepsilon E(a,b)}{2}$, where $N = 2^n$.
Now, using linearity, we just need to determine $\ghzmap{I_N}$ and $\ghzmap{E(a,b)}$.
Then, due to Lemma~\ref{lem:Eab}(a), we have $\ghzmap{M^T} = \frac{1}{2} \left( \ghzmap{I_N} + (-1)^{ab^T} \ghzmap{E(a,b)} \right)$.


\begin{theorem}
\label{thm:ghz_stabilizer_measurement}
Given $n$ copies of the GHZ state shared between Alice, Bob and Charlie, measuring $E(a,b)$ on Alice's $n$ qubits and obtaining the result $\varepsilon \in \{ \pm 1 \}$ is equivalent to measuring the following with results $+1$ on the qubits of Bob and Charlie: 
\begin{align*}
\varepsilon E(a,b)_{\text{B}}^T \otimes E(a,0)_{\text{C}} = \varepsilon (-1)^{ab^T} E(a,b)_{\text{B}} \otimes E(a,0)_{\text{C}} \ \ \text{and} \\ 
\{ Z_{\text{B}_i} Z_{\text{C}_i} = E(0,e_i)_{\text{B}} \otimes E(0,e_i)_{\text{C}} \ ; \ i = 1,2,\ldots,n \}, 
\end{align*}
where $Z_{\text{B}_i}$ (resp. $Z_{\text{C}_i}$) refers to $Z$ on $i$-th qubit of Bob (resp. Charlie), and $e_i$ has a $1$ in the $i$-th position and zeros elsewhere.
\end{theorem}
\begin{IEEEproof}
Using the discussion before the statement of the theorem, we will calculate $\ghzmap{I_N}$ and $\ghzmap{E(a,b)}$ to establish the result.
Recollect that $\dketbra{0}_{n=1} = \frac{I + Z}{2}$ and hence $\dketbra{0}^{\otimes n} = \frac{1}{2^n} \sum_{v \in \mathbb{F}_2^n} E(0,v)$.
Then, using Lemma~\ref{lem:Eab}, we have
\begin{align}
\ghzmap{I_N} & = \sum_{x \in \mathbb{F}_2^n} \dketbra{x} \otimes \dketbra{x} \\
  & = \sum_{x \in \mathbb{F}_2^n} \left[ E(x,0) \dketbra{0}^{\otimes n} E(x,0) \right]^{\otimes 2} \\
  & = \sum_{x \in \mathbb{F}_2^n} \left[ E(x,0) \cdot \frac{1}{2^n} \sum_{v \in \mathbb{F}_2^n} E(0,v) \cdot E(x,0) \right]^{\otimes 2} \\
  & = \frac{1}{2^{2n}} \sum_{x \in \mathbb{F}_2^n} \left[ \sum_{v \in \mathbb{F}_2^n} (-1)^{xv^T} E(0,v) \right] \otimes \left[ \sum_{w \in \mathbb{F}_2^n} (-1)^{xw^T} E(0,w) \right] \\
  & = \frac{1}{2^{2n}} \sum_{x \in \mathbb{F}_2^n} \sum_{z \in \mathbb{F}_2^{2n}} (-1)^{[x,x] z^T} E(0,z) \quad (\text{where}\ z = [v,w]) \\
  & = \frac{1}{2^{2n}} \sum_{z \in \mathbb{F}_2^{2n}} E(0,z) \cdot \left( \sum_{x \in \mathbb{F}_2^n} (-1)^{[x,x] z^T} \right) \\
  & = \frac{1}{2^{2n}} \sum_{z \in \mathbb{F}_2^{2n}} E(0,z) \cdot 2^n \mathbb{I}\left(z \perp [x,x] \ \forall \ x \in \mathbb{F}_2^n \right) \\
  & = \frac{1}{2^n} \sum_{z' \in \mathbb{F}_2^n} E([0,0],[z',z']) \\
  & = \bigotimes_{i=1}^n \frac{\left( I_N + E([0,0],[e_i,e_i]) \right)}{2},
\end{align}
where $e_i \in \mathbb{F}_2^n$ is the standard basis vector with $1$ in the $i$-th position and zeros elsewhere.
Note that $E([0,0],[e_i,e_i])_{\text{BC}} = Z_{\text{B}_i} \otimes Z_{\text{C}_i}$ is the GHZ stabilizer $I_{\text{A}} Z_{\text{B}} Z_{\text{C}}$ on the $i$-th triple of qubits between A, B and C~\eqref{eq:ghz_n_stabilizers}.
Next, we proceed to calculate $\ghzmap{E(a,b)}$ using a similar approach.
\begin{align}
\ghzmap{E(a,b)} & = \sum_{x,y \in \mathbb{F}_2^n} \dbra{x} E(a,b) \dket{y} \dketbra{x,x}{y,y} \\
  & = \sum_{x,y \in \mathbb{F}_2^n} \dbra{x} \imath^{ab^T} (-1)^{by^T} \dket{y \oplus a} \dketbra{x,x}{y,y} \\
  & = \sum_{x \in \mathbb{F}_2^n} \imath^{ab^T} (-1)^{b (x \oplus a)^T} \dketbra{x,x}{x \oplus a, x \oplus a} \\
  & = \sum_{x \in \mathbb{F}_2^n} \imath^{-ab^T} (-1)^{b x^T} \left[ E(x,0) \cdot \dketbra{0}^{\otimes n} \cdot E(x \oplus a, 0) \right]^{\otimes 2} \\
  & = \sum_{x \in \mathbb{F}_2^n} \imath^{-ab^T} (-1)^{b x^T} \left[ E(x,0) \cdot \frac{1}{2^n} \sum_{v \in \mathbb{F}_2^n} E(0,v) \cdot E(x \oplus a, 0) \right]^{\otimes 2} \\ 
  & = \frac{1}{2^{2n}} \sum_{x \in \mathbb{F}_2^n} \imath^{-ab^T} (-1)^{b x^T} \left[ E(a,0) \cdot \sum_{v \in \mathbb{F}_2^n} (-1)^{xv^T + av^T} E(0,v) \right]^{\otimes 2} \quad (\text{Lemma~\ref{lem:Eab}(c)}) \\
  & = E(a,0)^{\otimes 2} \cdot \frac{1}{2^{2n}} \sum_{x \in \mathbb{F}_2^n} \imath^{-ab^T} (-1)^{b x^T} \sum_{z \in \mathbb{F}_2^{2n}} (-1)^{[x \oplus a, x \oplus a] z^T} E(0,z) \\
  & = E([a,a],[0,0]) \sum_{z \in \mathbb{F}_2^{2n}} \imath^{-ab^T} (-1)^{[a,a] z^T} E(0,z) \cdot \left( \frac{1}{2^{2n}} \sum_{x \in \mathbb{F}_2^n} (-1)^{[x,x] (z + [b,0])^T} \right) \\
  & = E([a,a],[0,0]) \cdot \frac{1}{2^n} \sum_{z' \in \mathbb{F}_2^{n}} \imath^{-ab^T} (-1)^{z'a^T + z'a^T + ab^T} E([0,0],[z' \oplus b, z']) \\
  & = E([a,a],[0,0]) \cdot \frac{\imath^{ab^T}}{2^n} \sum_{z' \in \mathbb{F}_2^{n}} E([0,0],[b,0]) \,  E([0,0],[z', z']) \\
  & = \imath^{-ab^T} E([a,a],[b,0]) \cdot \imath^{ab^T} \cdot \ghzmap{I_N} \quad (\text{Lemma~\ref{lem:Eab}(b)}) \\
  & = \left( E(a,b) \otimes E(a,0) \right) \cdot \ghzmap{I_N}.
\end{align}
Thus, when Alice's measurement applies the projector $M = \frac{I_N + \varepsilon E(a,b)}{2}$, Bob's and Charlie's qubits experience the projector
\begin{align}
\ghzmap{M^T} & = \frac{\ghzmap{I_N} + \varepsilon (-1)^{ab^T} \ghzmap{E(a,b)}}{2} \\
  & = \frac{\left( I_N \otimes I_N \right) \cdot \ghzmap{I_N} + \varepsilon (-1)^{ab^T} \left( E(a,b) \otimes E(a,0) \right) \cdot \ghzmap{I_N}}{2} \\
  & = \frac{ \left( I_N \otimes I_N + \varepsilon (-1)^{ab^T} E(a,b) \otimes E(a,0) \right) }{2} \cdot \bigotimes_{i=1}^n \frac{\left( I_N + E([0,0],[e_i,e_i]) \right)}{2}.
\end{align}
Since the second term, $\ghzmap{I_N}$, only corresponds to already existing stabilizers $Z_{\text{B}_i} \otimes Z_{\text{C}_i}$ for $n$ copies of the GHZ state, the only new measurement corresponds to the Pauli operator $\varepsilon (-1)^{ab^T} E(a,b) \otimes E(a,0)$. \hfill \IEEEQEDhere
\end{IEEEproof}

\begin{example}
\normalfont
Consider $n=1$ and the case when Alice applies $M = \frac{I+Z}{2} = \frac{I+E(0,1)}{2}$, with $a = 0, b = 1$.
Then $\ghzmap{I} = \frac{I \otimes I + Z \otimes Z}{2}$ and $\ghzmap{E(0,1)^T} = (E(0,1)^T \otimes E(0,0)) \cdot \ghzmap{I} = (Z \otimes I) \cdot \ghzmap{I}$.
Therefore, the stabilizers for BC are $\langle Z \otimes I, Z \otimes Z \rangle$.

If we had an $X$-measurement for Alice, where $a=1, b=0$, then $E(a,b)^T \otimes E(a,0) = X \otimes X$.
Combined with the $Z \otimes Z$ from $\ghzmap{I}$, the qubits on BC are projected to the Bell state.

More interestingly, if we consider a $Y$-measurement for Alice, where $a=b=1$, then $E(a,b)^T \otimes E(a,0) = Y^T \otimes X = - Y \otimes X$.
Thus, assuming the measurement result is $+1$, the new BC stabilizers  are $\langle - Y \otimes X, Z \otimes Z \rangle$.
It can be verified that the post-measurement state for this case will be $\frac{(\dket{0} + \imath \dket{1})}{\sqrt{2}} \otimes \frac{(\dket{00} - \imath \dket{11})}{\sqrt{2}}$, which is fixed by the above stabilizer. \hfill \IEEEQEDhere
\end{example}

Naturally, this insight can be generalized to larger GHZ states as well.

\begin{theorem}
\label{thm:ghz_stabilizer_measurement_l}
Given $n$ copies of the $\ell$-qubit GHZ state with subsystems $\text{A}_1, \text{A}_2, \ldots, \text{A}_{\ell}$, measuring $E(a,b)$ on the $n$ qubits of subsystem $\text{A}_1$ and obtaining the result $\varepsilon \in \{ \pm 1 \}$ is equivalent to measuring the following with results $+1$ on the qubits of the remaining $(\ell-1)$ subsystems: 
\begin{align*}
\varepsilon (-1)^{\left( \sum_{i=1}^{\ell-2} \sum_{j > i}^{\ell-1} b_i \ast b_j \right) a^T} \bigotimes_{t=2}^{\ell} E(a,b_{t-1})_{\text{A}_t}^T 
 & = \varepsilon (-1)^{\left( b + \sum_{i=1}^{\ell-2} \sum_{j > i}^{\ell-1} b_i \ast b_j \right) a^T} \bigotimes_{t=2}^{\ell} E(a,b_{t-1})_{\text{A}_t} \ , \\
 \{ Z_{\text{A}_2,i} Z_{\text{A}_3,i} & = E(0,e_i)_{\text{A}_2} \otimes E(0,e_i)_{\text{A}_3} \ , \\ 
  Z_{\text{A}_3,i} Z_{\text{A}_4,i} & = E(0,e_i)_{\text{A}_3} \otimes E(0,e_i)_{\text{A}_4} \ , \ \ldots \ , \\
  Z_{\text{A}_{\ell-1},i} Z_{\text{A}_{\ell},i} & = E(0,e_i)_{\text{A}_{\ell-1}} \otimes E(0,e_i)_{\text{A}_{\ell}} \ ; \ i = 1,2,\ldots,n \},
\end{align*}
where $b_1, b_2, \ldots, b_{\ell-1} \in \mathbb{F}_2^n$ satisfy $b_1 \oplus b_2 \oplus \cdots \oplus b_{\ell-1} = b$, $x \ast y$ denotes the element-wise product of two vectors, $Z_{\text{A}_t,i}$ refers to $Z$ on $i$-th qubit of subsystem $\text{A}_t$, and $e_i$ is the standard basis vector with a $1$ in the $i$-th position and zeros elsewhere.
\end{theorem}

\begin{remark}
There are two special cases that eliminate the sign in the new joint stabilizer.
One can set $b_1 = b$ and $b_2 = b_3 = \cdots = b_{\ell-1} = 0$, in which case $b_i \ast b_j = 0$ always.
More generally, one can define $\{ b_i \colon b_i \neq 0 \}$ such that $b_i \ast b_j = 0$ while $b_1 \oplus b_2 \oplus \cdots \oplus b_{\ell-1} = b$ still holds, i.e., splitting the entries of $b$ into $(\ell-1)$ disjoint groups. 
\end{remark}

As we desired, the above result shows how a Pauli measurement on one subsystem, $\text{A}_1$, of (multiple copies of) the GHZ state affects the remaining subsystems.
All the GHZ stabilizers involving subsystems $\text{A}_2, \text{A}_3, \ldots, \text{A}_{\ell}$ are retained.
Hence, the post-measurement state is ``GHZ-like'' on these $(\ell-1)$ subsystems but with an additional globally entangling stabilizer. 
This is akin to the globally entangling all-$X$ stabilizer for the standard GHZ state, but it depends on the Pauli operator being measured on $\text{A}_1$.
Note that, since the Pauli measurement randomly projects onto a subspace, the induced stabilizers given by the theorem do not uniquely determine the post-measurement state on the $(\ell-1)$ subsystems.
The degrees of freedom for the state will be quantified shortly in a more general setting.
One might argue that this theorem can be obtained by directly applying the stabilizer formalism to $S_{\text{GHZ}}$.
However, some thought clarifies that arriving at the conclusions rigorously takes at least an equal amount of effort.

In the context of measuring a set of $(n-k)$ stabilizer generators of a code (on qubits $\text{A}_1$), the above result confirms that this induces a joint \emph{stabilizer} code on the remaining $(\ell-1)$ subsystems.
There are $n(\ell-1)$ qubits on these subsystems and each code stabilizer generator contributes a stabilizer generator for this induced code.
Besides, as stated in the theorem, there are $n(\ell-2)$ GHZ stabilizers on all pairs of adjacent subsystems, $\{ \text{A}_j \text{A}_{j+1} \, ; \, j = 2,\ldots,\ell-1 \}$, independent of the code stabilizers being measured.
Hence, the induced code has $(n-k) + n(\ell-2)$ stabilizer generators, which means it is an $\llbr n(\ell-1), k \rrbr$ code and the post-measurement state has $k$ logical degrees of freedom.
The minimum distance of the induced code will depend on the minimum distance of the $\text{A}_1$-code as well as the new GHZ stabilizers and the choice of $\{ b_i \}$.

\subsection{Protocol I}

We now have all the tools to investigate a natural stabilizer code based GHZ distillation protocol that attempts to generalize the Bell pair distillation protocol discussed in Section~\ref{sec:bell_distillation}.
The block diagram of this protocol was shown earlier in Fig.~\ref{fig:GHZprotocol1} and the protocol is summarized as an algorithm in Algorithm~\ref{algo:algo_ghz}.
Let us consider the $3$-qubit code with stabilizers $S = \langle YYI, IYY \rangle$ to understand the subtleties in the steps of the protocol.
First, similar to the Bell pair scenario, we have the following stabilizer group for $3$ copies of the GHZ state:
\begin{align}
\label{eq:ghz_stabilizers_std}
\mathcal{G}_3 & = \langle \ Z_{\text{A}_i} Z_{\text{B}_i} I_{\text{C}_i}, \ I_{\text{A}_i} Z_{\text{B}_i} Z_{\text{C}_i}, \ X_{\text{A}_i} X_{\text{B}_i} X_{\text{C}_i} \ ; \ i = 1,2,3 \ \rangle \\
  & = \langle \ E([0^{\text{A}},0^{\text{B}},0^{\text{C}}],[e_i^{\text{A}},e_i^{\text{B}},0^{\text{C}}]), \ E([0^{\text{A}},0^{\text{B}},0^{\text{C}}],[0^{\text{A}},e_i^{\text{B}},e_i^{\text{C}}]), \nonumber \\
  & \hspace{6cm} E([e_i^{\text{A}},e_i^{\text{B}},e_i^{\text{C}}],[0^{\text{A}},0^{\text{B}},0^{\text{C}}]) \ ; \ i = 1,2,3 \ \rangle \\
  & = \langle \ E([0^{\text{A}},0^{\text{B}},0^{\text{C}}],[e_i^{\text{A}},e_i^{\text{B}},0^{\text{C}}]), \ E([0^{\text{A}},0^{\text{B}},0^{\text{C}}],[0^{\text{A}},e_i^{\text{B}},e_i^{\text{C}}]), \nonumber \\
  & \hspace{6cm} - E([e_i^{\text{A}},e_i^{\text{B}},e_i^{\text{C}}],[e_i^{\text{A}},e_i^{\text{B}},0^{\text{C}}]) \ ; \ i = 1,2,3 \ \rangle \\
  & = \langle \ Z_{\text{A}_i} Z_{\text{B}_i} I_{\text{C}_i}, \ I_{\text{A}_i} Z_{\text{B}_i} Z_{\text{C}_i}, \ - Y_{\text{A}_i} Y_{\text{B}_i} X_{\text{C}_i} \ ; \ i = 1,2,3 \ \rangle.
\end{align}
Next, like the example for the Bell pair distillation protocol, we can evolve these stabilizers through the proposed steps of the protocol to understand its working.
In Appendix~\ref{sec:protocol1_3qubit}, we use such a tabular approach to elucidate the steps of this protocol.
This serves as an instructive example to understand how the GHZ property influences the construction of a purification protocol for GHZ states.
In particular, since the property implies that the $Z$-component of any non-purely-$X$-type stabilizer is lost in the induced code on qubits `B' and `C', we discuss how one can perform diagonal Clifford operations to ensure that all three subsystems obtain the same code.
The placement of these operations is critical and we detail its effects by simulating the protocol performance for the $5$-qubit code.


In this protocol, Alice starts by preparing $n$ GHZ states and measuring the $n$-qubit stabilizers of her code on qubits `A'.
Then, using Theorem~\ref{thm:ghz_stabilizer_measurement}, Bob proceeds by measuring the $2n$-qubit stabilizers of the code induced on qubits `B' and `C' by Alice's choice of code on qubits `A'.
Subsequently, he also measures the same $n$-qubit stabilizers as Alice but on qubits `B', so that there is a code induced just on qubits `C' and Charlie can use that code to correct errors from the channel.
If we imagine the three parties being on a linear network topology, then this protocol seems reasonable since each party retains his/her qubits and passes on all remaining qubits to the next hop in the chain.
However, there is an asymmetry in the operations since Bob needs to perform two rounds of measurements and one involves twice the number of qubits.
Furthermore, the protocol is (potentially) not scalable to larger number of parties with varied network topologies.

\subsection{Distillation-Inspired Algorithm to Generate Logical Pauli Operators}

While constructing and analyzing the protocol using the tabular approach, we realized that the evolution of the table under stabilizer measurements automatically reveals the logical Pauli operators of the code in an explicit manner in certain rows.
Indeed, each stabilizer measurement replaces one row and alters several others that anticommute with it using the rules of the stabilizer formalism for measurements (Section~\ref{sec:stabilizer_formalism}).
After all stabilizers are measured on qubits `A', one realizes that the non-replaced (but altered) rows in the top section of the table, i.e., the $ZZI$-type rows, are of the form $\overline{Z}_i^{\text{A}} \overline{Z}_i^{\text{B}} \overline{I}_i^{\text{C}}$ where $\overline{Z}_i$ denotes the logical $Z$ operator on the $i$-th logical qubit of the code.
Therefore, one can easily read off these logical operators (up to some subtleties that can be taken care of).
A similar approach is applied to the bottom section of the table, i.e., the $XXX$-type rows, to obtain the logical $X$ operators of the code.
The details of the algorithm are discussed in Appendix~\ref{sec:logical_paulis_ghz_msmt} and the algorithm itself is summarized in Algorithm~\ref{algo:logical_paulis_ghz_msmt}.

\begin{algorithm} 
\DontPrintSemicolon
\SetAlgoLined
\SetKwInOut{Input}{Input}
\SetKwInOut{Output}{Output}
\Input{$n$ GHZ states $\ket{\text{GHZ}}^{\otimes n}$ at Alice, 
$\llbr n,k,d \rrbr$ stabilizer code $\mathcal{Q}(S)$ defined by a stabilizer group $S$}
\Output{$k$ GHZ states of higher quality shared between Alice and Bob if channel introduces a correctable error}
Initialization: Rearrange the $3n$ qubits in $\ket{\text{GHZ}}^{\otimes n}$ to obtain $\ket{\text{GHZ}_n}$~\eqref{eq:ghz_state_rearranged} for processing by Alice and Bob, respectively\;
  \;
 Alice \;
 (a) measures the stabilizer generators $\{ E(a_i,b_i) \, ; \, i = 1,2,\ldots,r=n-k \}$ on her $n$ qubits and obtains syndrome $\{ \varepsilon_i^{\text{A}} \}$, \;
 (b) sends the remaining $2n$ qubits to Bob over a noisy quantum channel, \;
 (c) sends the stabilizers, syndrome and logical Pauli operators to Bob over a perfect classical channel. \;
 \;
{  Bob \;
 (a) uses Theorem~\ref{thm:ghz_stabilizer_measurement} to define the $2n$-qubit joint BC code and measures all the $(2n-k)$ stabilizer generators 
 $$\{ \varepsilon_i^{\text{A}} E(a_i,b_i)_{\text{B}}^T \otimes E(a_i,0)_{\text{C}} \, , \, Z_{\text{B}_j} Z_{\text{C}_j} = E(0,e_j)_{\text{B}} \otimes E(0,e_j)_{\text{C}} \},$$ 
 for $i = 1,2,\ldots,r=n-k$ and $j = 1,2,\ldots,n$, on the received $2n$ qubits, \;
 (b) performs necessary Pauli corrections on all qubits to bring them to the code space of the joint BC code, \;
 (c) measures the stabilizer generators $\{ E(a_i,b_i) \, ; \, i = 1,2,\ldots,r=n-k \}$ on the $n$ qubits of subsystem B and obtains syndrome $\{ \varepsilon_i^{\text{B}} \}$; for purely $Z$-type stabilizers $E(0,b_i)$ the sign is $\varepsilon_i^{\text{A}}$, so we set $\varepsilon_i^{\text{B}} \coloneqq +1$ for them, \;
(d) sends the stabilizers, syndrome and logical Pauli operators to Charlie over a perfect classical channel, \;
(e) performs appropriate (see Appendix~\ref{sec:diagonal_clifford}) local diagonal Clifford on qubits C, \;
(f) sends qubits C to Charlie over a noisy Pauli channel. \;
  \;
 Charlie \; 
 (a) uses $\mathcal{Q}(S)$, Theorem~\ref{thm:ghz_stabilizer_measurement}, and Bob's syndrome to determine the signs $\varepsilon_i^{\text{A}} \varepsilon_i^{\text{B}} (-1)^{a_i b_i^T}$ of his stabilizers, and then measures the generators $\{ \varepsilon_i^{\text{A}} \varepsilon_i^{\text{B}} (-1)^{a_i b_i^T} E(a_i,b_i) \, ; \, i = 1,2,\ldots,r=n-k \}$ on his $n$ qubits, \;
 (b) performs the necessary Pauli corrections on all qubits to bring them to the code space of his code. \;
  \;
 // If the channel error was correctable, triples of logical qubits of Alice's, Bob's and Charlie's codes form $k$ GHZ states \;
 // If channel error was NOT correctable, some triple of logical qubits form a GHZ state with an unknown Pauli error \;
 Alice, Bob, and Charlie respectively apply the inverse of the encoding unitary for their code on their $n$ qubits\;
 // The encoding unitary is determined by the logical Pauli operators obtained from Algorithm~\ref{algo:logical_paulis_ghz_msmt} \;
}
  \caption{Protocol I to convert $n$ GHZ states into $k$ GHZ states of higher quality, using an $\llbr n,k,d \rrbr$ stabilizer code}
 \label{algo:algo_ghz}
\end{algorithm}

\begin{algorithm}

\DontPrintSemicolon
\SetAlgoLined

\caption{Algorithm to generate logical Paulis of a stabilizer code through GHZ measurements (see Appendix~\ref{sec:logical_paulis_ghz_msmt})}
\label{algo:logical_paulis_ghz_msmt}


\SetKwInOut{Input}{Input}
\SetKwInOut{Output}{Output}

\;

\Input{An $\llbr n,k,d \rrbr$ stabilizer code defined by its stabilizer generators $\{ \varepsilon_i E(a_i,b_i) \, ; \, i=1,2,\ldots,r=n-k \}$}

\;

\Output{The logical $X$ generators $\{ \nu_j E(c_j,d_j) \, ; \, j=1,\ldots,k \}$ and logical $Z$ generators $\{ E(0,f_j) \, ; \, j=1,\ldots,k \}$}

\;

Initialization: Form a $r \times (2n+1)$ binary parity-check matrix $H$ for the code, whose rows are $[a_i,b_i, \ \varepsilon_i]$. 
Preprocess the matrix so that its first $2n$ columns take the form $H_{1 : 2n} = \begin{bsmallmatrix} 0 & H_Z \\ H_1 & H_2 \end{bsmallmatrix}$, where $H_Z$ is a $r_Z \times n$ matrix of full rank, and $H_1$ is a $r_X \times n$ matrix of full rank ($r_X + r_Z = r = n-k$). 
The rows of $H_Z$ provide the generators for all purely $Z$-type stabilizers of the code. 
While performing row operations on $H_{1 : 2n}$, care must be taken to adhere to Pauli multiplication arithmetic (Lemma~\ref{lem:Eab}(b)). \;

\;

Simulate the creation of $n$ copies of the GHZ state as follows.
Create a $2n \times (6n+1)$ GHZ stabilizer matrix $S_{\text{GHZ}}$ whose first $n$ rows take the form $[0, 0, 0,\ e_i, e_i, 0,\ +1]$ and the second $n$ rows take the form $[e_i, e_i, e_i,\ 0, 0, 0,\ +1]$, where $i=1,2,\ldots,n$.
This matrix is almost the same as Step (0) in Table~\ref{tab:ghz_protocol}, but we have omitted the middle section. \;

\;

\For{$p=1$ to $r$}
{
  Simulate the measurement of $H^{(p)}$, the $p$-th row of $H$, on subsystem A of the GHZ states, using Section~\ref{sec:stabilizer_formalism}: \;
  Replace the first anticommuting row of $S_{\text{GHZ}}$ with $H^{(p)}$ and multiply subsequent anticommuting rows by $H^{(p)}$, using Lemma~\ref{lem:Eab}(b) \;
}

\;

\For{$q$ in the set of non-replaced rows of $S_{\text{GHZ}}$ with (row) index at most $n$}
{
  \eIf{$S_{\text{GHZ}}^{(q)}$ (only the $2n$ columns of subsystem A) is linearly independent from all rows of $H$}{
    Define a new logical $Z$ generator $E(0,f_j)$ from the A-columns of $S_{\text{GHZ}}^{(q)}$, with sign $+1$ \;
    Append $[0,\ f_j,\ +1]$ as a new row to $H$ \;
  }{
    continue \;
  }
}

// Now, $H$ has $n$ rows where the last $k$ rows correspond to the logical $Z$ generators $E(0,f_j)$ \;

\end{algorithm}

\begin{algorithm}[h]

\DontPrintSemicolon
\SetAlgoLined

\setcounter{AlgoLine}{21}

\;

\For{$q'$ in the set of non-replaced rows of $S_{\text{GHZ}}$ with (row) index at least $(n+1)$}
{
  \eIf{$S_{\text{GHZ}}^{(q')}$ (only the $2n$ columns of subsystem A) is linearly independent from all rows of $H$}{
    Define a new logical $X$ generator $\nu_j E(c_j,d_j)$ from the A-columns of $S_{\text{GHZ}}^{(q')}$, with sign $\nu_j$ (last column of $S_{\text{GHZ}}^{(q')}$) \;
    Append $[c_j,\ d_j,\ \nu_j]$ as a new row to $H$ \;
  }{
    continue \;
  }
}

// Now, we have $k$ logical $Z$ and logical $X$ generators, but they might not pair up appropriately \;

\;

Compute the $k \times k$ symplectic inner product matrix $T$ with entries $T_{ij} = \syminn{[0,f_i]}{[c_j,d_j]}$ for $i,j \in \{1,\ldots,k\}$ \;

\eIf{$T$ is not the $k \times k$ identity matrix}{
  Compute the binary inverse $T^{-1}$ of $T$ \;
  Form a $k \times n$ matrix $F$ whose rows are $f_j$ \;
  Define the new $f_j$'s as the rows of $T^{-1} F$ \;
}{
  Retain the definitions of logical $Z$ generators $E(0,f_j)$ and logical $X$ generators $\nu_j E(c_j,d_j)$ \;
}

\;

return $\{ \overline{Z}_j = E(0,f_j) \, , \, \overline{X}_j = \nu_j E(c_j,d_j) \, ; \, j = 1,2,\ldots,k \}$ \;

\end{algorithm}

\subsection{Output Fidelity of GHZ Distillation Protocol}
\label{sec:output_density}

During the protocol, if error correction at Bob and/or Charlie miscorrects and introduces a logical error, then the final effect is a change in the signs of some of the logical GHZ stabilizers.
This in turn means that after the decoding step, some of the $k$ triples will be the standard GHZ state corrupted by an unknown Pauli operation.
Hence, the output of the protocol is correct with probability $(1-P_f)$, and produces at least one Pauli corrupted GHZ state with probability $P_f$, using the notation in Fig.~\ref{fig:GHZ_5qubit_code}.
To make this precise, denote by $\dket{\text{GHZ}_0}, \dket{\text{GHZ}_1}, \ldots, \dket{\text{GHZ}_7}$ the eight possible variants of the GHZ state under Pauli operations, i.e., each variant has the stabilizer group $\langle \alpha_1 \, ZZI, \ \alpha_2 \, IZZ, \ \alpha_3 \, XXX \rangle$ with $\alpha_1, \alpha_2, \alpha_3 \in \{ \pm 1 \}$.
Then, assuming all variants are equally likely conditioned on a failure event, the density matrix representing the output of the protocol is 
\begin{align}
\rho_{\text{out}} = (1 - P_f) \dketbra{\text{GHZ}_{00\cdots 0}} + P_f \sum_{i=1}^{8^k-1} \frac{1}{8^k-1} \dketbra{\text{GHZ}_{i_1 i_2 \cdots i_k}},
\end{align}
where $\dketbra{\text{GHZ}_{00\cdots 0}} = \dketbra{\text{GHZ}}^{\otimes k}$, $(i_1 i_2 \cdots i_k)$ is the base-$8$ expansion of $i$, and $\dketbra{\text{GHZ}_{i_1 i_2 \cdots i_k}} = \dketbra{\text{GHZ}_{i_1}} \otimes \dketbra{\text{GHZ}_{i_2}} \otimes \cdots \otimes \dketbra{\text{GHZ}_{i_k}}$.

Similar to the case of triorthogonal codes in magic state distillation~\cite{Bravyi-pra12}, it is likely useful to consider the reduced density matrix for one of the $k$ output triples, and relate its fidelity (with respect to $\dketbra{\text{GHZ}}$) to properties of the code and decoder.
In Ref.~\cite{Bravyi-pra12}, the authors adopted exactly such a strategy for distillation of $T$-states, under a purely $Z$-error model and relying on post-selection where non-trivial syndromes are discarded.
In recent work~\cite{Krishna-arxiv18}, it has been shown that performing error correction rather than just detection (and post-selection) leads to better performance of triorthogonal codes.
For our scenario of GHZ distillation, it is an interesting problem to construct codes and decoders for this protocol where we can relate the output fidelity to code properties and arrive at analytical scaling arguments with increasing code size.
This would be useful for comparing with fundamental limits of entanglement distillation~\cite{Fang-it19} and assessing the optimality of this protocol.

\subsection{Protocol II}
\label{sec:protocol2}

\begin{algorithm} 
\DontPrintSemicolon
\SetAlgoLined
\SetKwInOut{Input}{Input}
\SetKwInOut{Output}{Output}
\Input{$n$ GHZ states $\ket{\text{GHZ}}^{\otimes n}$ at Alice, 
$\llbr n,k,d \rrbr$ CSS code $\mathcal{Q}(S)$ defined by a stabilizer group $S$}
\Output{$k$ GHZ states of higher quality shared between Alice and Bob if channel introduces a correctable error}
Initialization: Rearrange the $3n$ qubits in $\ket{\text{GHZ}}^{\otimes n}$ to obtain $\ket{\text{GHZ}_n}$~\eqref{eq:ghz_state_rearranged} for processing by Alice and Bob, respectively\;
  \;
 Alice \;
 (a) measures the stabilizer generators $\{ E(a_i,b_i) \, ; \, i = 1,2,\ldots,r=n-k \}$ on the $n$ qubits `A', obtains syndrome $\{ \varepsilon_i^{\text{A}} \}$, \;
 (b) measures the stabilizer generators $\{ E(a_i,b_i) \, ; \, i = 1,2,\ldots,r=n-k \}$ on the $n$ qubits`B', obtains syndrome $\{ \varepsilon_i^{\text{B}} \}$, \;
 (c) sends the stabilizers, syndrome $\{ \varepsilon_i^{\text{B}} \}$ and logical Pauli operators to Bob over a perfect classical channel, \;
 (d) sends the stabilizers, syndrome $\{ \varepsilon_i^{\text{A}}, \varepsilon_i^{\text{B}} \}$ and logical Pauli operators to Charlie over a perfect classical channel, \;
 (e) sends qubits `B' to Bob and qubits `C' to Charlie over noisy quantum channel.
 \;
{  Bob \;
 (a) measures the stabilizer generators $\{ \varepsilon_i^{\text{B}} E(a_i,b_i) \, ; \, i = 1,2,\ldots,r=n-k \}$ on the $n$ qubits of subsystem `B' and obtains syndrome; for purely $Z$-type stabilizers $E(0,b_i)$ the sign is $\varepsilon_i^{\text{A}}$ \;
 (b) uses the syndrome to run a decoder that estimates a Pauli error\;
 (c) applies the estimate to qubits `B' as the recovery operation\;
  \;
 Charlie \; 
 (a) uses $\mathcal{Q}(S)$, Theorem~\ref{thm:ghz_stabilizer_measurement}, and syndrome to determine the signs $\varepsilon_i^{\text{A}} \varepsilon_i^{\text{B}}$ of his stabilizers, and then measures the generators $\{ \varepsilon_i^{\text{A}} \varepsilon_i^{\text{B}} E(a_i,b_i) \, ; \, i = 1,2,\ldots,r=n-k \}$ on the $n$ qubits `C'; for purely $Z$-type stabilizers $E(0,b_i)$ the sign is $\varepsilon_i^{\text{A}}$, \;
 (b) uses the syndrome to run a decoder that estimates a Pauli error\;
 (c) applies the estimate to qubits `C' as the recovery operation\;
  \;
 // If the channel error was correctable, triples of logical qubits of Alice's, Bob's and Charlie's codes form $k$ GHZ states \;
 // If channel error was NOT correctable, some triple of logical qubits form a GHZ state with an unknown Pauli error \;
 Alice, Bob, and Charlie respectively apply the inverse of the encoding unitary for their code on their $n$ qubits\;
 // The encoding unitary is determined by the logical Pauli operators obtained from Algorithm~\ref{algo:logical_paulis_ghz_msmt} \;
}
  \caption{Protocol II to convert $n$ GHZ states into $k$ GHZ states of higher quality, using an $\llbr n,k,d \rrbr$ CSS code}
 \label{algo:algo_ghz_2}
\end{algorithm}

To address the drawbacks of Protocol I, the protocol can be modified so that Alice starts by measuring qubits `A' and qubits `B' separately.
Though this does not circumvent the issue of performing twice the number of measurements at one of the nodes, this avoids the need of $2n$-qubit measurements.
Since the GHZ property implies the inducement of a $2n$-qubit code on qubits `B' and `C', it appears that this extra round of $n$-qubit measurements on qubits `B' is inevitable.
So, even now, when Alice measures $\{ \varepsilon_i E(a_i,b_i) \}$ on qubits `A', Theorem~\ref{thm:ghz_stabilizer_measurement} still dictates that there is a $2n$-qubit code $\{ \varepsilon_i E(a_i,b_i)_{\text{B}}^T \otimes E(a_i,0)_{\text{C}} \}$ jointly on qubits `B' and `C'.
But, when she measures the same stabilizers $\{ \varepsilon_i E(a_i,b_i) \}$ on qubits `B', one can multiply with the corresponding $2n$-qubit stabilizer to see that the joint stabilizers can be broken into purely `B' and purely `C' stabilizers.
Therefore, once Alice performs the two rounds of measurements, she can send qubits `B' to Bob and qubits `C' to Charlie, along with the necessary classical information.
As individual codes have been induced separately on qubits `B' and qubits `C', Bob and Charlie can still perform local $n$-qubit measurements to fix errors during qubit transmission.
Finally, this scheme suits other network topologies such as when Alice is connected to both Bob and Charlie but those parties do not have a direct connection between them.

While Protocol II can be generalized to arbitrary stabilizer codes using the diagonal Clifford correction discussed in Protocol I, Algorithm~\ref{algo:algo_ghz_2} describes Protocol II specifically for CSS codes, just for simplicity.
Note that for CSS codes, for any stabilizer generator $\varepsilon_i E(a_i,b_i)$, whenever $a_i \neq 0$ we have $b_i = 0$. 
Hence, the induced code from Theorem~\ref{thm:ghz_stabilizer_measurement} is automatically CSS and we do not need any diagonal Clifford operation mentioned earlier in Protocol I.
Since Protocol II relies on the same intuitions from Theorem~\ref{thm:ghz_stabilizer_measurement}, we do not elaborate further.
We also note that there can be further variations based on other practical considerations.

This simplified protocol was shown earlier in Fig.~\ref{fig:GHZprotocol2}.
We simulated the protocol by following a tabular approach, as in Appendix~\ref{sec:protocol1_3qubit} for Protocol I, using a state-of-the-art family of lifted product QLDPC codes with asymptotic rate $0.118$ and an efficient iterative decoder based on the min-sum algorithm (MSA) with normalization factor $0.8$.
The results were shown in Fig.~\ref{fig:GHZsimple_LP118_MSAseqvars80}, where we can see that the threshold under depolarizing noise is about $10.7\%$.
Comparing the results to Fig.~\ref{fig:LP118_MSAseqvars80}, it is apparent that the threshold matches that of the underlying logical error rate of the code on this channel (i.e., no distillation but standard quantum error correction simulation).
This is important because it shows that even when both Bob and Charlie run decoders to correct errors on their respective qubits, the overall threshold is unchanged from the single channel setting.
On the other hand, the comparison also shows that the protocol failure rate is significantly worse for each channel parameter compared to Fig.~\ref{fig:LP118_MSAseqvars80}.
This could be the effect of requiring both decoders to succeed, but it is a cause for concern when we extend the protocol to GHZ states with $\ell > 3$.
Indeed, we do not want the protocol failure rates to progressively get worse, albeit with the same threshold.
Therefore, we will study this phenomenon more carefully in future work and identify the best way to scale this protocol for larger $\ell$.
For completeness, we summarize the protocol for arbitrary $\ell$.

\subsection{Protocol II for Arbitrary $\ell$}

Initially, $\text{A}_1$ generates $n$ ideal copies of the $\ell$-qubit GHZ state, names the qubits of each copy $\text{A}_1$ through $\text{A}_{\ell}$, chooses some $\llbr n,k \rrbr$ code $\mathcal{Q}(\mathcal{S})$ defined by a stabilizer $\mathcal{S}$, and measures the $(n-k)$ generators of $\mathcal{S}$ on qubits $\text{A}_1$.
Then, $\text{A}_1$ applies Theorem~\ref{thm:ghz_stabilizer_measurement} to determine the induced code $\mathcal{Q}\left( \mathcal{S}^{(\ell-1)} \right)$ on the remaining subsystems.
Let us consider $\ell=4$ for simplicity.
For tracking the protocol, we initially create a table whose rows are the binary representations of the generators of $\mathcal{S}_{\text{GHZ}}^{\otimes n}$.
Group the $n$ $Z_{\text{A}_i} Z_{\text{B}_i} I_{\text{C}_i} I_{\text{D}_i}$ generators in the first part of the table, the $n$ $I_{\text{A}_i} Z_{\text{B}_i} Z_{\text{C}_i} I_{\text{D}_i}$ generators in the second part, the $n$ $I_{\text{A}_i} I_{\text{B}_i} Z_{\text{C}_i} Z_{\text{D}_i}$ in the third part, and finally the $n$ $X_{\text{A}_i} X_{\text{B}_i} X_{\text{C}_i} X_{\text{D}_i}$ in the fourth part.
If there is a purely $Z$-type generator, $E(0,b)_{\text{A}}$, for $\mathcal{S}$, then it will commute with the first three parts and only affect the last part based on the stabilizer formalism.
Moreover, by an appropriate linear combination of the rows of the first part, one can produce the element $E(0,b)_{\text{A}} \otimes E(0,b)_{\text{B}}$, which when multiplied by the new code stabilizer produces the stabilizer $E(0,b)_{\text{B}}$ on purely subsystem `B'.
By a similar trick in the second part and subsequently in the third part, one can produce single-subsystem stabilizers $E(0,b)_{\text{C}}$ and $E(0,b)_{\text{D}}$ as well.
Hence, it suffices to only consider non-purely-$Z$-type stabilizers $E(a,b)_{\text{A}}, a \neq 0$.

Such stabilizers transform into the multiple-subsystem stabilizers described by Theorem~\ref{thm:ghz_stabilizer_measurement}.
Now, qubits of `B', `C', and `D' need to be transmitted over a noisy channel to the respective nodes, based on the network topology.
For those nodes to be able to correct errors, a code needs to be imposed purely on each subsystem \emph{before} transmission of the respective qubits.
Let (node) A be connected to (node) B.
Then, based on the choice of $b_1, b_2, b_3$ in Theorem~\ref{thm:ghz_stabilizer_measurement}, A measures code stabilizers $E(a,b_1)_{\text{B}}$ on qubits `B'.
With some thought, one sees that these stabilizers only affect the second part of the table.
Now, since $\pm E(a,b_1)_{\text{B}} \otimes E(a,b_2)_{\text{C}} \otimes E(a,b_3)_{\text{D}}$ is already a stabilizer, by multiplying with $E(a,b_1)_{\text{B}}$ we obtain a code on `B' and a residual code jointly on `C' and `D'.
The qubits of `B' can be transmitted to node B (along with necessary classical sign information of stabilizers), which can perform error correction.

If A is not connected to C and D, then A has to send those qubits to B.
Thus, it appears that A has to perform stabilizer measurements as above not only on `B' but on `C' and `D' as well.
However, this can be relegated to subsequent nodes to reduce the burden on A.
Let A also send qubits `C' and `D' to node B along with qubits `B'.
There is some joint Pauli error on `B', `C', and `D', and the error correction of B only fixes the error part on `B'.
If B measures code stabilizers on `C', then the preexisting Pauli error can be transformed into an effective Pauli error \emph{after} the code was imposed on `C'.
This enables node C to correct this error as well as any error encountered while B sends qubits `C'.
A similar statement holds for D as well.
Thus, the protocol can be stated as follows: for every edge connected to a node, the node performs stabilizer measurements on the respective subsystem to impose a code on the qubits of the recipient on that edge.
The correctness of the protocol relies on carefully tracking signs of stabilizers based on such measurements at each node.
Once all qubits are distributed, each node uses the logical Paulis of their respective codes to determine and invert the encoding unitary. 
This converts the $k$ logical GHZ states into $k$ perfect physical GHZ states, provided all error corrections were successful.
The average output density matrix and average output fidelity still take the form discussed in Section~\ref{sec:output_density}.

\section{Conclusion and Future Work}
\label{sec:conclusion}

In this work, we began by describing the Bell pair distillation protocol introduced in Ref.~\cite{Wilde-isit10}, and used the stabilizer formalism to understand its working.
We identified that the Bell state matrix identity (Appendix~\ref{sec:bell_state_identity}) plays a critical role in that protocol.
As our first result, we proved the equivalent matrix identity for GHZ states, where we introduced the GHZ-map and showed that it is an algebra homomorphism.
Using the GHZ-map, we proved our main result (Theorem~\ref{thm:ghz_stabilizer_measurement}) that describes the effect of Alice's stabilizer measurements (on qubits `A') on qubits `B' and qubits `C'.
Then, we constructed a natural GHZ distillation protocol whose steps were guided by the aforementioned main result.
We demonstrated that the placement of a certain local Clifford on qubits `C' in the protocol has an immense effect on the performance of the protocol.
We described the relation between the probability of failure of the protocol and the output fidelity of the GHZ states.
As part of our protocol, we also developed a new algorithm to generate logical Pauli operators for an arbitrary stabilizer code.
To circumvent some drawbacks of the protocol, we described an alternate protocol and produced performance results using state-of-the-art QLDPC codes and an efficient iterative decoder.
Finally, we discussed the scalability of the protocol for larger GHZ states involving more than $3$ parties and arbitrary network topologies.

In future work, we plan to study the scaling of the logical error rate with the increase in number of parties.
Since a key motivation for this work was distributed quantum computing (DQC), we will investigate a complete architecture for a distributed implementation of the recently proposed optimal families of QLDPC codes.
As part of the architecture, we envisage that the QLDPC-based GHZ purification scheme proposed in this paper will play a critical role in supplying logical GHZ states encoded in the same QLDPC codes that are used for DQC.
We will study the implications for fault-tolerance of such an architecture.

\section*{Acknowledgements}

The authors would like to thank Kaushik Seshadreesan and Saikat Guha for helpful discussions, and Mark Wilde for describing his protocol to distill Bell pairs.
This work is funded by the NSF Center for Quantum Networks (CQN), under the grant NSF-ERC 1941583, and also by the NSF grants CIF-1855879, CIF-2106189,
CCF-2100013 and ECCS/CCSS-2027844. 
This research was carried out in part at the Jet Propulsion Laboratory, California Institute of Technology, under a contract with the National Aeronautics and Space Administration and funded through JPL’s Strategic University Research Partnerships (SURP) program.
Bane Vasi\'c has disclosed an outside interest in his startup company Codelucida to the University of Arizona.  
Conflicts of interest resulting from this interest are being managed by The University of Arizona in accordance with its policies.


\appendix

\section{Notation and Background}
\label{sec:background}


\subsection{Pauli Matrices}
\label{sec:paulis}

We will use the standard Dirac notation to represent pure quantum states. 
An arbitrary $n$-qubit state will be denoted as a ket $\dket{\psi} = \sum_{v \in \mathbb{F}_2^n} \alpha_v \dket{v}$, where $\{ \alpha_v \in \mathbb{C}\, ; \, v \in \mathbb{F}_2^n \}$ satisfy $\sum_{v \in \mathbb{F}_2^n} |\alpha_v|^2 = 1$ as required by the Born rule~\cite{Wilde-2013}.
Here, $\dket{v} = \dket{v_1} \otimes \dket{v_2} \otimes \cdots \otimes \dket{v_n}$ is a standard basis vector for $v = [v_1,v_2,\ldots,v_n]$, with $v_i \in \mathbb{F}_2 = \{0,1\}$, and $\otimes$ denotes the Kronecker (or tensor) product.
Define $\imath \coloneqq \sqrt{-1}$.
Then, the well-known $n$-qubit Pauli group $\mathcal{P}_n$ consists of tensor products of the single-qubit Pauli matrices
\begin{align}
I = 
\begin{bmatrix}
1 & 0 \\ 0 & 1
\end{bmatrix}, \ 
X = 
\begin{bmatrix}
0 & 1 \\ 1 & 0
\end{bmatrix}, \ 
Z = 
\begin{bmatrix}
1 & 0 \\ 0 & -1
\end{bmatrix}, \ 
Y = \imath XZ =
\begin{bmatrix}
0 & -\imath \\ \imath & 0
\end{bmatrix},
\end{align}
multiplied by scalars $\imath^\kappa, \kappa \in \{0,1,2,3\}$, i.e., 
\begin{align}
\mathcal{P}_n \coloneqq \{ \imath^\kappa E_1 \otimes E_2 \otimes \cdots \otimes E_n, \ E_i \in \{I,X,Z,Y\}, \ \kappa \in \mathbb{Z}_4 = \{0,1,2,3\} \}.
\end{align}

Given two binary row vectors $a = [a_1,a_2,\ldots,a_n], b = [b_1,b_2,\ldots,b_n] \in \mathbb{F}_2^n$, we will write $E(a,b)$ to denote an arbitrary Hermitian (and unitary) Pauli matrix, where $a$ represents the ``$X$-component'' and $b$ represents the ``$Z$-component''~\cite{Rengaswamy-pra19}:
\begin{align}
\label{eq:Eab}
E(a,b) \coloneqq \left( \imath^{a_1 b_1} X^{a_1} Z^{b_1} \right) \otimes \left( \imath^{a_2 b_2} X^{a_2} Z^{b_2} \right) \otimes \cdots \otimes \left( \imath^{a_n b_n} X^{a_n} Z^{b_n} \right) = \imath^{ab^T} \bigotimes_{i=1}^n (X^{a_i} Z^{b_i}).
\end{align}
For example, $E([1,0,1],[0,1,1]) = X \otimes Z \otimes Y$.
It can be verified that $E(a,b)^2 = I_N = I^{\otimes n}$, where $N \coloneqq 2^n$.
Hence,
\begin{align}
\label{eq:Pauli_group}
\mathcal{P}_n = \{ \imath^\kappa E(a,b) \colon \ a,b \in \mathbb{F}_2^n, \ \kappa \in \mathbb{Z}_4 = \{0,1,2,3\} \}.
\end{align}
Using the properties of the Kronecker product, primarily the identities $(A \otimes B) (C \otimes D) = (AC) \otimes (BD)$ and $(A \otimes B)^T = A^T \otimes B^T$, we can show the following.
We represent standard addition by ``$+$'' and modulo $2$ addition by ``$\oplus$''.

\begin{lemma}
\label{lem:Eab}
For any $a,b \in \mathbb{F}_2^n$, the Pauli matrix $E(a,b)$ satisfies the following properties:
\begin{enumerate}
    
\item[(a)] $E(a,b)^T = (-1)^{ab^T} E(a,b)$;
    
\item[(b)] $E(a,b) \cdot E(c,d) = \imath^{bc^T - ad^T} E(a+c, b+d)$, where the exponent and the sums $(a+c), (b+d)$ are performed modulo $4$ and the definition in~\eqref{eq:Eab} is directly extended
 to $a,b \in \mathbb{Z}_4^n$;
    
\item[(c)] $E(a,b) \cdot E(c,d) = (-1)^{\syminn{[a,b]}{[c,d]}} E(c,d) \cdot E(a,b)$, where 
\begin{align}
\label{eq:syminn}
\syminn{[a,b]}{[c,d]} \coloneqq ad^T + bc^T \ (\bmod\ 2)
\end{align}
is the symplectic inner product between $[a,b]$ and $[c,d]$ in $\mathbb{F}_2^{2n}$, as indicated by the subscript `s'.
    
\end{enumerate}
\end{lemma}

Hence, the map $\gamma \colon (\mathcal{P}_n, \cdot) \rightarrow (\mathbb{F}_2^{2n}, \oplus)$ defined by $E(a,b) \mapsto [a,b]$ is a homomorphism with kernel $\{ \imath^\kappa I_N,\ \kappa \in \mathbb{Z}_4 \}$.
For details about extending the definition of $E(a,b)$ to $\mathbb{Z}_4$-valued arguments, see~\cite{Rengaswamy-pra19}.

\subsection{Stabilizer Codes and Encoding Unitaries}
\label{sec:stabilizer_codes}

A stabilizer group $S$ is a commutative subgroup of $\mathcal{P}_n$ that does not contain $-I_N$.
If the group has $r \leq n$ independent generators $\varepsilon_i E(a_i,b_i)$, where $\varepsilon_i \in \{ \pm 1 \}$, then $S = \langle \varepsilon_i E(a_i,b_i); \ i = 1,\ldots,r \rangle$ has size $|S| = 2^r$.
Since the generators are Hermitian and unitary, they have eigenvalues $\pm 1$. 
Recollect that commuting matrices can be diagonalized simultaneously.
The stabilizer code defined by $S$ is the common $+1$ eigenspace of all generators, i.e., it is the $2^k$-dimensional subspace, $k = n-r$, fixed by all elements of $S$:
\begin{align}
\mathcal{Q}(S) \coloneqq \{ \dket{\psi} \in \mathbb{C}^N \colon g \dket{\psi} = \dket{\psi}\ \forall\ g \in S \}.
\end{align}

Using the homomorphism $\gamma$, we can write a $r \times (2n+1)$ generator matrix $G_S$ for the stabilizer group: the $i^{\text{th}}$ row of $G_S$ is $[a_i,b_i,\ \varepsilon_i] \in \mathbb{F}_2^{2n} \times \{ \pm 1 \}$.
Since $S$ must be a commutative group, the symplectic inner product between any pair of rows must be zero.
Hence, the subspace of binary mappings of all elements of $S$, denoted $\gamma(S)$, is given by the rowspace of $G_S$.

A CSS (Calderbank-Shor-Steane) code is a special type of stabilizer code for which there exists a set of generators where either $b_i = 0$ or $a_i = 0$ in each generator, i.e., the generators are purely $X$-type and purely $Z$-type operators.
Clearly, for such a code, $G_S$ has a block diagonal form where we can express the $X$-type (resp. $Z$-type) operators as the rowspace of a matrix $[H_X, 0]$ (resp. $[0, H_Z]$), and $0$ represents the all-zeros matrix (of appropriate size).
In this case, the commutativity condition for stabilizers is equivalent to the condition $H_X H_Z^T = 0$.
Therefore, $H_X$ and $H_Z$ can be thought of as generating two classical linear codes $\MCC_X$ and $\MCC_Z$.

The projector onto the $+1$ eigenspace of a Pauli matrix $E(a,b)$ is $\frac{I_N + E(a,b)}{2}$.
Therefore, since $\mathcal{Q}(S)$ is the simultaneous $+1$ eigenspace of $r$ commuting matrices $\varepsilon_i E(a_i,b_i)$, the projector onto the code subspace $\mathcal{Q}(S)$ is
\begin{align}
\label{eq:stabilizer_projector}
\Pi_S = \prod_{i=1}^r \frac{(I_N + \varepsilon_i E(a_i,b_i))}{2} = \frac{1}{2^r} \sum_{m = [m_1,\ldots,m_r] \in \mathbb{F}_2^r} \prod_{i=1}^r \left( \varepsilon_i E(a_i,b_i) \right)^{m_i} = \frac{1}{2^r} \sum_{\varepsilon E(a,b) \in S} \varepsilon E(a,b).
\end{align}

While the stabilizer group $S$ defines the code space, an encoding unitary $\mathcal{U}_{\text{Enc}}(S)$ fully specifies the mapping from logical $k$-qubit states to physical $n$-qubit code states in $\mathcal{Q}(S)$.
The $n$ input qubits to $\mathcal{U}_{\text{Enc}}(S)$ can be split into $k$ logical qubits, whose joint state is arbitrary, and $r = n-k$ ancillary qubits, each of which is initialized in some specific state such as $\dket{0}$.
If ancillas are initialized in the $\dket{0}$ state, then the stabilizer group for these $n$ input qubits is generated by $\{ Z_i \, ; \, i = k+1,k+2,\ldots,n \}$, since $Z \dket{0} = \dket{0}$.
If we conjugate each of these $r$ generators by the encoding unitary $\mathcal{U}_{\text{Enc}}(S)$, then we will obtain $r$ generators $\mathcal{U}_{\text{Enc}}(S) \, Z_i \, \mathcal{U}_{\text{Enc}}(S)^\dagger$ of $S$.
Similarly, if we conjugate the $X_i$ and $Z_i$ operations on the $k$ logical qubits --- which can be used to express arbitrary operations on them since Pauli operators form a basis --- by $\mathcal{U}_{\text{Enc}}(S)$, then we will obtain the generators of logical $X$ and $Z$ operators compatible with the chosen $\mathcal{U}_{\text{Enc}}(S)$.
Therefore, an alternative method to specify $\mathcal{U}_{\text{Enc}}(S)$ is to specify the generators of $S$ as well as the generators of logical $X$ and $Z$ operators.
Since we are requiring $\mathcal{U}_{\text{Enc}}(S)$ to map Paulis to Paulis, it is always Clifford~\cite{Gottesman-icgtmp98}.
Note that $\mathcal{U}_{\text{Enc}}(S)$ is still not unique since we are not specifying how $X_i$ on the ancillas must be mapped, but we do not care about these additional mappings.

There are at least two algorithms provided in the literature for generating the logical Pauli operators of stabilizer codes.
One is by Gottesman~\cite{Gottesman-phd97,Nielsen-2010}, where the idea is to construct the normalizer of the stabilizer group inside the Pauli group, and then perform suitable row operations on the generators of the normalizer.
The other is by Wilde~\cite{Wilde-physreva09}, where he performs a symplectic Gram-Schmidt orthogonalization procedure to arrive at the generators of logical $X$ and logical $Z$ operators.
In this work, as part of our GHZ distillation protocol, we provide a new algorithm to generate logical $X$ and $Z$ operators for any stabilizer code (see Algorithm~\ref{algo:logical_paulis_ghz_msmt}).
The output of the algorithm is compatible with the way logical Paulis must be defined for our analysis of the protocol.
Additionally, the logical $Z$ operators from our algorithm are always guaranteed to be purely $Z$-type operators for any stabilizer code.
If the code is CSS, then the logical $X$ operators are always purely $X$-type.

\subsection{Stabilizer Formalism}
\label{sec:stabilizer_formalism}

When a stabilizer group on $n$ qubits has $n$ independent generators, $\mathcal{Q}(S)$ is a $1$-dimensional subspace that corresponds to a unique quantum state $\dket{\psi(S)}$ (up to an irrelevant global phase), commonly referred to as a stabilizer state~\cite{Gottesman-phd97}.
The actions of unitary operations and measurements on $\dket{\psi(S)}$ can be tracked by updating these $n$ generators accordingly~\cite{Gottesman-icgtmp98,Aaronson-pra04}.
For any element $g$ of $S$, and an arbitrary unitary operation $U$ on $\dket{\psi(S)}$, we observe that
\begin{align}
U \dket{\psi(S)} = U \cdot g \cdot\dket{\psi(S)} = (UgU^\dagger) \cdot U \dket{\psi(S)},
\end{align}
so the stabilizer element $g$ has evolved into the element $g' = UgU^\dagger$ after the action of $U$.
Of course, only if $U$ is a Clifford operation we have that $g'$ is also a Pauli matrix.
Thus, in this case the evolution of the state can be tracked efficiently by simply transforming $G_S$ (and the associated signs) through binary operations (see ``CHP'' algorithm~\cite{Aaronson-pra04}).

The stabilizer formalism also provides a method to systematically update the stabilizers under Pauli measurements of the state $\dket{\psi(S)}$.
Assume that we have $n$ generators for the stabilizer group, namely $\varepsilon_i E(a_i,b_i), i = 1,\ldots,n$, and that we are measuring the Pauli operator $\mu E(u,v)$ to obtain the measurement $(-1)^m, m \in \{0,1\}$. Then, we have the following cases.


\begin{enumerate}
\item If $\syminn{[u,v]}{[a_i,b_i]} = 0$ for all $i$, then either $E(u,v)$ or $-E(u,v)$ already belongs to $S$, so there is nothing to update.

\item If $\syminn{[u,v]}{[a_j,b_j]} = 1$ for exactly one $j \in \{1,\ldots,n\}$, then we replace $\varepsilon_j E(a_j,b_j)$ by $(-1)^m \mu E(u,v)$.

\item If $\syminn{[u,v]}{[a_i,b_i]} = 1$ for $i \in \mathcal{I} \subseteq \{1,\ldots,n\}$, then we replace $\varepsilon_j E(a_j,b_j)$ by the operator $(-1)^m \mu E(u,v)$ for any one $j \in \mathcal{I}$, and update $\varepsilon_i E(a_i,b_i) \mapsto \varepsilon_i E(a_i,b_i) \cdot \varepsilon_j E(a_j,b_j)$ for all $i \in \mathcal{I} \setminus \{j\}$ (using Lemma~\ref{lem:Eab}(b)).

\end{enumerate}

\begin{example}
\label{eg:measurement}
\normalfont
Consider the standard Bell state $\dket{\Phi^+} = \frac{\dket{00} + \dket{11}}{\sqrt{2}}$, whose stabilizer group is $S = \langle X \otimes X, Z \otimes Z \rangle= \langle E(11,00), E(00,11) \rangle$.
If we measure $Z \otimes I = E(00,10)$, and obtain the result $-1$, then the new stabilizers are $S = \langle -E(00,10), E(00,11) \rangle \equiv \langle -E(00,10), -E(00,01) \rangle$.
This group perfectly stabilizes the post-measurement state $\dket{11}$.

If we instead measure $Y \otimes I = E(10,10)$, and obtain the result $+1$, then the new stabilizers are $S = \langle E(10,10), -E(11,11) \rangle \equiv \langle E(10,10), -E(01,01) \rangle$.
This group perfectly stabilizes the post-measurement state $\frac{(\dket{0} + \imath \dket{1})}{\sqrt{2}} \otimes \frac{(\dket{0} - \imath \dket{1})}{\sqrt{2}}$. \hfill \IEEEQEDhere
\end{example}

\subsection{Bell State Matrix Identity}
\label{sec:bell_state_identity}

Let $n$ standard Bell pairs be shared between Alice and Bob. 
We rearrange the $2n$ qubits to keep Alice's qubits together and Bob's qubits together. 
We can write the joint state as 
\begin{align}    
\label{eq:bell_state_rearranged}
\dket{\Phi^{+}_n}_{\text{AB}} & = \left(\frac{\dket{00}_{\text{AB}} + \dket{11}_{\text{AB}}}{\sqrt{2}}\right)^{\otimes n} = \frac{1}{\sqrt{2^n}} \sum_{x \in \mathbb{F}_2^n} \dket{x}_{\text{A}} \dket{x}_{\text{B}}. 
\end{align}
Let $M = \sum_{x,y \in \mathbb{F}_2^n} M_{xy} \dketbra{x}{y} \in \mathbb{C}^{2^n \times 2^n}$ be any matrix acting on Alice's qubits.
Then, it has been observed that \cite{Bennett-pra96,Wilde-isit10}
\begin{align}
(M \otimes I)\ket{\Phi^{+}_n} & = \frac{1}{\sqrt{2^n}} \sum_{x,y \in \mathbb{F}_2^n} M_{xy} \dket{x}_{\text{A}} \dket{y}_{\text{B}} \\
  & = \frac{1}{\sqrt{2^n}} \sum_{x,y \in \mathbb{F}_2^n} \dket{x}_{\text{A}} (M^T)_{yx} \dket{y}_{\text{B}} \\
  & = (I \otimes M^T) \ket{\Phi^{+}_n}. 
\end{align}

\begin{example}
\label{eg:Y_measurement}
\normalfont
As in Example~\ref{eg:measurement}, consider the standard Bell pair and measure $Y \otimes I$. 
If the measurement result is $+1$, then the projector $P_Y = \frac{I_2 + Y}{2}$ gets applied to the first qubit.
Then, according to the above identity, this is equivalent to applying $P_Y^T = \frac{I_2 - Y}{2}$ on the second qubit.
This exactly agrees with the post-measurement state $\frac{(\dket{0} + \imath \dket{1})}{\sqrt{2}} \otimes \frac{(\dket{0} - \imath \dket{1})}{\sqrt{2}}$. \hfill \IEEEQEDhere
\end{example}

If Alice measures the generators of a stabilizer group $S = \langle \varepsilon_i E(a_i,b_i); \ i = 1,\ldots,r \rangle$, and obtains results $(-1)^{m_i}, m_i \in \{0,1\},$ then $M = \Pi_{S'}$ is the projector onto the subspace $\mathcal{Q}(S')$ of the stabilizer code defined by $S' = \langle (-1)^{m_i} \varepsilon_i E(a_i,b_i); \ i = 1,\ldots,r \rangle$. 
According to the above identity, this is equivalent to projecting Bob's qubits onto the stabilizer code defined by 
\begin{align}
S'' = \langle (-1)^{m_i} \varepsilon_i E(a_i,b_i)^T; \ i = 1,\ldots,r \rangle = \langle (-1)^{m_i + a_i b_i^T} \varepsilon_i E(a_i,b_i); \ i = 1,\ldots,r \rangle,
\end{align}
where we have applied Lemma~\ref{lem:Eab}(a).
Note that, in such cases where $M$ is a projector, we can write
\begin{align}
\label{eq:projector_bell_pair}
(M \otimes I)\ket{\Phi^{+}_n} = (M^2 \otimes I)\ket{\Phi^{+}_n} = (M \otimes M^T)\ket{\Phi^{+}_n},
\end{align}
so that the action of Alice can be interpreted as \emph{both} Alice and Bob projecting their own qubits simultaneously.

\section{Logical Bell Pairs for Arbitrary CSS Codes}
\label{sec:logical_bell_CSS}

In this appendix, we show that when $n$ raw Bell pairs are projected onto the subspace of a CSS code through stabilizer measurements, the induced logical state is that of $k$ Bell pairs.
We take a meet-in-the-middle approach where we first consider $k$ Bell pairs and show how their encoded state looks like, and then we project $n$ Bell pairs to prove that the resulting state is the same as the aforesaid encoded state.

Let $\MCC_1, \MCC_2$ be two binary linear codes such that $\MCC_2 \subset \MCC_1$.
For the $\llbr n,k \rrbr$ CSS code defined by these codes, $\MCC_2$ produces the $X$-stabilizers, $\MCC_1$ produces the logical $X$ operators, $\MCC_1^{\perp}$ produces the $Z$-stabilizers, and $\MCC_2^{\perp}$ produces the logical $Z$ operators.
Let $G_{\MCC_1/\MCC_2}$ denote a generator matrix for the quotient group $\MCC_1/\MCC_2$ that represents the ``pure'' logical $X$ operators that do not have any $X$-stabilizer component.
In other words, the rows of $G_{\MCC_1/\MCC_2}$ give the generators of logical $X$ operators for the CSS code.
Let $\mathcal{U}_{\text{Enc}}$ denote an encoding unitary for the code.
Then, the encoded state of $k$ Bell pairs is~\cite{Calderbank-physreva96,Nielsen-2010}
\begin{align}
& \left( (\mathcal{U}_{\text{Enc}})_{\text{A}} \otimes (\mathcal{U}_{\text{Enc}})_{\text{B}} \right) \left( \dket{\Phi_k^+}_{\text{AB}} \otimes \dket{00}_{\text{AB}}^{\otimes (n-k)} \right) \nonumber \\
  & = \frac{1}{\sqrt{2^k}} \sum_{x \in \mathbb{F}_2^k} \mathcal{U}_{\text{Enc}} \left( \dket{x}_{\text{A}} \dket{0}_{\text{A}}^{\otimes (n-k)} \right) \otimes \mathcal{U}_{\text{Enc}} \left( \dket{x}_{\text{B}} \dket{0}_{\text{B}}^{\otimes (n-k)} \right) \\
\label{eq:encoded_bell_pairs}
  & = \frac{1}{\sqrt{2^k}} \sum_{x \in \mathbb{F}_2^k} \left[ \frac{1}{\sqrt{|\MCC_2|}} \sum_{y \in \MCC_2} \dket{xG_{\MCC_1/\MCC_2} \oplus y}_{\text{A}} \right] \otimes \left[ \frac{1}{\sqrt{|\MCC_2|}} \sum_{y' \in \MCC_2} \dket{xG_{\MCC_1/\MCC_2} \oplus y'}_{\text{B}} \right].
\end{align}

For the other direction, we start with $\dket{\Phi_n^+}_{\text{AB}}$ and then apply the projector $\Pi_{\text{CSS}}$ for the code on Alice's qubits.
By the Bell state matrix identity (Section~\ref{sec:bell_state_identity}), this means that we are effectively simultaneously applying $\Pi_{\text{CSS}}^T = \Pi_{\text{CSS}}$ on Bob's qubits as well.
Here, the transpose has no effect because the stabilizer generators for CSS codes are purely $X$-type or purely $Z$-type, and only such operators appear in the expression for the code projector~\eqref{eq:stabilizer_projector}.
Let $G_2$ and $G_1^{\perp}$ represent generator matrices for the codes $\MCC_2$ and $\MCC_1^{\perp}$, respectively.
Then, we have
\begin{align}
\Pi_{\text{CSS}} = \prod_{u \in \text{rows}(G_2)} \frac{I_N + E(u,0)}{2} \cdot \prod_{v \in \text{rows}(G_1^{\perp})} \frac{I_N + E(0,v)}{2} \eqqcolon \Pi_X \cdot \Pi_Z.
\end{align}
For any $z \in \mathbb{F}_2^n$, since $E(0,v) \dket{z} = (-1)^{zv^T} \dket{z}$, we have $(I_N + E(0,v)) \dket{z} = 2 \dket{z}$ if $zv^T = 0$ and $(I_N + E(0,v)) \dket{z} = 0$ otherwise.
This implies that $\Pi_Z \dket{z} = \dket{z}$ or $0$ depending on whether $z \in \MCC_1$ or not, respectively.
Similarly, it is easy to check that $\Pi_X \dket{z} = \frac{1}{|\MCC_2|} \sum_{y \in \MCC_2} \dket{z \oplus y}$.
Putting these together, we observe that
\begin{align}
& \left( \left( \Pi_{\text{CSS}} \right)_{\text{A}} \otimes \left( \Pi_{\text{CSS}} \right)_{\text{B}} \right) \dket{\Phi_n^+} \nonumber \\ 
  & = \frac{1}{\sqrt{2^n}} \sum_{z \in \mathbb{F}_2^n} \Pi_X \Pi_Z \dket{z}_{\text{A}} \otimes \Pi_X \Pi_Z \dket{z}_{\text{B}} \\
  & = \frac{1}{\sqrt{2^n}} \sum_{z \in \MCC_1} \Pi_X \dket{z}_{\text{A}} \otimes \Pi_X \dket{z}_{\text{B}} \\
  & = \frac{1}{\sqrt{2^n}} \sum_{z \in \MCC_1} \frac{1}{|\MCC_2|} \sum_{y \in \MCC_2} \dket{z \oplus y}_{\text{A}} \otimes \frac{1}{|\MCC_2|} \sum_{y' \in \MCC_2} \dket{z \oplus y'}_{\text{B}} \\
  & = \frac{1}{\sqrt{2^n}} \sum_{x \in \mathbb{F}_2^k} \sum_{y'' \in \MCC_2} \frac{1}{|\MCC_2|} \sum_{y \in \MCC_2} \dket{(x G_{\MCC_1/\MCC_2} \oplus y'') \oplus y}_{\text{A}} \otimes \frac{1}{|\MCC_2|} \sum_{y' \in \MCC_2} \dket{(x G_{\MCC_1/\MCC_2} \oplus y'') \oplus y'}_{\text{B}} \\
  & = \frac{1}{\sqrt{2^n}} \sum_{x \in \mathbb{F}_2^k} |\MCC_2| \frac{1}{|\MCC_2|} \sum_{y \in \MCC_2} \dket{x G_{\MCC_1/\MCC_2} \oplus y}_{\text{A}} \otimes \frac{1}{|\MCC_2|} \sum_{y' \in \MCC_2} \dket{x G_{\MCC_1/\MCC_2} \oplus y'}_{\text{B}} \\
\label{eq:projected_bell_pairs}
  & = \frac{1}{\sqrt{2^n}} \sum_{x \in \mathbb{F}_2^k} \frac{1}{\sqrt{|\MCC_2|}} \sum_{y \in \MCC_2} \dket{x G_{\MCC_1/\MCC_2} \oplus y}_{\text{A}} \otimes \frac{1}{\sqrt{|\MCC_2|}} \sum_{y' \in \MCC_2} \dket{x G_{\MCC_1/\MCC_2} \oplus y'}_{\text{B}}.
\end{align}
This state must be normalized by the square root of the probability that we get the all $+1$ syndrome, which corresponds to the subspace of the considered CSS code.
It can be checked that all syndromes are equally likely, so the probability is $1/2^{n-k}$.
Dividing~\eqref{eq:projected_bell_pairs} by $1/\sqrt{2^{n-k}}$, we arrive at exactly the same state in~\eqref{eq:encoded_bell_pairs}.
This establishes that when CSS stabilizer measurements are performed on $n$ Bell pairs, the resulting code state corresponds to $k$ logical Bell pairs.

\section{Protocol I  with the $3$-Qubit Code}
\label{sec:protocol1_3qubit}

{\tiny
\begin{xltabular}{0.8\linewidth}{c c || *{3}{c} | *{3}{c} || c}
\caption{\label{tab:ghz_protocol} Steps of the GHZ distillation protocol based on the $\llbr 3,1,1 \rrbr$ code defined by $S = \langle YYI, IYY \rangle$. Each `$0$' below represents $000$, and $e_i \in \mathbb{F}_2^3$ is the standard basis vector with a $1$ in the $i$-th position and zeros elsewhere. Code stabilizers are typeset in boldface. An additional left arrow indicates which row is being replaced with a code stabilizer, i.e., the first row that anticommutes with the stabilizer. Other updated rows are highlighted in gray. Classical communications: A $\rightarrow$ B, B $\rightarrow$ C.} \\
\toprule
Step & Sign & \multicolumn{3}{c|}{$X$-Components} & \multicolumn{3}{c||}{$Z$-Components} & Pauli Representation \\
     &  ($\pm 1$)   &   $A$ & $B$ & $C$   &   $A$ & $B$ & $C$   &   \\
\midrule
\midrule
%
%
%
$(0)$ & $+1$ &    $0$ & $0$ & $0$   &   $e_1$ & $e_1$ & $0$    & $Z_{\text{A}_1} Z_{\text{B}_1}$ \\
     & $+1$ &    $0$ & $0$ & $0$   &   $e_2$ & $e_2$ & $0$    & $Z_{\text{A}_2} Z_{\text{B}_2}$ \\
     & $+1$ &    $0$ & $0$ & $0$   &   $e_3$ & $e_3$ & $0$    & $Z_{\text{A}_3} Z_{\text{B}_3}$ \\
\cmidrule(lr){2-9}
     & $+1$ &    $0$ & $0$ & $0$   &   $0$ & $e_1$ & $e_1$    & $Z_{\text{B}_1} Z_{\text{C}_1}$ \\
     & $+1$ &    $0$ & $0$ & $0$   &   $0$ & $e_2$ & $e_2$    & $Z_{\text{B}_2} Z_{\text{C}_2}$ \\
     & $+1$ &    $0$ & $0$ & $0$   &   $0$ & $e_3$ & $e_3$    & $Z_{\text{B}_3} Z_{\text{C}_3}$ \\
\cmidrule(lr){2-9}
     & $-1$ &   $e_1$ & $e_1$ & $e_1$   &   $e_1$ & $e_1$ & $0$    & $- Y_{\text{A}_1} Y_{\text{B}_1} X_{\text{C}_1}$ \\
     & $-1$ &   $e_2$ & $e_2$ & $e_2$   &   $e_2$ & $e_2$ & $0$    & $- Y_{\text{A}_2} Y_{\text{B}_2} X_{\text{C}_2}$ \\
     & $-1$ &   $e_3$ & $e_3$ & $e_3$   &   $e_3$ & $e_3$ & $0$    & $- Y_{\text{A}_3} Y_{\text{B}_3} X_{\text{C}_3}$ \\
     &     &       &     &       &       &     &       &   \\
\midrule
%
%
%
$(1)$ & $\boldsymbol{\varepsilon_1^{\text{A}}}$ &    $\boldsymbol{110}$ & $\boldsymbol{000}$ & $\boldsymbol{000}$   &   $\boldsymbol{110}$ & $\boldsymbol{000}$ & $\boldsymbol{000}$    & \qquad \ \ $\boldsymbol{\varepsilon_1^{\text{A}} \, Y_{\text{A}_1} Y_{\text{A}_2} I_{\text{A}_3}} \ \ \boldsymbol{\longleftarrow}$ \\
\rowcolor{lightgray}
     & $+1$ &    $0$ & $0$ & $0$   &   $e_1+e_2$ & $e_1+e_2$ & $0$    & $Z_{\text{A}_2} Z_{\text{B}_2} Z_{\text{A}_1} Z_{\text{B}_1}$ \\
     & $+1$ &    $0$ & $0$ & $0$   &   $e_3$ & $e_3$ & $0$    & \makecell{$Z_{\text{A}_3} Z_{\text{B}_3}$} \\
\cmidrule(lr){2-9}
     & $+1$ &    $0$ & $0$ & $0$   &   $0$ & $e_1$ & $e_1$    & $Z_{\text{B}_1} Z_{\text{C}_1}$ \\
     & $+1$ &    $0$ & $0$ & $0$   &   $0$ & $e_2$ & $e_2$    & $Z_{\text{B}_2} Z_{\text{C}_2}$ \\
     & $+1$ &    $0$ & $0$ & $0$   &   $0$ & $e_3$ & $e_3$    & $Z_{\text{B}_3} Z_{\text{C}_3}$ \\
\cmidrule(lr){2-9}
\rowcolor{lightgray}
     & $\varepsilon_1^{\text{A}}$ &   $0$ & $e_1+e_2$ & $e_1+e_2$   &   $0$ & $e_1+e_2$ & $0$    & \makecell{$\varepsilon_1^{\text{A}} \, (Y_{\text{B}_1} Y_{\text{B}_2} I_{\text{B}_3})$ \\ $\cdot \ (X_{\text{C}_1} X_{\text{C}_2} I_{\text{C}_3})$} \\
     & $-1$ &   $e_2$ & $e_2$ & $e_2$   &   $e_2$ & $e_2$ & $0$    & $- Y_{\text{A}_2} Y_{\text{B}_2} X_{\text{C}_2}$ \\
     & $-1$ &   $e_3$ & $e_3$ & $e_3$   &   $e_3$ & $e_3$ & $0$    & $- Y_{\text{A}_3} Y_{\text{B}_3} X_{\text{C}_3}$ \\
     &     &       &     &       &       &     &       &   \\
\midrule
%
%
$(2)$ & $\boldsymbol{\varepsilon_1^{\text{A}}}$ &    $\boldsymbol{110}$ & $\boldsymbol{000}$ & $\boldsymbol{000}$   &   $\boldsymbol{110}$ & $\boldsymbol{000}$ & $\boldsymbol{000}$    & $\boldsymbol{\varepsilon_1^{\text{A}} \, Y_{\text{A}_1} Y_{\text{A}_2} I_{\text{A}_3}}$ \\
     & $\boldsymbol{\varepsilon_2^{\text{A}}}$ &    $\boldsymbol{011}$ & $\boldsymbol{000}$ & $\boldsymbol{000}$   &   $\boldsymbol{011}$ & $\boldsymbol{000}$ & $\boldsymbol{000}$    & \qquad \ \ $\boldsymbol{\varepsilon_2^{\text{A}} \, I_{\text{A}_1} Y_{\text{A}_2} Y_{\text{A}_3}} \ \ \boldsymbol{\longleftarrow}$ \\
\rowcolor{lightgray}
     & $+1$ &    $0$ & $0$ & $0$   &   $e_1+e_2+e_3$ & $e_1+e_2+e_3$ & $0$    & \makecell{$Z_{\text{A}_3} Z_{\text{B}_3} Z_{\text{A}_2} Z_{\text{B}_2}$\\ $\cdot \ Z_{\text{A}_1} Z_{\text{B}_1}$} \\
\cmidrule(lr){2-9}
     & $+1$ &    $0$ & $0$ & $0$   &   $0$ & $e_1$ & $e_1$    & $Z_{\text{B}_1} Z_{\text{C}_1}$ \\
     & $+1$ &    $0$ & $0$ & $0$   &   $0$ & $e_2$ & $e_2$    & $Z_{\text{B}_2} Z_{\text{C}_2}$ \\
     & $+1$ &    $0$ & $0$ & $0$   &   $0$ & $e_3$ & $e_3$    & $Z_{\text{B}_3} Z_{\text{C}_3}$ \\
\cmidrule(lr){2-9}
     & $\varepsilon_1^{\text{A}}$ &   $0$ & $e_1+e_2$ & $e_1+e_2$   &   $0$ & $e_1+e_2$ & $0$    & \makecell{$\varepsilon_1^{\text{A}} \, (Y_{\text{B}_1} Y_{\text{B}_2} I_{\text{B}_3})$ \\ $\cdot \ (X_{\text{C}_1} X_{\text{C}_2} I_{\text{C}_3})$} \\
\rowcolor{lightgray}
     & $\varepsilon_2^{\text{A}}$ &   $0$ & $e_2+e_3$ & $e_2+e_3$   &   $0$ & $e_2+e_3$ & $0$    & \makecell{$\varepsilon_2^{\text{A}} \, (I_{\text{B}_1} Y_{\text{B}_2} Y_{\text{B}_3})$ \\ $\cdot \ (I_{\text{C}_1} X_{\text{C}_2} X_{\text{C}_3})$} \\
     & $-1$ &   $e_3$ & $e_3$ & $e_3$   &   $e_3$ & $e_3$ & $0$    & $- Y_{\text{A}_3} Y_{\text{B}_3} X_{\text{C}_3}$ \\
     &     &       &     &       &       &     &       &   \\
\midrule
%
%
%
$(3)$ & $\boldsymbol{\varepsilon_1^{\text{A}}}$ &    $\boldsymbol{110}$ & $\boldsymbol{000}$ & $\boldsymbol{000}$   &   $\boldsymbol{110}$ & $\boldsymbol{000}$ & $\boldsymbol{000}$    & $\boldsymbol{\varepsilon_1^{\text{A}} \, Y_{\text{A}_1} Y_{\text{A}_2} I_{\text{A}_3}}$ \\
     & $\boldsymbol{\varepsilon_2^{\text{A}}}$ &    $\boldsymbol{011}$ & $\boldsymbol{000}$ & $\boldsymbol{000}$   &   $\boldsymbol{011}$ & $\boldsymbol{000}$ & $\boldsymbol{000}$    & $\boldsymbol{\varepsilon_2^{\text{A}} \, I_{\text{A}_1} Y_{\text{A}_2} Y_{\text{A}_3}}$ \\
\rowcolor{lightgray}
     & $\eta$ &    $0$ & $0$ & $0$   &   $e_1+e_2+e_3$ & $e_1+e_2+e_3$ & $0$    & \makecell{$\eta \, Z_{\text{A}_3} Z_{\text{B}_3} Z_{\text{A}_2} Z_{\text{B}_2}$\\ $\cdot \ Z_{\text{A}_1} Z_{\text{B}_1}$} \\
\cmidrule(lr){2-9}
\rowcolor{lightgray}
     & $\nu_1$ &    $0$ & $0$ & $0$   &   $0$ & $e_1$ & $e_1$    & $\nu_1 \, Z_{\text{B}_1} Z_{\text{C}_1}$ \\
\rowcolor{lightgray}
     & $\nu_2$ &    $0$ & $0$ & $0$   &   $0$ & $e_2$ & $e_2$    & $\nu_2 \, Z_{\text{B}_2} Z_{\text{C}_2}$ \\
\rowcolor{lightgray}
     & $\nu_3$ &    $0$ & $0$ & $0$   &   $0$ & $e_3$ & $e_3$    & $\nu_3 \, Z_{\text{B}_3} Z_{\text{C}_3}$ \\
\cmidrule(lr){2-9}
\rowcolor{lightgray}
     & $\mu_1 \varepsilon_1^{\text{A}}$ &   $\boldsymbol{000}$ & $\boldsymbol{110}$ & $\boldsymbol{110}$   &   $\boldsymbol{000}$ & $\boldsymbol{110}$ & $\boldsymbol{110}$    & \tiny \makecell{$\boldsymbol{\mu_1 \varepsilon_1^{\text{A}} \, (Y_{\text{B}_1} Y_{\text{B}_2} I_{\text{B}_3})}$ \\ $\cdot \ \boldsymbol{(Y_{\text{C}_1} Y_{\text{C}_2} I_{\text{C}_3})}$} \\
\rowcolor{lightgray}
     & $\mu_2 \varepsilon_2^{\text{A}}$ &   $\boldsymbol{000}$ & $\boldsymbol{011}$ & $\boldsymbol{011}$   &   $\boldsymbol{000}$ & $\boldsymbol{011}$ & $\boldsymbol{011}$    & \tiny \makecell{$\boldsymbol{\mu_2 \varepsilon_2^{\text{A}} \, (I_{\text{B}_1} Y_{\text{B}_2} Y_{\text{B}_3})}$ \\ $\cdot \ \boldsymbol{(I_{\text{C}_1} Y_{\text{C}_2} Y_{\text{C}_3})}$} \\
\rowcolor{lightgray}
     & $\mu_3$ &   $e_3$ & $e_3$ & $e_3$   &   $e_3$ & $e_3$ & $e_3$    & $\mu_3 \, Y_{\text{A}_3} Y_{\text{B}_3} Y_{\text{C}_3}$ \\
     &     &       &     &       &       &     &       &   \\
\midrule
%
%
%
$(4)$ & $\boldsymbol{\varepsilon_1^{\text{A}}}$ &    $\boldsymbol{110}$ & $\boldsymbol{000}$ & $\boldsymbol{000}$   &   $\boldsymbol{110}$ & $\boldsymbol{000}$ & $\boldsymbol{000}$    & $\boldsymbol{\varepsilon_1^{\text{A}} \, Y_{\text{A}_1} Y_{\text{A}_2} I_{\text{A}_3}}$ \\
     & $\boldsymbol{\varepsilon_2^{\text{A}}}$ &    $\boldsymbol{011}$ & $\boldsymbol{000}$ & $\boldsymbol{000}$   &   $\boldsymbol{011}$ & $\boldsymbol{000}$ & $\boldsymbol{000}$    & $\boldsymbol{\varepsilon_2^{\text{A}} \, I_{\text{A}_1} Y_{\text{A}_2} Y_{\text{A}_3}}$ \\
     & $+1$ &    $0$ & $0$ & $0$   &   $e_1+e_2+e_3$ & $e_1+e_2+e_3$ & $0$    & \qquad \quad $\overline{Z}_{\text{A}} \overline{Z}_{\text{B}} \overline{I}_{\text{C}}$ (logical) \\
\cmidrule(lr){2-9}
     & $\boldsymbol{\varepsilon_1^{\text{B}}}$ &    $\boldsymbol{000}$ & $\boldsymbol{110}$ & $\boldsymbol{000}$   &   $\boldsymbol{000}$ & $\boldsymbol{110}$ & $\boldsymbol{000}$    & \qquad \ \ $\boldsymbol{\varepsilon_1^{\text{B}} \, Y_{\text{B}_1} Y_{\text{B}_2} I_{\text{B}_3}} \ \ \boldsymbol{\longleftarrow}$ \\
     & $\boldsymbol{\varepsilon_2^{\text{B}}}$ &    $\boldsymbol{000}$ & $\boldsymbol{011}$ & $\boldsymbol{000}$   &   $\boldsymbol{000}$ & $\boldsymbol{011}$ & $\boldsymbol{000}$    & \qquad \ \ $\boldsymbol{\varepsilon_2^{\text{B}} \, I_{\text{B}_1} Y_{\text{B}_2} Y_{\text{B}_3}} \ \ \boldsymbol{\longleftarrow}$ \\
\rowcolor{lightgray}
     & $\beta$ &    $0$ & $0$ & $0$   &   $0$ & $e_1+e_2+e_3$ & $e_1+e_2+e_3$    & $\beta \, \overline{I}_{\text{A}} \overline{Z}_{\text{B}} \overline{Z}_{\text{C}}$ (logical) \\
\cmidrule(lr){2-9}
\rowcolor{lightgray}
     & $\boldsymbol{\alpha_1 \varepsilon_1^{\text{A}} \varepsilon_1^{\text{B}}}$ &   $\boldsymbol{000}$ & $\boldsymbol{000}$ & $\boldsymbol{110}$   &   $\boldsymbol{000}$ & $\boldsymbol{000}$ & $\boldsymbol{110}$    & $\boldsymbol{\alpha_1 \varepsilon_1^{\text{A}} \varepsilon_1^{\text{B}} \, Y_{\text{C}_1} Y_{\text{C}_2} I_{\text{C}_3}}$ \\
\rowcolor{lightgray}
     & $\boldsymbol{\alpha_2 \varepsilon_2^{\text{A}} \varepsilon_2^{\text{B}}}$ &   $\boldsymbol{000}$ & $\boldsymbol{000}$ & $\boldsymbol{011}$   &   $\boldsymbol{000}$ & $\boldsymbol{000}$ & $\boldsymbol{011}$    & $\boldsymbol{\alpha_2 \varepsilon_2^{\text{A}} \varepsilon_2^{\text{B}} \, I_{\text{C}_1} Y_{\text{C}_2} Y_{\text{C}_3}}$ \\
\rowcolor{lightgray}
     & $\alpha_3$ &   $e_3$ & $e_3$ & $e_3$   &   $e_3$ & $e_3$ & $e_3$    & $\alpha_3 \, \overline{X}_{\text{A}} \overline{X}_{\text{B}} \overline{X}_{\text{C}}$ (logical) \\
     &     &       &     &       &       &     &       &   \\
%
%
\bottomrule
\bottomrule
\end{xltabular}
}

The steps of the protocol, with this particular $\llbr 3,1,1 \rrbr$ code as an example, are shown in Table~\ref{tab:ghz_protocol}.
Again, we use the stabilizer formalism for measurements from Section~\ref{sec:stabilizer_formalism}.
We will explain each step below and discuss the potential subtleties that can arise.
It could be useful to imagine the three parties as being three nodes A --- B --- C on a linear network chain.
\begin{enumerate}

\item[(0)] Alice locally prepares $3$ copies of the perfect GHZ state and groups her qubits together for further processing. She keeps aside the grouped qubits of Bob's and Charlie's but does not send those to them yet. She also writes down the parity check matrix for the $9$ qubits, based on only GHZ stabilizers, along with signs, as shown in Step (0) of Table~\ref{tab:ghz_protocol}. \\

\item[(1)] Alice measures the stabilizer $Y_{\text{A}_1} Y_{\text{A}_2} I_{\text{A}_3} = E([(e_1+e_2)^{\text{A}},0^{\text{B}},0^{\text{C}}],[(e_1+e_2)^{\text{A}},0^{\text{B}},0^{\text{C}}])$ and the group $\mathcal{G}_3$ gets updated as shown in Step (1) of Table~\ref{tab:ghz_protocol}, assuming that the measurement result is $\varepsilon_1^{\text{A}} \in \{ \pm 1 \}$.
Based on the stabilizer formalism (Section~\ref{sec:stabilizer_formalism}), the measured stabilizer replaces the first row (as indicated by the left arrow) and the second row is multiplied with the previous first row.
For visual clarity, code stabilizer rows are boldfaced and binary vectors are written out in full.
Furthermore, as per Theorem~\ref{thm:ghz_stabilizer_measurement}, this measurement of $E(e_1+e_2, e_1+e_2)$ by Alice should imply that 
$$ \varepsilon_1^{\text{A}} E(e_1+e_2, e_1+e_2)_{\text{B}}^T \otimes E(e_1+e_2, 0)_{\text{C}} = \varepsilon_1^{\text{A}} E(e_1+e_2, e_1+e_2)_{\text{B}} \otimes E(e_1+e_2, 0)_{\text{C}} $$ 
automatically belongs to the (new) stabilizer group. 
Indeed, this element can be produced by multiplying the elements $- E([e_i^{\text{A}},e_i^{\text{B}},e_i^{\text{C}}],[e_i^{\text{A}},e_i^{\text{B}},0^{\text{C}}])$ for $i=1,2$ along with $Y_{\text{A}_1} Y_{\text{A}_2} I_{\text{A}_3} = E([(e_1+e_2)^{\text{A}},0^{\text{B}},0^{\text{C}}],[(e_1+e_2)^{\text{A}},0^{\text{B}},0^{\text{C}}])$ using Lemma~\ref{lem:Eab}(b).
This is exactly how the seventh row gets updated. \\

\item[(2)] Alice measures $I_{\text{A}_1} Y_{\text{A}_2} Y_{\text{A}_3} = E([(e_2+e_3)^{\text{A}},0^{\text{B}},0^{\text{C}}],[(e_2+e_3)^{\text{A}},0^{\text{B}},0^{\text{C}}])$, the second stabilizer, and the group gets updated as shown in Step (2) of Table~\ref{tab:ghz_protocol}. 
The procedure is very similar to that in Step (1). \\

Since Alice has measured all her stabilizer generators, and the stabilizer formalism preserves the commutativity of the elements in the group, the third row in the first block of 3 rows must necessarily commute with Alice's stabilizers.
Thus, the Alice component of the third row must form a logical operator for Alice's code, and we define it to be the logical $Z$ operator, i.e., $\overline{Z}_{\text{A}} = ZZZ = E(0,e_1+e_2+e_3)$.
We will see shortly that Bob's qubits get the same code (possibly with sign changes for the stabilizers), so this third row can be written as the logical GHZ stabilizer $\overline{Z}_{\text{A}} \overline{Z}_{\text{B}} \overline{I}_{\text{C}}$. \\

This phenomenon also generalizes to any $\llbr n,k,d \rrbr$ stabilizer code, with some caveats when the code has some purely $Z$-type stabilizers, and we determine the logical $Z$ operators either after Alice's set of measurements or apriori using some linear algebraic arguments (see Appendix~\ref{sec:logical_paulis_ghz_msmt} for details).
Note that it is convenient to choose the logical $Z$ operators such that they respect the GHZ structure of our analysis, e.g., $\overline{Z} = IIY$ will not be compatible here. \\

\item[(3)] If we consider the parity-check matrix after Step (2), we see that rows 4 through 8 are the stabilizers promised by Theorem~\ref{thm:ghz_stabilizer_measurement} that act only on B and C systems.
However, due to the same result, the C parts of rows 7 and 8 only have $X$-s instead of $Y$-s.
So, to change them back to $Y$-s, Alice applies the inverse of the Phase (i.e., $\sqrt{Z}$) gate to all $3$ qubits of the C system.
She specifically applies the inverse, rather than $\sqrt{Z}$ itself, to get rid of the $-1$ sign for the last row. \\

For a general stabilizer code, the appropriate diagonal Clifford must be chosen as discussed in Appendix~\ref{sec:diagonal_clifford}.
This operation converts $\varepsilon_i^{\text{A}} (-1)^{a_ib_i^T} E(a_i,b_i)_{\text{B}} \otimes E(a_i,0)_{\text{C}}$, the BC stabilizers, into $\varepsilon_i^{\text{A}} (-1)^{a_ib_i^T} E(a_i,b_i)_{\text{B}} \otimes E(a_i,b_i)_{\text{C}}$, which ensures that Charlie gets the same code (up to signs of stabilizers) as Alice and Bob.
Since the Clifford is guaranteed to be diagonal, it leaves purely $Z$-type stabilizers unchanged.
Later, in Section~\ref{sec:local_clifford}, we show that it is better for Bob to perform this Clifford on Charlie's qubits, rather than Alice. \\

Though we have used the $-YYX$ GHZ stabilizer here for convenience, for a general code we can simply continue to use $XXX$.
After Alice has measured her stabilizer generators, this last block of 3 rows could have changed but they still commute with the generators.
Since the middle block never gets affected by Alice's measurements, we can guarantee using Theorem~\ref{thm:ghz_stabilizer_measurement} that two of the last 3 rows must be the joint BC stabilizers induced by Alice's two generators.
Hence, the remaining row's Alice component must form a logical operator for Alice's code, and will be distinct from the previously defined logical $Z$ operator.
We define this to be the logical $X$ operator, i.e., $\overline{X}_{\text{A}} = IIY = Y_3 = E(e_3,e_3)$.
As we will see shortly, both Bob and Charlie get the same code, so this last row of the third block can be written as the logical GHZ stabilizer $\overline{X}_{\text{A}} \overline{X}_{\text{B}} \overline{X}_{\text{C}}$.
The generalization to arbitrary stabilizer codes is discussed in Appendix~\ref{sec:logical_paulis_ghz_msmt}. \\

Then, she sends Bob both his qubits as well as Charlie's qubits over a noisy Pauli channel, which introduces the signs $\eta, \nu_i, \mu_i \in \{ \pm 1 \}, i = 1,2,3$.
She also classically communicates the code stabilizers, her syndromes $\{\varepsilon_1^{\text{A}}, \varepsilon_2^{\text{A}}\}$, and the logical $Z$ and $X$ operators to him. \\

\item[(4)] Now, based on Alice's classical communication, Bob applies Theorem~\ref{thm:ghz_stabilizer_measurement} to obtain the stabilizer generators
\begin{align*}
\varepsilon_1^{\text{A}} (Y_{\text{B}_1} Y_{\text{B}_2} I_{\text{B}_3}) \, (Y_{\text{C}_1} Y_{\text{C}_2} I_{\text{C}_3}) & = \varepsilon_1^{\text{A}} E(e_1+e_2,e_1+e_2)_{\text{B}}^T \otimes E(e_1+e_2,e_1+e_2)_{\text{C}} , \\ 
\varepsilon_2^{\text{A}} (I_{\text{B}_1} Y_{\text{B}_2} Y_{\text{B}_3}) \, (I_{\text{C}_1} Y_{\text{C}_2} Y_{\text{C}_3}) & = \varepsilon_2^{\text{A}} E(e_2+e_3,e_2+e_3)_{\text{B}}^T \otimes E(e_2+e_3,e_2+e_3)_{\text{C}}
\end{align*}
for the induced joint code on his as well as Charlie's qubits.
As per Theorem~\ref{thm:ghz_stabilizer_measurement}, he also includes $\{ Z_{\text{B}_i} Z_{\text{C}_i} = E([0^{\text{A}},0^{\text{B}},0^{\text{C}}],[0^{\text{A}},e_i^{\text{B}},e_i^{\text{C}}]) \ ; \ i = 1,2,3 \}$ i.e., the $IZZ$-type GHZ stabilizers, as stabilizer generators for the $\llbr 6,1 \rrbr$ joint code on BC systems.
He measures these $5$ stabilizers to deduce and correct the error introduced by the channel on the $6$ qubits sent by Alice.
Assuming perfect error correction, the signs will be back to the ones in Alice's final parity-check matrix. \\

When Bob sends Charlie's qubits to him, the channel might introduce errors on those $3$ qubits.
To deduce and correct these errors, there must have been a code induced on Charlie's qubits \emph{even before the transmission}.
Hence, after correcting errors on the $6$ qubits of BC systems, Bob measures the same stabilizers as Alice's code but on his qubits.
This produces rows $4$ and $5$ in Step (4) of Table~\ref{tab:ghz_protocol}, and row $6$ gets updated as per the stabilizer formalism.
When looking at rows $7$ and $8$ of Step (3), it is evident that one can correspondingly multiply them with these new rows $4$ and $5$ to produce the same code just on Charlie's qubits. \\

This phenomenon extends to general stabilizer codes as well, where the joint stabilizers $\varepsilon_i^{\text{A}} (-1)^{a_ib_i^T} E(a_i,b_i)_{\text{B}} \otimes E(a_i,b_i)_{\text{C}}$ are multiplied with Bob's stabilizers $\varepsilon_i^{\text{B}} E(a_i,b_i)_{\text{B}}$ to obtain Charlie's stabilizers $\varepsilon_i^{\text{A}} \varepsilon_i^{\text{B}} (-1)^{a_ib_i^T} E(a_i,b_i)_{\text{C}}$ (see Algorithm~\ref{algo:algo_ghz}).
Any purely $Z$-type stabilizer directly carries over to Charlie (without the above argument) as follows. 
At the beginning of the protocol, we can rewrite the $E([0^{\text{A}},0^{\text{B}},0^{\text{C}}],[e_i^{\text{A}},e_i^{\text{B}},0^{\text{C}}])$ rows such that a subset of them correspond to $E([0^{\text{A}},0^{\text{B}},0^{\text{C}}],[z^{\text{A}},z^{\text{B}},0^{\text{C}}])$, where $E(0,z)$-s are the purely $Z$-type stabilizer generators of the code.
This subset of rows will never be replaced by stabilizer measurements since they commute with other stabilizers.
After Alice's measurements, there will be rows corresponding to $\varepsilon_z^{\text{A}} E(0,z)_{\text{A}}$, which can be multiplied respectively with the aforesaid subset of rows to obtain $\varepsilon_z^{\text{A}} E(0,z)_{\text{B}}$.
Just like we rewrote a subset of the first $n$ rows, we can rewrite a subset of the second $n$ rows to obtain $E([0^{\text{A}},0^{\text{B}},0^{\text{C}}],[0^{\text{A}},z^{\text{B}},z^{\text{C}}])$, which when multiplied with $\varepsilon_z^{\text{A}} E(0,z)_{\text{B}}$ produces the desired $\varepsilon_z^{\text{A}} E(0,z)_{\text{C}}$ for Charlie's code.
See Appendix~\ref{sec:logical_paulis_ghz_msmt} for some related discussion.
In order to merge this phenomenon for purely $Z$-type operators with the general signs $\varepsilon_i^{\text{A}} \varepsilon_i^{\text{B}} (-1)^{a_ib_i^T}$ for Charlie, we set $\varepsilon_i^{\text{B}} \coloneqq +1$ whenever $a_i=0$, as mentioned in Algorithm~\ref{algo:algo_ghz}. \\

Now, Bob has the parity-check matrix shown in Step (4) but without the new signs $\beta, \alpha_1, \alpha_2, \alpha_3$, which will be introduced by the channel during transmission of Charlie's qubits.
He sends Charlie his qubits (over a noisy Pauli channel), the code stabilizer generators, along with the corresponding signs $\{ \varepsilon_1^{\text{A}} \varepsilon_1^{\text{B}}, \varepsilon_2^{\text{A}} \varepsilon_2^{\text{B}} \}$, and the logical $Z$ and $X$ operators. \\


\end{enumerate}

Finally, Charlie measures these generators and fixes errors based on discrepancies in signs with respect to $\{ \varepsilon_1^{\text{A}} \varepsilon_1^{\text{B}}, \varepsilon_2^{\text{A}} \varepsilon_2^{\text{B}} \}$ (the additional signs $(-1)^{a_i b_i^T}$ do not make a difference for this example).
In the matrix in Step (4) of Table~\ref{tab:ghz_protocol}, after excluding the three sets of code stabilizers, we see that there are $3$ rows left which exactly correspond to the logical GHZ stabilizers, where we have defined the logical operators $\overline{Z} = ZZZ, \overline{X} = IIY = Y_3$ for the code.
Therefore, we have shown that after all steps of the protocol, the logical qubits of A, B, and C are in the GHZ state.
Since the signs of the stabilizer generators can be different for each of the three parties, although their logical $X$ and $Z$ operators are the same, the encoding unitary can be slightly different.
If they each perform the inverse of their respective encoding unitaries on their qubits, then the logical GHZ state is converted into a physical GHZ state.

It might seem like this last step requires coordination among all three of them, which would require two-way communications between parties.
However, this is not necessary as Alice can perform the unitary on her qubits once she sends the $6$ qubits to Bob, and Bob can perform the unitary on his qubits once he sends the $3$ qubits to Charlie.
Subsequent operations will necessarily commute with these local unitaries as those qubits are not touched by the remaining parties in the protocol.

Hence, we have illustrated a complete GHZ distillation protocol, although much care must be taken while executing the steps for an arbitrary code.
For example, the local Clifford on C must be determined by solving a set of linear equations and finding a binary symmetric matrix that specifies the diagonal Clifford, via the connection to binary symplectic matrices~\cite{Dehaene-physreva03,Rengaswamy-tqe20}.
This is discussed in detail in Appendix~\ref{sec:diagonal_clifford}.
Similarly, the logical operators of the code that are compatible with our analysis of the protocol must be determined by simulating Alice's part of the protocol and applying some linear algebraic arguments. 
For a general $\llbr n,k,d \rrbr$ code, there will be $3k$ non-code-stabilizer rows at the end, and one needs to identify $k$ pairs of logical $X$ and $Z$ operators for the code from these rows.
Although any valid definition of logical Paulis would likely suffice, we use Algorithm~\ref{algo:logical_paulis_ghz_msmt} to define them so that they naturally fit our analysis.
The explanations for the steps involved in this algorithm are given in Appendix~\ref{sec:logical_paulis_ghz_msmt}.
In order to keep the main paper accessible, we have moved the discussion on implementation details to Appendix~\ref{sec:ghz_implementation}.

\subsection{Placement of Local Clifford and Distillation Performance}
\label{sec:local_clifford}

In our description of the protocol above, we mentioned that Alice performs the local Clifford $\left( \sqrt{Z}^\dagger \right)^{\otimes 3}$ on the qubits of system C in order to make Charlie's code the same as Alice's and Bob's.
However, due to this operation, the joint BC code (in Step (4)) cannot distinguish between Bob's qubits and Charlie's qubits.
Indeed, consider the two-qubit operator $X_{\text{B}_1} X_{\text{C}_1}$.
This commutes with all $5$ stabilizer generators of this code, although just $X_{\text{B}_1}$ or just $X_{\text{C}_1}$ would have anticommuted with the first generator $(Y_{\text{B}_1} Y_{\text{B}_2} I_{\text{B}_3}) \, (Y_{\text{C}_1} Y_{\text{C}_2} I_{\text{C}_3})$.
Therefore, if the true error is $X_{\text{C}_1}$, then the maximum likelihood decoder will correct it with $X_{\text{B}_1}$, which results in a logical error.
Of course, the $3$-qubit code only has distance $1$, but even if we consider the same $5$-qubit code as in Section~\ref{sec:bell_distillation}, the above phenomenon will still occur.
In effect, the induced joint BC code has distance dropping to $2$ whenever Alice's code does not have any purely $Z$-type stabilizer.
If we do not perform the diagonal Clifford at all, then in such cases Charlie's code will have only distance $1$.

To mitigate this, we can instead make \emph{Bob} perform the same diagonal Clifford operation on Charlie's qubits.
This ensures that the stabilizers for the BC code induced by Alice's code are of the form $E(a,b)_{\text{B}} \otimes E(a,0)_{\text{C}}$, or just $E(0,b)_{\text{B}}$ whenever the stabilizer is purely $Z$-type.
If Alice's code has good distance properties, then this joint BC code will have at least that much protection for Bob's qubits.
Although the C parts of the stabilizers are purely $X$-type, the additional GHZ stabilizers $Z_{\text{B}_i} Z_{\text{C}_i}$ help in detecting $X$-errors on system C as well.
Alternatively, one could make Alice perform one type of diagonal Clifford and Bob perform another diagonal Clifford, both on system C, to make both the BC code as well as Charlie's code as good as possible.
In future work, we will investigate these interesting degrees of freedom.

\begin{figure}


\begin{center}
     \scalebox{0.75}{%
         \input{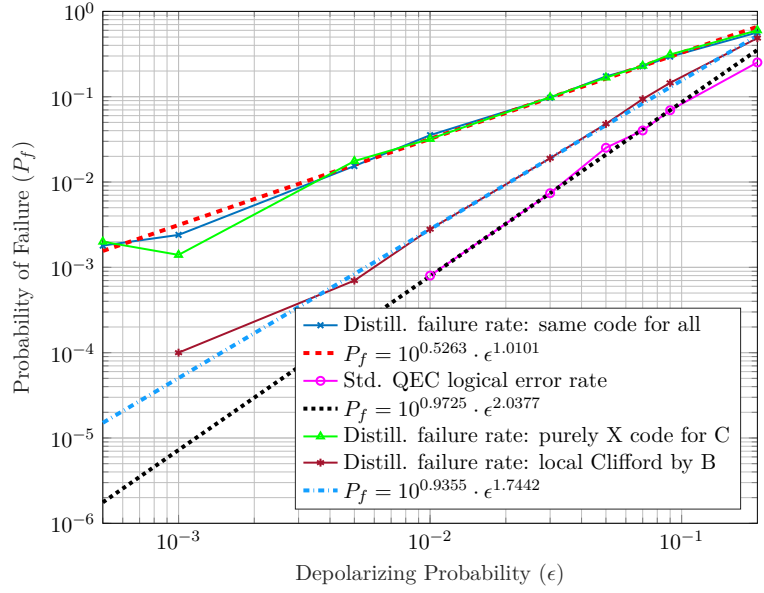}
     }
\end{center}

\caption{\label{fig:GHZ_5qubit_code}Performance of variations of the GHZ distillation protocol using the $\llbr 5,1,3 \rrbr$ perfect code, and comparison with the standard QEC performance of the same code on the depolarizing channel. The decoder employs a maximum likelihood decoding scheme that identifies a minimal weight error pattern matching the syndrome.}


\end{figure}

We developed a MATLAB simulation of this protocol%
\footnote{Implementation available online: \url{https://github.com/nrenga/ghz_distillation_qec}}
and tested it using the $\llbr 5,1,3 \rrbr$ perfect code defined by $S = \langle \, XZZXI,\, IXZZX,\, XIXZZ,\, ZXIXZ \, \rangle$.
The result is shown as the green curve marked `purely X code for C' in Fig.~\ref{fig:GHZ_5qubit_code}.
When compared to the standard QEC performance of this code on the depolarizing channel, we see that the exponent is worse.
This is because, by the arguments above, all non-purely $Z$-type stabilizers of $\mathcal{Q}(S)$ get converted into a purely $X$-type stabilizer for Charlie's code.
To mitigate this, we make Alice perform a local diagonal Clifford operation on qubits C to transform $E(a_i,0)_{\text{C}}$ into $E(a_i,b_i)_{\text{C}}$ later.
The performance of this is shown as the solid blue curve marked `same code for all', which is equally worse.
This time the reason is that the BC code $\mathcal{Q}(S')$ has stabilizers of the form $E(a_i,b_i)_{\text{B}} \otimes E(a_i,b_i)_{\text{C}}$, which means that the code cannot distinguish the $i$-th qubit of B and the $i$-th qubit of C.
Finally, we make \emph{Bob} perform the aforesaid diagonal Clifford on qubits C, and this produces the solid dark red curve marked `local Clifford by B'.
Clearly, the exponent now approaches the ``fundamental limit'' set by standard QEC on the depolarizing channel.

\section{Implementation Details of Protocol I}
\label{sec:ghz_implementation}

In this appendix we provide additional details regarding the implementation of Protocol I, which is also relevant for Protocol II.
Specifically, we discuss how to identify the appropriate diagonal Clifford to be applied on qubits C, and explain the reasoning behind Algorithm~\ref{algo:logical_paulis_ghz_msmt} that generates logical Pauli operators for any stabilizer code.
The discussion related to Algorithm~\ref{algo:logical_paulis_ghz_msmt} will also clarify some aspects of the distillation protocol.

\subsection{Diagonal Clifford on qubits C}
\label{sec:diagonal_clifford}

A diagonal Clifford unitary on $n$ qubits can be described using an $n \times n$ binary symmetric matrix $R$ as~\cite{Dehaene-physreva03,Rengaswamy-tqe20} 
\begin{align}
U_R = \sum_{v \in \mathbb{F}_2^n} \imath^{vRv^T \bmod 4} \dketbra{v} = \text{diag}\left( \{ \imath^{vRv^T \bmod 4} \}_{v \in \mathbb{F}_2^n} \right).
\end{align}
Its action on a Pauli matrix $E(a,b)$ is given by~\cite{Rengaswamy-pra19,Rengaswamy-tqe20} 
\begin{align}
U_R \, E(a,b) \, U_R^{\dagger} & = E(a, b+aR) \\
  & = E(a, (b \oplus aR) + 2(b \ast aR)) \\
  & = \imath^{2 a (b \ast aR)^T} E(a, b \oplus aR) \\
  & = (-1)^{a (b \ast aR)^T} E(a, b \oplus aR),
\end{align}
where $b \ast aR$ is the entrywise product of the two binary vectors.
It is well-known that any diagonal Clifford operator can be formed using the phase gate, $P$, and the controlled-$Z$ gate, CZ, defined as
\begin{align}
P = 
\begin{bmatrix}
1 & 0 \\
0 & \imath
\end{bmatrix} \quad , \quad
\text{CZ} = 
\begin{bmatrix}
I_2 & 0 \\
0 & Z
\end{bmatrix} =
\begin{bmatrix}
1 & 0 & 0 & 0 \\
0 & 1 & 0 & 0 \\
0 & 0 & 1 & 0 \\
0 & 0 & 0 & -1
\end{bmatrix}.
\end{align}
Set $n=2$.
Then, the $R$ matrices for $P_1 = (P \otimes I), P_2 = (I \otimes P),$ and CZ are respectively 
$$ R_{P_1} = \begin{bmatrix} 1 & 0 \\ 0 & 0 \end{bmatrix}, 
R_{P_2} = \begin{bmatrix} 0 & 0 \\ 0 & 1 \end{bmatrix}, \text{ and }
R_{\text{CZ}} = \begin{bmatrix} 0 & 1 \\ 1 & 0 \end{bmatrix}. $$
This generalizes naturally to more than $2$ qubits.
The diagonal entries of $R$ describe which qubits get acted upon by $P$, and the pairwise off-diagonal entries describe which pairs of qubits get acted upon by CZ~\cite{Rengaswamy-tqe20}.

In our protocol, from Theorem~\ref{thm:ghz_stabilizer_measurement} we observed that Alice's stabilizers of the form $E(a_i,b_i)$, with $a_i \neq 0$, become the stabilizer $E(a_i,0)$ for Charlie.
This also means that any logical $X$ operator of the form $E(c_j,d_j)$ would have transformed into $E(c_j,0)$ (e.g., see last row of Step (2) in Table~\ref{tab:ghz_protocol} and compare it to the last row of Step (3)). 
Therefore, the purpose of the diagonal Clifford on qubits C is to convert the stabilizers $E(a_i,0)$ back into $E(a_i,b_i)$ and the logical $X$ operators $E(c_j,0)$ back into $E(c_j,d_j)$.
Given the above insight into diagonal Clifford operators, we want to find a binary symmetric matrix $R$ such that $U_R E(a_i,0) U_R^\dagger = E(a_i,b_i)$ for all $i=1,2,\ldots,r_X$ (using notation in Algorithm~\ref{algo:logical_paulis_ghz_msmt}) and $U_R E(c_j,0) U_R^\dagger = E(c_j,d_j)$ for all $j=1,2,\ldots,k$.
Thus, we need a feasible solution $R$ for $\{ a_i R = b_i,\ i=1,2,\ldots,r_X \}$ and $\{ c_j R = d_j,\ j=1,2,\ldots,k \}$.

We solve this system of linear equations on a binary symmetric matrix as follows.
First, let $A$ be the matrix whose rows are $\{ a_i \}$ and $\{ c_j \}$, and let $B$ be the matrix whose rows are $\{ b_i \}$ and $\{ d_j \}$.
Then, we have the system $A R = B$.
We recall the vectorization property of matrices, which implies that 
\begin{align}
\text{vec}(QUV) = (V^T \otimes Q) \text{vec}(U).
\end{align}
Here, vectorization of a matrix is the operation of reading the matrix entries columnwise, top to bottom, and forming a vector (e.g., this is done through the command \texttt{U(:)} in MATLAB).
Setting $Q=A, U=R, V=I$, we get $(I \otimes A) \text{vec}(R) = \text{vec}(B)$, which is a standard linear algebra problem for the unknown vector $\text{vec}(R)$.
However, we desire a binary \emph{symmetric} matrix $R$.
We impose this constraint as $(I - W) \text{vec}(R) = 0$, where $0$ denotes the all-zeros vector of length $n^2$, and $W$ is the permutation matrix which transforms $\text{vec}(Q)$ into $\text{vec}(Q^T)$ for any matrix $Q$.
In summary, we obtain the desired $R$ (or equivalently the diagonal Clifford $U_R$) by solving
\begin{align}
\text{find} \quad R \quad \text{s.t.} \quad (I \otimes A) \text{vec}(R) & = \text{vec}(B), \nonumber \\
  (I - W) \text{vec}(R) & = 0 .
\end{align}
Since $R$ is symmetric, it has $n(n+1)/2$ degrees of freedom, which accounts for the second constraint.
The matrix $A$ has $r_X + k < n$ rows and the Kronecker product with $I$ results in $n(r_X + k)$ constraints on $n(n+1)/2$ variables.
It remains to be shown if there is always a feasible solution for any valid $A$ and $B$.
Note that $[A,B]$ represents a matrix whose rows are stabilizers and logical $X$ operators.
This means any pair of rows must be orthogonal with respect to the symplectic inner product, which implies that $AB^T + BA^T = 0$.
Thus, a given $A$ and $B$ is valid if and only if $AB^T$ is symmetric.

\subsection{Logical Paulis from GHZ Measurements}
\label{sec:logical_paulis_ghz_msmt}

The procedure in Algorithm~\ref{algo:logical_paulis_ghz_msmt} to determine logical $X$ and $Z$ generators of a stabilizer code is inspired by the stabilizer measurements on Bell or GHZ states, viewed through the lens of the stabilizer formalism for measurements (Section~\ref{sec:stabilizer_formalism}).
Though the algorithm could have been constructed just using measurements on Bell states, we preferred GHZ states because there can be an additional non-trivial sign for the logical $X$ operators due to an odd number of subsystems in the GHZ state.
Of course, logical operators obtained using GHZ states will also apply to the Bell protocol since a negative sign on an even number of subsystems (A and B in Bell states) leads to an overall positive sign for $\overline{X}_A \overline{X}_B$ and $\overline{Z}_A \overline{Z}_B$.

We have a code $\mathcal{Q}(S)$ defined by its stabilizer group $S = \langle \varepsilon_i E(a_i,b_i) \, ; \, i = 1,2,\ldots,r=n-k \rangle$.
Define the $r \times (2n+1)$ stabilizer (or parity-check) matrix $H'$ whose $i$-th row is $[a_i,b_i,\ \varepsilon_i]$.
First, we bring the first $2n$ columns of the stabilizer (or parity-check) matrix of the code to the following standard form:
\begin{align}
H_{1:2n} = 
\begin{bmatrix}
0 & H_Z \\
H_1 & H_2
\end{bmatrix}.
\end{align}
Here, the $r_Z$ rows of $H_Z$ form all generators for the purely $Z$-type stabilizers of the code.
The bottom part of the matrix is such that the $r_X \times n$ matrix $H_1$ has full rank ($r_X + r_Z = r = n-k$).
While performing row operations on the initial parity-check matrix $H'$, one has to account for the Pauli multiplication rule in Lemma~\ref{lem:Eab}(b), and not simply perform binary sums of (the first $2n$ columns of the) rows, i.e., the last column of $H$ must be updated to reflect changes in signs.

Next, we simulate the creation of $n$ GHZ states by creating a $2n \times (6n+1)$ GHZ stabilizer matrix $S_{\text{GHZ}}$, whose first $n$ rows are $[0,0,0,\ e_i,e_i,0,\ +1]$ and the second $n$ rows are $[e_i,e_i,e_i,\ 0,0,0,\ +1]$.
This matrix is the same as Step (0) of Table~\ref{tab:ghz_protocol}, but we have omitted the middle section since the measurements on subsystem $A$ trivially commute with entries $I_{\text{A}} Z_{\text{B}} Z_{\text{C}}$ of this section.
Now, we use the stabilizer formalism for measurements (Section~\ref{sec:stabilizer_formalism}) to simulate measurements of the rows of $H$ on subsystem A of $S_{\text{GHZ}}$.
Clearly, the stabilizers from $[0,H_Z]$ commute with the first $n$ rows, so these will only replace $r_Z$ rows in the bottom half of $S_{\text{GHZ}}$.
The stabilizers from $[H_1,H_2]$ will necessarily anticommute with at least one of the first $n$ rows of $S_{\text{GHZ}}$, and these $r_X$ rows get replaced.
This can be established by counting the dimension of purely $Z$-type operators with which each row of $[H_1,H_2]$ can commute, one after the other.
Crucially, the stabilizer formalism guarantees that all rows of the evolved $S_{\text{GHZ}}$ remain linearly independent and always commute.

The $(n-r_X)$ non-replaced rows within the first $n$ rows can be divided into two types. 
Before we simulate any stabilizer measurements, the first $n$ rows have standard basis vectors $e_i$ for the $Z$-parts of A and B. 
These can be rewritten such that we have $r_Z$ rows of the form $[0,0,0,\ z,z,0,\ +1]$, where $z$ corresponds to rows of $H_Z$, all of which are linearly independent by assumption of the standard form.
Since these correspond to code stabilizers (on A as well as B), the measurement of rows of $[H_1,H_2]$ will not replace these.
After the measurements, when $r_Z$ rows in the bottom half have been replaced by $[0,0,0,\ z,0,0,\ \varepsilon_z]$, we can multiply with the corresponding rows of the top half, i.e., $[0,0,0,\ z,z,0,\ +1]$, to produce purely $Z$-type stabilizers on subsystem B, which later define Bob's code.
These $Z$-operators on B in the top half form the first type of $r_Z$ rows.
The remaining $(n-r_X)-r_Z = k$ rows form the second type, and they have to form logical $Z_{\text{A}_j} Z_{\text{B}_j}$, for $j=1,2,\ldots,k$, since they commute with all code stabilizers and the columns of subsystem C remain zero.
Thus, the $Z$-component of subsystem A of these $k$ rows produce the logical $Z$ generators of the code, and they always have sign $+1$.

A similar argument applies to the bottom half of the evolved $S_{\text{GHZ}}$ matrix.
The stabilizer measurements from $H_Z$ replace $r_Z$ rows out of the $n$ rows.
The remaining $(n-r_Z)$ rows can again be divided into two types.
The first type of rows give operators that can be rewritten as the BC stabilizers guaranteed by Theorem~\ref{thm:ghz_stabilizer_measurement}.
Specifically, these can be identified by the fact that their A-parts will be linearly dependent on the A-parts of the other rows of the evolved $S_{\text{GHZ}}$ matrix.
Indeed, this is how one can cancel the A-parts of these $r_X$ rows to produce the $r_X$ BC stabilizers corresponding to $[H_1,H_2]$.
The remaining $(n-r_Z)-r_X = k$ rows of the bottom half form the second type, and they have to form logical $X_{\text{A}_j}' X_{\text{B}_j}'$, for $j=1,2,\ldots,k$, since they commute with all code stabilizers and are linearly independent from all other rows.
The A-parts of these rows are used to define the logical $X'$ operators of the code.
The primes on these logical operators indicate that they might not exactly pair up with the corresponding logical $Z$ operators defined earlier.
This is because they are only guaranteed to be logical operators independent of the logical $Z$ operators, but not to be the appropriate pairs $\{ \overline{X}_j \}$ of the previously determined $\{ \overline{Z}_j \}$.

Once these pseudo logical $X$ operators are determined, we can easily find the necessary pairs for the logical $Z$ operators.
Let the logical $Z$ operators be $E(0,f_i), i = 1,2,\ldots,k$, and let these pseudo logical $X$ operators be $\nu_j E(c_j,d_j), j = 1,2,\ldots,k$.
If they are the correct pairs, then we would get $\syminn{[0,f_i]}{[c_j,d_j]} = \delta_{ij}$ for all $i,j \in \{ 1,2,\ldots,k \}$, where $\delta_{ij} = 1$ if $i=j$ and $0$ otherwise.
The symplectic inner product can be expressed as 
$$ \syminn{[0,f_i]}{[c_j,d_j]} = [0,f_i] \, \Omega \, [c_j,d_j]^T,\ \text{where}\ \Omega = \begin{bmatrix} 0 & I_n \\ I_n & 0 \end{bmatrix}. $$
Therefore, if $F$ is the matrix whose rows are $f_i$ and $[C,D]$ is the matrix whose rows are $[c_j,d_j]$, then we need 
$$ [0,F] \, \Omega \, [C,D]^T \eqqcolon T = I_k. $$
If $T \neq I_k$, then we can simply pre-multiply the equation by $T^{-1}$ (mod $2$) to achieve the desired result.
In this case, we redefine the logical $Z$ operators to be given by the rows of $T^{-1}\, [0,F]$.
This completes the reasoning behind Algorithm~\ref{algo:logical_paulis_ghz_msmt}.

\end{document}

%% file: figures/LP118_12_16_20_MSAseqvars80.tex
%
%
\definecolor{mycolor1}{rgb}{1.00000,0.00000,1.00000}%
\begin{tikzpicture}

\begin{axis}[%
width=4.521in,
height=3.563in,
at={(0.758in,0.484in)},
scale only axis,
xmode=log,
xmin=0.01,
xmax=0.108,
xminorticks=true,
xlabel style={font=\color{white!15!black}},
xlabel={Depolarizing Probability},
ymode=log,
ymin=1e-06,
ymax=1,
yminorticks=true,
ylabel style={font=\color{white!15!black}},
ylabel={Logical Error Rate},
axis background/.style={fill=white},
title style={font=\bfseries},
xmajorgrids,
xminorgrids,
ymajorgrids,
yminorgrids,
legend style={at={(0.04,0.8)}, anchor=south west, legend cell align=left, align=left, draw=white!15!black}
]
\addplot [color=blue, mark=o, mark options={solid, blue}]
  table[row sep=crcr]{%
0.01	2e-06\\
0.03	0.000851999216160721\\
0.05	0.0184128153194623\\
0.07	0.132\\
0.09	0.539\\
0.1	0.744\\
0.104	0.832\\
0.108	0.862\\
};
\addlegendentry{$\llbr n=544, k=80, d=12 \rrbr$}

\addplot [color=red, mark=x, mark options={solid, red}]
  table[row sep=crcr]{%
0.01	0\\
0.03	0.000212065239750357\\
0.05	0.00799808046068943\\
0.07	0.0771010023130301\\
0.09	0.473\\
0.1	0.726\\
0.104	0.82\\
0.108	0.868\\
};
\addlegendentry{$\llbr n=714, k=100, d=16 \rrbr$}

\addplot [color=mycolor1, mark=square, mark options={solid, mycolor1}]
  table[row sep=crcr]{%
0.01	0\\
0.03	2.9e-05\\
0.05	0.00169678459319589\\
0.07	0.0364963503649635\\
0.09	0.369\\
0.1	0.695\\
0.104	0.804\\
0.108	0.887\\
};
\addlegendentry{$\llbr n=1020, k=136, d=20 \rrbr$}

\end{axis}

\begin{axis}[%
width=5.833in,
height=4.375in,
at={(0in,0in)},
scale only axis,
xmin=0,
xmax=1,
ymin=0,
ymax=1,
axis line style={draw=none},
ticks=none,
axis x line*=bottom,
axis y line*=left
]
\end{axis}
\end{tikzpicture}%

%% file: figures/LP118_12_16_20_MSAseqvars80_threshold.tex
%
%
\definecolor{mycolor1}{rgb}{1.00000,0.00000,1.00000}%
\begin{tikzpicture}

\begin{axis}[%
width=4.521in,
height=3.566in,
at={(0.758in,0.481in)},
scale only axis,
xmin=0.1,
xmax=0.108,
every minor tick,
  x tick label style={
    /pgf/number format/.cd,
    fixed,
    fixed zerofill,
    precision=3
  },
xlabel style={font=\color{white!15!black}},
xlabel={Depolarizing Probability},
ymin=0.65,
ymax=0.9,
ylabel style={font=\color{white!15!black}},
ylabel={Logical Error Rate},
axis background/.style={fill=white},
title style={font=\bfseries},
xmajorgrids,
ymajorgrids,
legend style={at={(0.04,0.8)}, anchor=south west, legend cell align=left, align=left, draw=white!15!black}
]
\addplot [color=blue, mark=o, mark options={solid, blue}]
  table[row sep=crcr]{%
0.01	2e-06\\
0.03	0.000851999216160721\\
0.05	0.0184128153194623\\
0.07	0.132\\
0.09	0.539\\
0.1	0.744\\
0.104	0.832\\
0.108	0.862\\
};
\addlegendentry{$\llbr n=544, k=80, d=12 \rrbr$}

\addplot [color=red, mark=x, mark options={solid, red}]
  table[row sep=crcr]{%
0.01	0\\
0.03	0.000212065239750357\\
0.05	0.00799808046068943\\
0.07	0.0771010023130301\\
0.09	0.473\\
0.1	0.726\\
0.104	0.82\\
0.108	0.868\\
};
\addlegendentry{$\llbr n=714, k=100, d=16 \rrbr$}

\addplot [color=mycolor1, mark=square, mark options={solid, mycolor1}]
  table[row sep=crcr]{%
0.01	0\\
0.03	2.9e-05\\
0.05	0.00169678459319589\\
0.07	0.0364963503649635\\
0.09	0.369\\
0.1	0.695\\
0.104	0.804\\
0.108	0.887\\
};
\addlegendentry{$\llbr n=1020, k=136, d=20 \rrbr$}

\end{axis}

\begin{axis}[%
width=5.833in,
height=4.375in,
at={(0in,0in)},
scale only axis,
xmin=0,
xmax=1,
ymin=0,
ymax=1,
axis line style={draw=none},
ticks=none,
axis x line*=bottom,
axis y line*=left
]
\end{axis}
\end{tikzpicture}%

%% file: figures/GHZsimple_LP118_12_16_20_MSAseqvars80.tex
%
%
\definecolor{mycolor1}{rgb}{1.00000,0.00000,1.00000}%
\begin{tikzpicture}

\begin{axis}[%
width=4.521in,
height=3.566in,
at={(0.758in,0.481in)},
scale only axis,
xmin=0.09,
xmax=0.11,
xminorticks=true,
every minor tick,
  x tick label style={
    /pgf/number format/.cd,
    fixed,
    fixed zerofill,
    precision=3
  },
xlabel style={font=\color{white!15!black}},
xlabel={Depolarizing Probability},
ymin=0.591016548463357,
ymax=0.990967741935484,
yminorticks=true,
ylabel style={font=\color{white!15!black}},
ylabel={GHZ Failure Rate},
axis background/.style={fill=white},
title style={font=\bfseries},
xmajorgrids,
xminorgrids,
ymajorgrids,
yminorgrids,
legend style={at={(0.04,0.8)}, anchor=south west, legend cell align=left, align=left, draw=white!15!black}
]
\addplot [color=blue, mark=o, mark options={solid, blue}]
  table[row sep=crcr]{%
0.09	0.779423226812159\\
0.1	0.938262338149747\\
0.104	0.967941888619855\\
0.108	0.983349753694581\\
0.11	0.989306930693069\\
};
\addlegendentry{$\llbr n=544, k=80, d=12 \rrbr$}

\addplot [color=red, mark=x, mark options={solid, red}]
  table[row sep=crcr]{%
0.09	0.706214689265537\\
0.1	0.925583117363939\\
0.104	0.96463768115942\\
0.108	0.983349753694581\\
0.11	0.989009900990099\\
};
\addlegendentry{$\llbr n=714, k=100, d=16 \rrbr$}

\addplot [color=mycolor1, mark=square, mark options={solid, mycolor1}]
  table[row sep=crcr]{%
0.09	0.591016548463357\\
0.1	0.90481360839667\\
0.104	0.959903846153846\\
0.108	0.98512315270936\\
0.11	0.990967741935484\\
};
\addlegendentry{$\llbr n=1020, k=136, d=20 \rrbr$}

\end{axis}

\begin{axis}[%
width=5.833in,
height=4.375in,
at={(0in,0in)},
scale only axis,
xmin=0,
xmax=1,
ymin=0,
ymax=1,
axis line style={draw=none},
ticks=none,
axis x line*=bottom,
axis y line*=left
]
\end{axis}
\end{tikzpicture}%

%% file: figures/GHZsimple_LP118_12_16_20_MSAseqvars80_threshold.tex
%
%
\definecolor{mycolor1}{rgb}{1.00000,0.00000,1.00000}%
\begin{tikzpicture}

\begin{axis}[%
width=4.521in,
height=3.566in,
at={(0.758in,0.481in)},
scale only axis,
xmin=0.1,
xmax=0.11,
every minor tick,
  x tick label style={
    /pgf/number format/.cd,
    fixed,
    fixed zerofill,
    precision=3
  },
xlabel style={font=\color{white!15!black}},
xlabel={Depolarizing Probability},
ymin=0.9,
ymax=1,
ylabel style={font=\color{white!15!black}},
ylabel={GHZ Failure Rate},
axis background/.style={fill=white},
title style={font=\bfseries},
xmajorgrids,
ymajorgrids,
legend style={at={(0.04,0.8)}, anchor=south west, legend cell align=left, align=left, draw=white!15!black}
]
\addplot [color=blue, mark=o, mark options={solid, blue}]
  table[row sep=crcr]{%
0.09	0.779423226812159\\
0.1	0.938262338149747\\
0.104	0.967941888619855\\
0.108	0.983349753694581\\
0.11	0.989306930693069\\
};
\addlegendentry{$\llbr n=544, k=80, d=12 \rrbr$}

\addplot [color=red, mark=x, mark options={solid, red}]
  table[row sep=crcr]{%
0.09	0.706214689265537\\
0.1	0.925583117363939\\
0.104	0.96463768115942\\
0.108	0.983349753694581\\
0.11	0.989009900990099\\
};
\addlegendentry{$\llbr n=714, k=100, d=16 \rrbr$}

\addplot [color=mycolor1, mark=square, mark options={solid, mycolor1}]
  table[row sep=crcr]{%
0.09	0.591016548463357\\
0.1	0.90481360839667\\
0.104	0.959903846153846\\
0.108	0.98512315270936\\
0.11	0.990967741935484\\
};
\addlegendentry{$\llbr n=1020, k=136, d=20 \rrbr$}

\end{axis}

\begin{axis}[%
width=5.833in,
height=4.375in,
at={(0in,0in)},
scale only axis,
xmin=0,
xmax=1,
ymin=0,
ymax=1,
axis line style={draw=none},
ticks=none,
axis x line*=bottom,
axis y line*=left
]
\end{axis}
\end{tikzpicture}%